\documentclass[10pt]{asme2ej}

\usepackage{epsfig} 
\usepackage{graphicx}
\usepackage{caption}
\usepackage{subcaption}
\usepackage{multirow}
\usepackage{multicol}
\usepackage{etoolbox}
\usepackage{setspace}
\usepackage{lscape}
\usepackage{amssymb,mathtools}
 
\usepackage[nocfg]{nomencl}
\usepackage{nomencl}
\usepackage{ifthen}
\renewcommand\nomgroup[1]{%
  \ifthenelse{\equal{#1}{A}}{%
    \item[\textbf{Acronyms}]}{
  \ifthenelse{\equal{#1}{R}}{%
    \item[\textbf{Roman Symbols}]}{
  \ifthenelse{\equal{#1}{G}}{%
    \item[\textbf{Greek Symbols}]}{
  \ifthenelse{\equal{#1}{S}}{%
    \item[\textbf{Superscripts}]}{
  \ifthenelse{\equal{#1}{U}}{%
    \item[\textbf{Subscripts}]}{
  \ifthenelse{\equal{#1}{X}}{%
    \item[\textbf{Other Symbols}]}{
  {}}}}}}}}

\makenomenclature
\makenomenclature

\usepackage[utf8]{inputenc}
\title{Effects of Semitransparent Window's Aspect Ratio on Interaction of Collimated Beam with Natural 
Convection: Part I
}
\author{G.Chanakya
}
\author{Pradeep Kumar\thanks{pradeepkumar@iitmandi.ac.in}
}
\author{
    \affiliation{Numerical Experiment Laboratory \\ 
	(Radiation \& Fluid Flow Physics)\\
	School of Engineering\\
	Indian Institute of Technology Mandi\\
	Mandi, Himachal Pradesh, India 175075\\
    chandrasekhara\_pratap@students.iitmandi.ac.in
    }
}

\begin{document}

\maketitle   

\begin{abstract}
{\it 
The effects of the semitransparent winodw's aspect ratio on the interaction of the collimated beam with natural convection have been investigated numerically in the present work. The combination of geometrical parameters of the semitransparent window, i.e., height ratio ($h_r$) and window width ratio ($w_r$) and Planck numbers of the medium have been considered. The other parameters, like flow parameter (Ra$=10^5$), fluid parameter (Pr=0.71), thermal parameter (N), Irradiation (G=1000 $W/m^2$), Angle of incidence ($\phi=135^0$) and geometrical parameter of the geometry ($A_r$=1) and the wall conditions have been kept constant. A collimated beam is irradiated with irradiation value (G=1000 $W/m^2$) on the semitransparent window at an azimuthal angle ($\phi) 135^0$. The cavity is convectively heated from the bottom with heat transfer coefficient 50 $W/m^2 K$ and free stream temperature 305 $K$. A semitransparent window is created on the left wall and isothermal conditions (T=296 $K$) is applied on the semitransparent, left and right vertical walls, wherein adiabatic conditions are applied on upper wall of the cavity. The dynamics of two vortices inside the cavity change considerably by combinations these semitransparent window's aspect ratio and Planck number (Pl) of the medium. The left vortex breaks into two parts and remains confined in upper and lower left corners for some combination of aspect ratios and Planck numbers of the medium. The thermal plume flickers depending on the situation of dynamics of two vortices inside the cavity. The localized hating of the fluid happens mostly for large height ratio of semitransparent window.  The conduction; radiation and total Nusselt number are also greatly affected by the semitransparent window's aspect ratio and the Planck number of the medium.\\
 
 \it
\textbf{Keywords: Semitransparent wall; Natural convection; Collimated beam; Symmetrical cooling; Aspect Ratio; Irradiation; } 
 
}
\end{abstract}

\nomenclature[a]{$a$}{Co-efficient}
    \nomenclature[a]{$C_{p}$}{Specific heat capacity ($J/kg-K$)}
    \nomenclature[a]{$g$}{Acceleration due to gravity ($m/s^2$)}
    \nomenclature[a]{$G$}{Irradiation ($W/m^2$)}
    \nomenclature[a]{$H$}{Height ($m$)}
    \nomenclature[a]{$I$}{Intensity ($W/m^2$)}
    \nomenclature[a]{$I_{b}$}{Black body intensity ($W/m^2$)}
    \nomenclature[a]{$k$}{Thermal conductivity ($W/mK$)}
    \nomenclature[a]{$L$}{Length of the domain of study ($m$)}
    \nomenclature[a]{$N$}{Conduction-radiation parameter}
    \nomenclature[a]{$Nu$}{Nusselt number}
    \nomenclature[a]{$p$}{Pressure ($N/m^2$)}
    \nomenclature[a]{$Pl$}{Planck number}
    \nomenclature[a]{$Pr$}{Prandtl number}
	\nomenclature[a]{$q$}{Flux ($W/m^2$)}
	\nomenclature[a]{$Ra$}{Rayleigh number}
	\nomenclature[a]{$U$}{Velocity ($m$)}
    \nomenclature[g]{$\beta_{T}$}{Thermal expansion coefficient ($1/K$)}
    \nomenclature[g]{$\epsilon$}{Emissivity}
	\nomenclature[g]{$\kappa_{a}$}{Absorption coefficient ($1/m$)} 
	\nomenclature[g]{$\rho$}{Density of the fluid ($kg/m^3$)}
    \nomenclature[g]{$\tau$}{Optical thickness}
    \nomenclature[g]{$\theta$}{Polar angle}
	\nomenclature[g]{$\phi$}{Azimuthal angle}
    \nomenclature[U]{$C$}{Conduction}	
	\nomenclature[U]{$c$}{Cold wall}
	\nomenclature[U]{$co$}{collimated beam}
	\nomenclature[U]{$conv$}{Convection}
	\nomenclature[U]{$f$}{Face centre}
	\nomenclature[U]{$free$}{Free stream}
	\nomenclature[U]{$i,j$}{Tensor indices}
	\nomenclature[U]{$nb$}{Neighbour cell}
	\nomenclature[U]{$p$}{Cell centre}		
	\nomenclature[U]{$R$}{Radiation}
	\nomenclature[U]{$ref$}{Reference}
	\nomenclature[U]{$t$}{Total}
	\nomenclature[U]{$w$}{Wall}
    
\printnomenclature

\section{Introduction}
Natural convection flows in square/rectangular enclosures have been extensively investigated by many experimental and numerical studies due to its wide variety of engineering applications like, nuclear reactors, cooling electronic equipments and heat exchangers, HVAC system, Building energy management etc. The flow in these applications are majorly buoyancy-driven. The buoyancy-driven flow exhibits a complex fluid flow phenomenon which mainly depends on the temperature gradient, the aspect ratios, geometric shapes and sizes like square/circular/rectangular/triangular and position of energy source inside the enclosures. Though, the practical applications have complex geometries but, most of the complex fluid flow and heat transfer phenomenon of natural convection can be understood in simplified geometries like square, rectangular etc. 

Two heated cylindrical geometries (elliptical and cylindrical) inside a square enclosure with different aspect ratios (0.25 to 4.00) were simulated for natural convection for the range of Rayleigh numbers $10^4$ to $10^6$ using immersed boundary method by Cho et al. \cite{Cho1, Cho2}. A second order accurate central difference scheme and fractional time step method were used to simulate the flow field. The advection and diffusion terms were treated by Adams-Bashforth second order scheme and the Cranck-Nicolson scheme, respectively. The authors observed the formation of secondary vortices above the top cylinder for Rayleigh number $10^5$ due to increase in the convection rate. The flow and the thermal fields were unsteady state for Rayleigh number (Ra=$10^6$). 

Cheong et al. \cite{Cheong} numerically studied the natural convection in an inclined enclosure with sinusoidal temperature profile on the left wall and isothermal condition on right wall, whereas the top and bottom walls were adiabatic. They found that the convection was dominated for the aspect ratio ($A_r$) of 0.25 $\leq$ $A_r$ $\leq$ 5 and the heat transfer was mostly by conduction for the enclosure for $A_r$=10 for all Rayleigh number range under study. Yigit et al. \cite{Yigit} has investigated the effect of enclosure aspect ratio of a cavity which was heated from the bottom, on the yield stress for Bingham fluid in natural convection case. It was reported that the number of convective vortices formation were greatly affected by aspect ratio  of the geometry. A comprehensive review on the natural convection in different non-square enclosures and practical geometries for engineering applications were complied by Das et al. \cite{Das} and Rahimi et al. \cite{Rahimi}, respectively.

Webb and Viskanta \cite{Webb} have performed an experiment to observe the distribution of the temperature and the flow field of water inside a rectangular enclosures that subjected to radiation flux on a wall and rest walls were kept adiabatic. Their experimental results showed that the formation of thin hydrodynamic boundary layer at the vertical walls, stagnant and stably stratified temperature field at core of the cavity. A numerical study have been reported for the interaction of the thermal radiation with buoyancy driven flows in different geometrical configurations with heated cylinder at its centre in \cite{Mezrhab,Hua,Mukul}. It was reported that the radiation exchange homogenized the temperature field inside the cavity. The average Nusselt numbers for the square and the triangular geometries are higher compared to the cylindrical geometry. The effect of the orientation of the cavity on the combined modes of heat transfer was studied in \cite{Kumar}. The flow and the temperature field were significantly altered by the radiation. The conducted Nusslet number increased with increase of optical thickness, whereas the radiative and the total Nusselt number have decreased. Sun et al. \cite{Yujia} had investigated the performance of P$_{1}$, SP$_{3}$, P$_{3}$, Discreate ordinate method (DOM) and finite volume method (FVM) for the two-dimensional combined natural convection with the radiation for an absorbing emitting medium and compared with the results by the monte carlo method. The effects of Rayleigh numbers, Planck numbers, and optical thickness on the flow field and the heat transfer rates were analyzed in details. It was reported that the FVM took twice CPU time than DOM. The SP$_{3}$ and P$_{3}$ produced high accuracy than FVM for optically thick medium but they could predict the radiative heat flux accurately because of oscillating behaviour of convergences at lower Planck numbers.

Mondal and Mishra \cite{Bittagopal} have used lattice Boltzman method (LBM) to simulate natural convection in a square cavity coupled with radiation to study the effects of various parameters such as the extinction coefficient and the scattering albedo on the flow filed and the temperature distributions inside the cavity. The FVM was used for the RTE. It was observed that flow field was symmetric and scattering coefficient had no much significant effect. The extinction coefficient had a pronounced effect on the temperature distributions. A coupled numerical investigation on natural convection with volumetric radiation with gray and isotropic scattering medium in two-dimensional rectangular cavity were analysed for Planck numbers, scattering albedo of the medium for various tilt angles and the aspect ratios of the cavity in \cite{Fu}. They observed that emissivity of the horizontal wall and the scattering albedo have significant effect on the flow and temperature patterns. The heat transfer has decreased with increase in the scattering albedo. Hakan and Derbentil \cite{Hakan} investigated the combined natural convection with radiation for the rectangular enclosure with different aspect ratios. They also proposed the correlations for the mean values Nusselt number and observed that the mean Nusselt number has increased from for higher aspect ratio of the cavity.

Nia and Nassab \cite{Nia} have made a numerical study of the natural convection due to the temperature and the concentration gradients in a square cavity with radiation for the range of optical thickness of the medium. It was stated that the optical thicknesses had affected the thermal and the mass transfer in the cavity. They also stated that thermal field reached to steady state faster than the concentration field. To get a better insight of the flow and the concentrations field, the time evolution of the isotherm, stream and iso-concentration lines were presented for $Ra=10^4$ and optical thickness 10. A transient simulations of the effect of the solar radiation on the reservoir sidearm has been performed by Lei et al. \cite{Lei}. Three distinct regimes 1) initial stage was dominated by the conduction at the bottom, 2) a transitional stage, where circulation was established and the presence of instabilities and 3) and quasi-steady state, have been observed. Ming and Zang \cite{Ming} implemented modified finite volume method for the combined natural convection and the radiation for the hybrid grids and validated their solver with the benchmark cases.

Wang et al. \cite{Wang1} have compared the first and second order formulations of the radiative transfer equation (RTE) by diffuse approximation meshless (DAM) method without using upwinding treatment for interpolation. Two and three dimensional geometries were considered to investigate the accuracy and the computational resource utilization. It was observed that first order formulation of RTE is faster than the second order.They used the moving least square meshless method to investigate the accuracy for coupled natural convection with radiation in semitransparent medium and considered the vorticity-stream function formulation and vorticity-vector potential formulation for 2-D and 3-D geometries, respectively,  and the discrete ordinate method to solve the radiative transfer equation. Their results showed that the moving least square meshless method was stable and accurate to deal with the natural convection coupled with the radiation.

In all the above works either pure natural convection or combined natural convection with diffuse radiation was considered, however a little work is available on the collimated beam radiation like, Discrete transfer method (DTM) was used to solve the radiative transfer equation in varying refractive index in participating media by Ben and Dez \cite{Ben}. Anand and Mishra \cite{Anand} used DTM to solve RTE in participating media and derived the exact radiative flux field expression for the linearly varying refractive index. Ilyushin \cite{Ya} studied numerically the collimated beam in the refractive index medium.

Rath and Mahapatra \cite{Rath} formulated the transient divergence of heat flux by solving the transient radiative transfer equation coupled with energy equation by using finite volume method They observed that the transient divergence of radiative flux predicted the temperature of the medium accurately than the conventional steady, where the conventional divergence of flux underpredicted the temperature field. The thermal damage in the proximity of laser source was evaluated by using transient radiation transfer equation by Anil et al. \cite{Anil}. The presence of absorbing, emitting and anisotropically scattering medium within a two-dimensional rectangular domain have been considered. A collimated beam is irradiated on top wall over a small width whereas remaining top, left, right, and bottom walls were maintained at the constant temperature. The collimated beam feature was only applied on the wall whereas diffuse radiation treatment was done inside the cavity.
Recently, a collimated beam feature has been developed in OpenFOAM by Chanakya and Kumar \cite{chanakya2020effects} and its effects on the natural convection also have been investigated. They further investigated the thermal adiabatic boundary condition \cite{chanakya2020investigation} on the semitransparent wall of the cavity.

From the above literature it is noticed that there is no significant work on the collimated beam radiation inside the cavity available to the best of author's knowledge at present. The collimated beam has wide range of application like, laser treatment, solar cavity receiver, laser solidification and melting, illumination from the head lamp of a car, etc. In above application, the collimated beam travel through a optical window which is also known as semitransparent window. The effects of semitransparent windows aspect ratio on the interaction of the collimated beam with natural convection have been studied numerically in OpenFOAM framework in the present work.

This paper is outlined as follows: the problem statement is defined in section 2, followed by mathematical modeling and numerical scheme in section 3. The verification and independent test for grids, respectively, are explained in section 4 and 5. Section 6 elaborates the results and discussion for Planck numbers and all aspect ratios.  Finally, the conclusions of this numerical study are provided in section 7. 

\section{Problem Description}

\begin{figure}[!t]
    \centering
    \includegraphics[width=7cm]{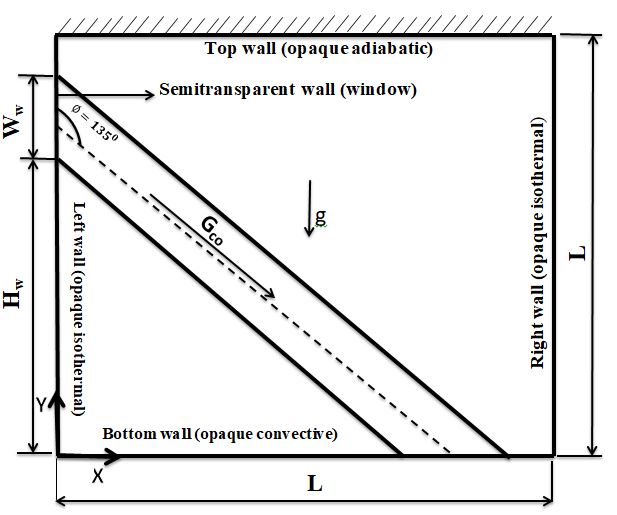}
    \caption{The schematic diagram of the present problem where a collimated beam is incidented on semitransparent window at an angle $135^0$}
    \label{semi_prob}
\end{figure}

Consider the buoyancy driven flow of Newtonian fluid within the square enclosure of as depicted in Fig. \ref{semi_prob}. The enclosures bottom wall is subjected to convective boundary with free stream temperature 305 K and heat transfer coefficient 50 $W/m^2 K$, top wall is taken as adiabatic, the right and the left walls are subjected to isothermal (296 K) boundary condition.  The Euclidean co-ordinate axes are along the bottom and the left vertical walls of the enclosure and the origin is at the junction of these two walls. The acceleration due to gravity acts vertically downward direction (negative direction). All walls of enclosure are treated as opaque with emissivity 0.9 for radiation expect semitransparent window on the left wall where collimated beam is applied. The four cases of height aspect ratios as height ratio ($h_r$=$H_{w}/L$) and window width ratio ($w_r$=$W_{w}/L$) have been considered below.

\hspace{-7mm} case A: $h_r$ = 0.8 and $w_r$ = 0.2\\
case B: $h_r$ = 0.8 and  $w_r$ = 0.4\\
case C: $h_r$ = 0.4 and  $w_r$ = 0.2\\
case D: $h_r$ = 0.4 and  $w_r$ = 0.4\\
In all the cases (i.e A, B, C and D) a collimated beam of irradiation value 1000 $W/m^2$ is applied on the semitransparent window at an angle of $135^0$. The semitransparent window is also isothermal with temperature 296 K. The simulations are carried out for the constant flow parameter (Ra=$10^5$) and fluid parameter (Pr=0.71), thermal parameter (Pl=0, 1, 10 and 50) and various semitransparent window's aspect ratios. 

\section{Mathematical formulation and Numerical procedures}

The following assumptions have been considered for the mathematical modeling of the above problem 
\begin{enumerate}
    \item Flow is two-dimensional, steady, laminar and incompressible.
    \item Flow is driven by buoyancy force that is modeled by Boussinesq approximation.
    \item The thermophysical properties of the fluid are constant.
    \item The fluid medium absorbs and emits but does not scatters the radiation energy.
   \item The transmissivity of the semitransparent window is one for the incoming radiation and zero for the other walls.
\end{enumerate}
Based on the above assumptions, the governing equations in the Cartesian coordinate system are given as 
\vspace{-0.5cm}
\begin{equation} \label{mass:equN} 
\frac{\partial u_i}{\partial x_i} = 0  
\end{equation}
\vspace{-2cm}
\begin{equation} \label{momentum}
\frac{\partial  u_i u_j}{\partial x_j}=-\frac{1}{\rho}\frac{\partial p}{\partial x_i} + \nu\frac{\partial^2u_i}{\partial x_j\partial x_j}+g \beta_{T}(T-T_{c})\delta_{i2}  
\end{equation}
\vspace{-2cm}
\begin{equation}\label{energy}
\frac{\partial u_jT}{\partial x_j} = \frac{k}{\rho C_p}\frac{\partial^2T}{\partial x_j\partial x_j} - \frac{1}{\rho C_p} \frac{\partial q_{R}}{\partial x_i} 
\end{equation}
where $u$ is velocity, $p$ is pressure, $\rho$ is density, $\beta_{T}$ is thermal expansion coefficient, $g$ is gravity, $c_{p}$ is specific heat capacity at constant pressure, $\kappa$ is thermal conductivity of fluid. $i, j$ are tensor indices which vary 1-3 in Cartesian co-ordinates system. The $\delta_{i2}$ is Kronecker delta and given as
\begin{equation*}
  \delta_{i2}=
    \begin{cases}
            0, &         \text{if} \text{\hspace{0.2cm}} i\neq 2,\\
            1, &         \text{if} \text{\hspace{0.2cm}} i=2.
    \end{cases}  
\end{equation*}
The $\frac{\partial q_{R_{i}}}{\partial x_{i}}$  in eq (\ref{energy}) is
the divergence of radiative flux which is calculated as
\vspace{-0.7cm}
\begin{eqnarray}
\frac{\partial q_{R_{i}}}{\partial x_{i}}=\kappa_{a}(4\pi I_b-G)\label{div_eq}
\end{eqnarray}
where $\kappa_{a}$ is the absorption coefficient, $I_b$ is the black body intensity and $G$ is the irradiation which is evaluated by integrating the radiative intensity ($I$) in all directions, i.e.,
\begin{eqnarray}
G=\int_{4\pi} I d\Omega
\end{eqnarray}
The intensity field inside the cavity can be obtained by solving the following radiative transfer equation (RTE) 
\vspace{-0.6cm}
\begin{equation} 
\label{equN:radiation_1}
\frac{\partial I(\bf \hat{r},\bf \hat{s})}{\partial s}=\kappa_{a} (I_{b}(\bf\hat{r},\bf \hat{s})- \text{I} (\bf\hat{r},\bf \hat{s}))
\end{equation}
where $\bf \hat{r}$ and $\bf \hat{s}$ are position and direction vectors, respectively, and s is path length in the beam direction.\\

The non-dimensional form of equations (\ref{mass:equN})-(\ref{energy}) are
\begin{equation} \label{NonMasseqn} 
\frac{\partial U_i}{\partial X_i} = 0  
\end{equation}
\vspace{-1.5cm}
\begin{equation} \label{NonMomentumeqn}
\frac{\partial  U_i U_j}{\partial X_j}=-\frac{\partial P}{\partial X_i} + \sqrt{\frac{Pr}{Ra}}\frac{\partial^2U_i}{\partial X_j\partial X_j}+ \theta \delta_{i2}  
\end{equation}
\vspace{-2cm}
\begin{equation}\label{Nonenergyeqn}
\frac{\partial U_j \theta}{\partial X_j} =\sqrt{\frac{1}{Ra.Pr}}\frac{\partial^2 \theta}{\partial X_j\partial X_j} - \frac{1}{ N \sqrt{Ra.Pr}} \frac{\partial q^{*}_{R}}{\partial X_i} 
\end{equation}

The non-dimensional quantities and parameters involved in the above equations are as follows,
\begin{eqnarray*}
U_{i} =\frac{u_{i}}{u_{o}},  \hspace{0.2cm}  \hspace{0.2cm} X_{i} =\frac{x_{i}}{L}  
\end{eqnarray*}
\vspace{-2cm}
\begin{eqnarray*}
\theta =\frac{T-T_{c}}{T_{free}-T_{c}}, \hspace{0.2cm} Ra=\frac{g \beta_{T} (T_{free}-T_{c})L^{3}}{\nu \alpha}, \hspace{0.2cm}  Pr = \frac{\nu}{\alpha} 
\end{eqnarray*}
\vspace{-2cm}
\begin{eqnarray*}
N=\frac{\kappa }{\sigma T_{free}^{3} L}, \hspace{1cm} \tau=\kappa_{a} L, \hspace{1cm} Pl=N\tau, \hspace{1cm} q^{*}_R=\frac{q_{r}}{\sigma T_{free}^{4}}
\end{eqnarray*}

The scales for length, velocity, temperature, conductive and radiative fluxes are L, u$_o$, (T$_{free}$-T$_{c}$), $\kappa$(T$_{free}$-T$_{c}$)/L and $\sigma$T$_{free}^{4}$ respectively, where $u_{o}=\sqrt {L g \beta_{T} (T_{free}-T_{c})}$ is convective velocity scale. $N$ is the conduction-radiation parameter, $Pl$ is the Planck number and $\tau$ is the optical thickness of the medium.

The non-dimensional irradiation is given as
\begin{eqnarray}
\overline{G}=\frac{G}{\sigma T^{4}_{free}} \hspace{1cm}
\end{eqnarray}

the radiative transfer equation is always solved in dimensional form, this may be due to fact that radiation quanties depends on absolute values of temperature, irradiation rather than scaled values.

The Navier-Stokes equation (\ref{momentum}), and temperature equations (\ref{energy}) are subjected to following boundary conditions 
\begin{enumerate}
 \item[]\textit{Flow boundary condition}
    \item[] Cavity walls: \hspace{0.2cm} $u_i$ $ = 0$
 \item[] \textit{Thermal boundary conditions}
    \item[1.] Left wall  (a) Semitransparent window: Isothermal at T=296 K 
  \newline  \text{\hspace{0.75cm}}  (b) Left wall: Isothermal at
  \newline \text{\hspace{4.5cm}} T=296K
    \item[2.] Right wall at \textit{x}=1:\hspace{0.2cm} T=296K
    \item[3.] Bottom wall \textit{y}=0: \hspace{0.2cm} $q_{conv}= h_{free} (T_{free}-T_{w})$
    \item[4.] Top wall at \textit{y}=1: \hspace{0.2cm} $q_{c}+q_{r}=0$
\item[]where $q_c=-k \frac{\partial T}{\partial n}$ and \\
$ q_{r}=\int_{4\pi}I({\bf r_{w}}, \bf \hat{s})({\bf \hat{n}\cdot \hat{s}}) \mathrm{d}\Omega $
\end{enumerate}

\begin{figure}[t]
 \begin{subfigure}{8cm}
    \centering\includegraphics[width=5cm]{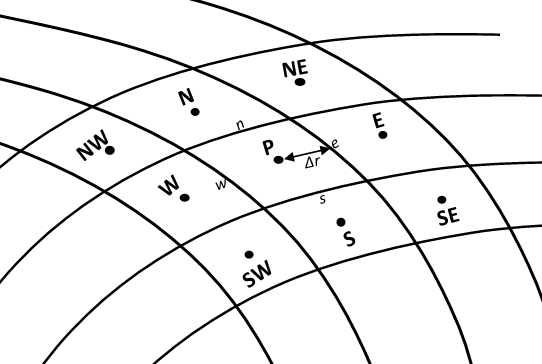}
    \caption{}
    \label{2d_fvm}
  \end{subfigure}
   \begin{subfigure}{8cm}
    \centering\includegraphics[width=5cm]{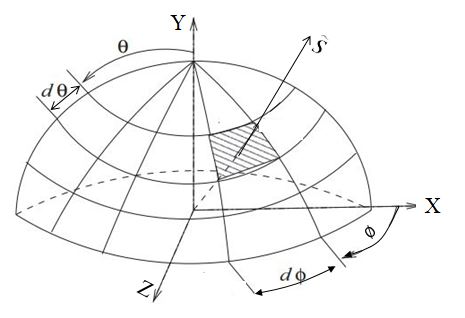}
    \caption{ }
    \label{angular}
  \end{subfigure}
  \caption{The pictorial representation of (a) cell arrangement for finite volume method for partial differential equations and (b) Angular discretization for the radiative transfer equation}
\label{diffuse}
\end{figure}

The radiative transfer equation (\ref{equN:radiation_1}) is subjected to following boundary condition on all the walls except semitransparent window

\begin{eqnarray}
\noindent I({\bf r_w,\hat{s}})=\epsilon_w I_b({\bf r_w})+\frac{1-\epsilon_w}{\pi}\int_{\bf \hat{n}\cdot\hat{s}>0}I({\bf r_w,\hat{s}})|{\bf \hat{n}\cdot\hat{s}}|\mathrm{d}\Omega.\nonumber\\
\mbox{for}~~{\bf \hat{n}\cdot\hat{s}}<0 \label{rte:bound2}
\end{eqnarray}

where $\bf\hat{n}$ is the unit area surface normal and the $\epsilon$ is emissivity of the walls and considered be 0.9 for present study.

The semitransparent window is subjected to collimated irradiation $(G_{co})$ of value 1000 $W/m^2$. The boundary condition for RTE on the semitransparent window is 
\begin{figure}[b]
\centering
 \begin{subfigure}{5cm}
    \includegraphics[width=4cm]{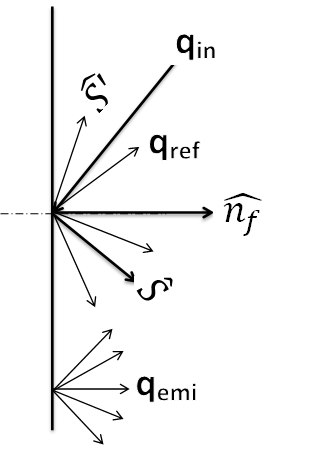}
    \caption{}
    \label{Diff_BC}
  \end{subfigure}
   \begin{subfigure}{3cm}
   \hspace{0cm}
    \includegraphics[width=4cm]{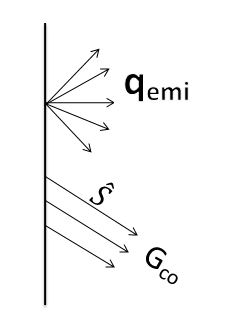}
    \caption{}
    \label{semi_colli}
  \end{subfigure}
  \caption{The pictorial representation of (a) Diffuse reflection of an incident ray and diffuse emission due to wall temperature (b) Diffuse emission and collimated transmission from a semitransparent wall}
\label{Colli_diffuseBC}
\end{figure}

\begin{eqnarray}
\noindent I({\bf r_w,\hat{s}})=I_{co}({\bf r_w,\hat{s}}) \delta (\phi-135^{0})+\epsilon_w I_b({\bf r_w})\\
+\frac{1-\epsilon_w}{\pi}\int_{\bf \hat{n}\cdot\hat{s}>0}I({\bf r_w,\hat{s}})|{\bf \hat{n}\cdot\hat{s}}|\mathrm{d}\Omega.\nonumber\\
\mbox{for}~~{\bf \hat{n}\cdot\hat{s}}<0 \nonumber \label{rte:bound3}
\end{eqnarray}

where $\delta (\phi-135^{0})$ is Dirac-delta function, and defined as
\begin{equation*}
  \delta(\phi-135^{0})=
    \begin{cases}
            1, &         \text{if} \text{\hspace{0.2cm}} \phi=135^{0},\\
            0, &         \text{if} \text{\hspace{0.2cm}} \theta \neq 135^{0}.
    \end{cases}  
\end{equation*}

$I_{co}$ is intensity of collimated irradiation and calculated from the irradiation value as below
\begin{equation}
    I_{co}=\frac{G_{co}}{\mathrm{d}\Omega}
\end{equation}
where $d\Omega$ is the collimated beam width. 

In the current work, the solid angle of discretized angular space (Fig. \ref{angular}) in collimated direction is considered as beam width. The pictorial representation of  diffuse emission, reflection and collimated beam radiation from the wall are shown in Fig. \ref{Colli_diffuseBC}. 

The collimated feature has been developed in OpenFOAM framework-an open source software, and coupled with the other fluid flow and heat transfer libraries. The combined application has been used for numerical simulation of the present problem. 

The OpenFOAM uses the finite volume method (FVM) to solve the Navier-Stokes (eq. \ref{momentum}) and the energy equations (eq. \ref{energy}). The FVM integrates an partial differential equation over a control volume (Fig. \ref{2d_fvm}) to convert the partial differential equation into a set of algebraic equations of the form
\begin{equation}
    a_{p}\phi_{p}= \sum_{nb}a_{nb}\phi_{nb}+S
\end{equation}
where $\phi_{p}$ is any scalar, $a_{p}$ is central coefficient, $a_{nb}$ coefficients of neighbouring cells and S is the source values. Whereas, RTE (eq (\ref{equN:radiation_1})) is converted into a set of algebraic equations by double integration over a control volume (Fig. \ref{2d_fvm}) and over a control angle (Fig. \ref{angular}) . The set of above algebraic equations are solved by preconditioned bi-conjugate gradient (PBiCG) and the details of the algorithm and its implementation in OpenFOAM can be found in the book by Patankar \cite{patankar} and Moukalled \cite{Moukalled}, respectively. \\

In the present simulation, linear upwind scheme which is second order accurate has been used for interpolate face centered values for the cell centred values. The linear upwind scheme is given mathematically as

\begin{equation}
   \phi_{f}=
    \begin{cases}
            \phi_{p}+ \nabla\phi_{p} \cdot \vec{r}, &         \text{if} f_{\phi} > 0,\\
            \phi_{nb}+ \nabla\phi_{nb} \cdot \vec{r}, &         \text{if} f_{\phi} < 0.
    \end{cases}   
\end{equation}
where $f_{\phi}$ is the flux of the scalar $\phi$ on a face (Fig \ref{2d_fvm}), and $p, nb$ (includes E, W, N, S, NE, NW, SE and SW i.e., East, West, North, South, North-East, North-West, South-East and South-West cells) indicate present and neighboring cells and $f$ indicates face value of scalar.

The conductive and radiative fluxes on the walls are converted into Nusselt number as

\begin{eqnarray*}
Nu_{C}=\frac{q_{Cw}L}{k(T_{free}-T_{c})}, \hspace{1cm} Nu_{R}=\frac{q_{Rw}L}{k(T_{free}-T_{c})} \hspace{1cm} 
\end{eqnarray*}
where, Nu$_{C}$ and Nu$_{R}$, are conductive and radiative Nusselt numbers, respectively $q_{Cw}$ and $q_{Rw}$ are conductive and radiative fluxes respectively and $L$ is the characteristic dimension of the present problem. Further, the total Nusselt number is defined as below
\begin{eqnarray*}
 Nu_{tot}=Nu_{C}+Nu_{R} \hspace{0.5cm}  \hspace{0.5cm}  
\end{eqnarray*}
where $Nu_{tot}$ is the total Nusselt number.

\section{Verification}

\begin{figure}[!t]
 \begin{subfigure}{8cm}
    \centering\includegraphics[width=8cm]{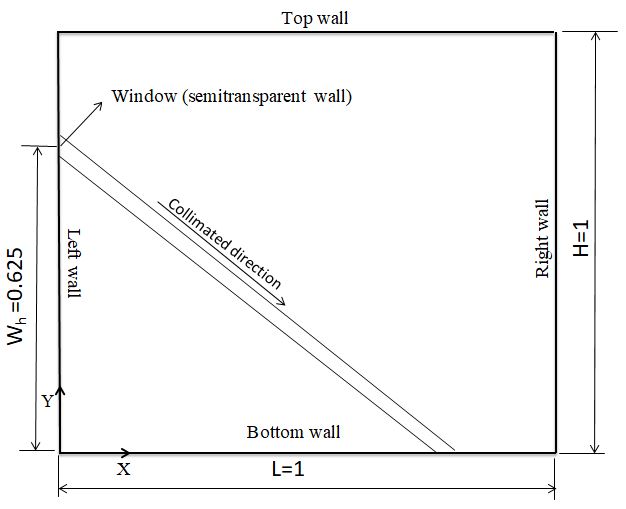}
    \caption{}
    \label{col_geometry}
  \end{subfigure}\hfill
   \begin{subfigure}{8cm}
    \centering\includegraphics[width=6cm]{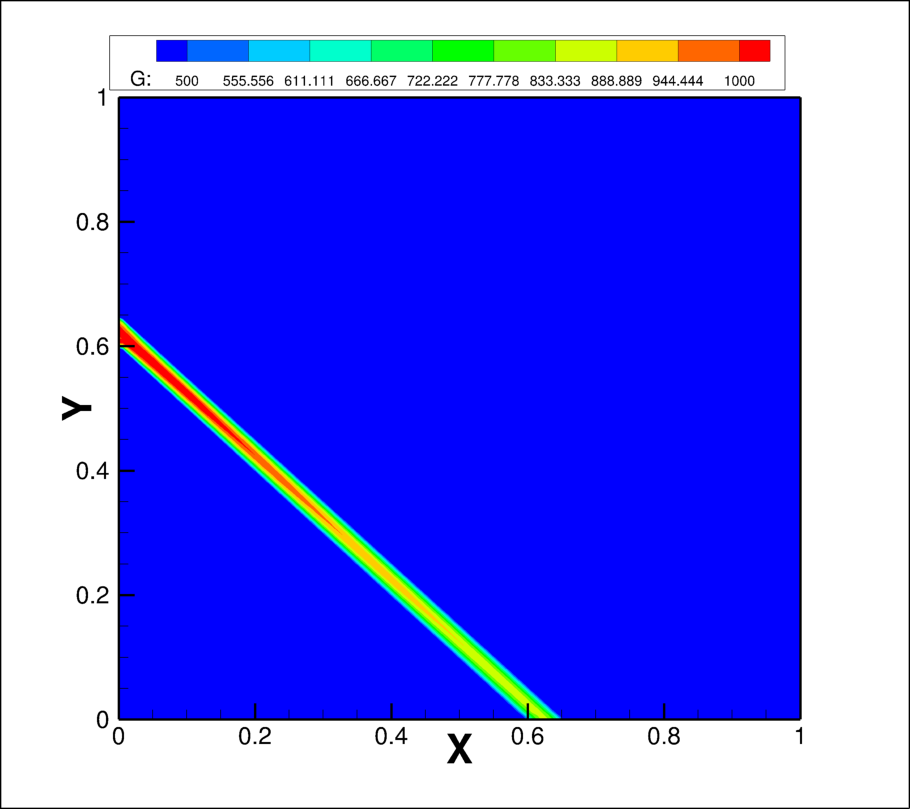}
    \caption{}
    \label{col_con}
  \end{subfigure}\hfill
  \caption{The verification of collimated beam feature (a) test geometry (b) contour of irradiation showing the travel of the beam in transparent medium}
\label{Collimated_valid}
\end{figure}

\begin{figure}[!t]
	\centering
	\includegraphics[width=80mm,scale=1]{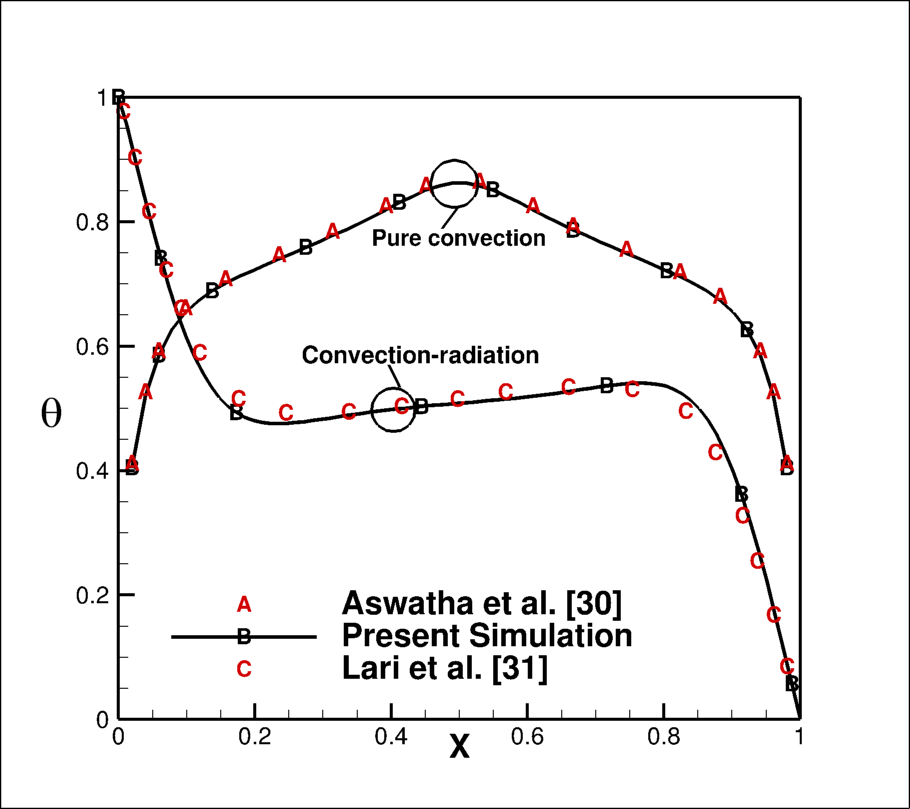}
	\caption{The verification of results for pure convection in a square enclosure which is heated from the bottom (symbol A) and the combined diffuse radiation with natural convection in a square enclosure where differential temperature is applied on the two vertical walls (symbol C)}
	\label{valid_temp}
\end{figure}

In the absence of any standard benchmark test case for the present problem, the validation has been performed in three steps, first, the standalone feature of collimated beam irradiation problem, in second step, pure natural convection problem which is heated from the bottom and in the third and last step , the combined natural convection and radiation in defferentially heated cavity have been verified. The collimated irradiation feature \cite {RAD19} has been tested in a square cavity as shown in the Fig. (\ref{col_geometry}). The left side of the wall has a small window of non-dimensional size 0.05 at a non-dimensional height of 0.6. The walls of square cavity are black and cold and also medium is non-participating. A collimated beam is irradiated on the window at azimuthal $135^0$ direction. It is expected that the beam would travel in oblique direction of $135^0$ angle without any attenuation and hit exactly non-dimensional distance of 0.6 from left wall. Figure (\ref{col_con}) shows the contour of irradiation which clearly shows the travel of collimated without any attenuation. For second step, fluid flow with heat transfer (without any radiation) is validated with Aswatha et al. \cite{Aswatha} and combined diffuse radiation and natural convection in a cavity whose top and bottom walls are adiabatic and vertical walls are isothermal at differential temperatures and radiatively opaque has been validated with Lari et al. \cite{Lari}. The present results for both the cases (see Fig. \ref{valid_temp}) are in good agreement with the published results.

\section{Grid Independent Tests}

Numerical solutions of Navier-Stokes, energy equations and radiation transfer equation are sensitive to the spatial discretization. Additionally, radiative transfer equation also requires angular space discretization which provides directions along which radiation transfer equation is being solved. Thus, optimum number of grids and directions have been obtained through independent test study in two steps,
\begin{enumerate}
    \item Spatial grids independence test:
    Three spatial grid sizes are chosen to calculate the area average total Nusselt number on the bottom wall as shown in Table \ref{spatial_grid} for the present problem of the natural convection. The percentage error between the first and second grid sizes is 0.8$\%$, whereas between second and third grid sizes is 0.15$\%$. Thus, the spatial grid points, i.e., 80$\times$80 is selected for further study.
    
\begin{table}[!b]
\centering
\caption{The area average total Nusselt number on the bottom wall}
\label{spatial_grid}
\begin{tabular}{|c|c|c|c|}
\hline
Nusselt number & 60 $\times$ 60 & 80 $\times$ 80 & 100  $\times$100   \\ \hline
Conduction & 6.42 & 6.615 & 6.78 \\ \hline
Radiation & -3.3 & -3.47 & -3.63  \\ \hline
Total & 3.12 &  3.145 & 3.15  \\ \hline
\end{tabular}
\end{table}

\item Angular direction independence test: 
The polar ($n_{\theta}$) discretization does not have any effect in two-dimensional analysis, thus, it has been fixed to 2 for polar angle of $180^0$ in OpenFOAM. The effect of angular discretization in one quadrant angular space on the area average total Nusselt number on the bottom wall is shown in Table \ref{angular_grid}. The percentage difference in area average Nusselt number in the first and second azimuthal discretization is 0.09$\%$, whereas in second and third angular discretization is 0.22$\%$. Thus, finally $n_\theta \times n_\phi =2 \times 5$ in one quadrant angular space is selected for the study of the present problem. 

\begin{table}[!t]
\centering
\caption{The area average total Nusselt number for angular discretization on the bottom wall}
\label{angular_grid}
\begin{tabular}{|c|c|c|c|}
\hline
Nusselt number & 2 $\times$ 3 & 2 $\times$ 5 & 2 $\times$ 7 \\ \hline
Conduction & 6.617 & 6.513 & 6.687 \\ \hline
Radiation & -3.47 & -3.363 & -3.53 \\ \hline
Total & 3.147 & 3.15 & 3.157 \\ \hline
\end{tabular}
\end{table}
\end{enumerate}

\section{Results and Discussion}

In the present numerical simulation, parameters such as, Rayleigh number, Prandtl number, Planck numbers, collimated irradiation and angle of collimated irradiation, respectively fixed to values $10^5$, 0.71, Pl= 0, 1, 10 and 50, $G_{co}=1000 W/m^2$, $135^0$ have been fixed. The simulations have been performed for the different aspect ratios of semitransparent window and Planck numbers and correspondingly, the fluid flow and heat transfer characteristics were studied.

\subsection{Case A: $h_r=$0.8 and $w_r=$0.2}
\begin{figure}[!b]
\begin{subfigure}{8cm}
    \centering\includegraphics[width=8cm]{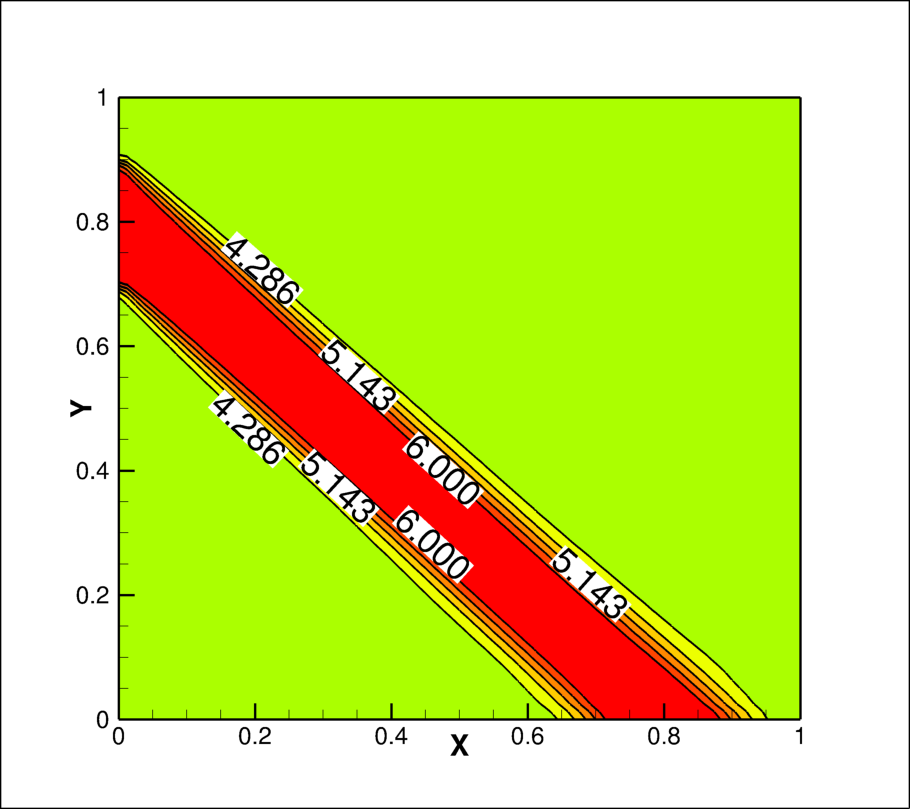}
    \caption{}
    \label{G_C_N0}
  \end{subfigure}
   \begin{subfigure}{8cm}
    \centering\includegraphics[width=8cm]{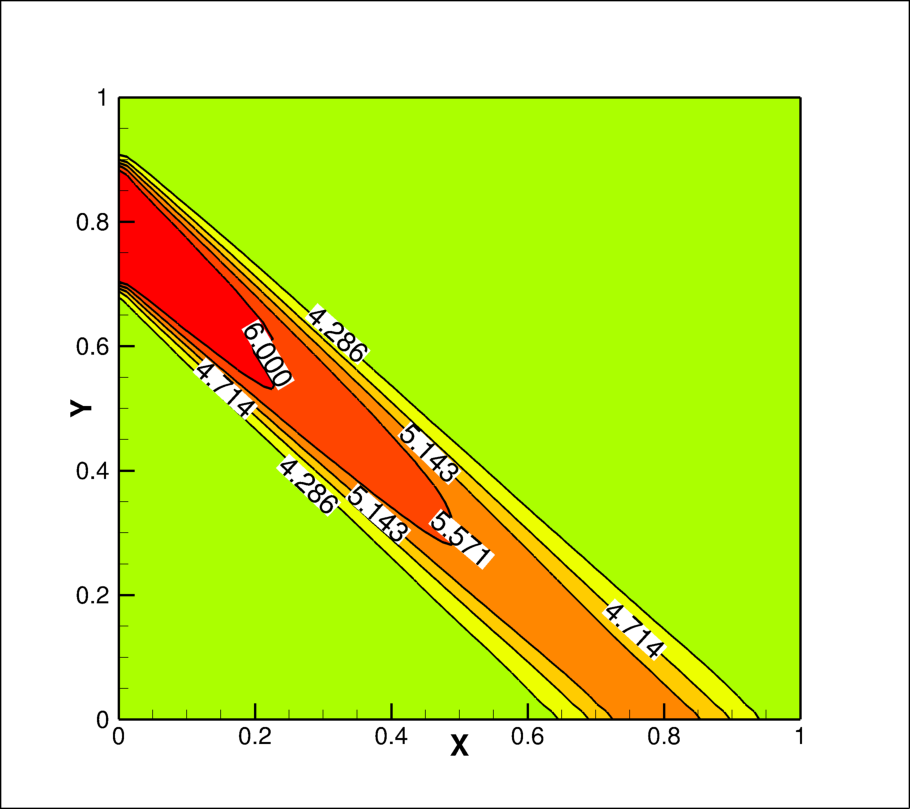}
    \caption{}
    \label{G_C_N1}
  \end{subfigure}
  \begin{subfigure}{8cm}
    \centering\includegraphics[width=8cm]{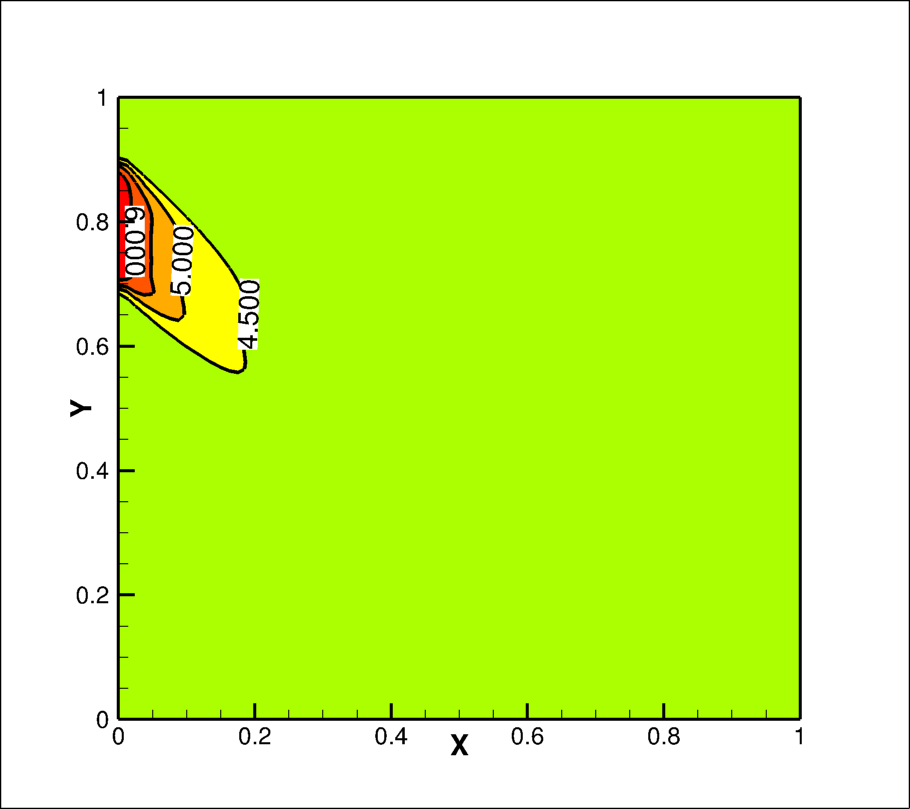}
    \caption{}
    \label{G_C_N10}
  \end{subfigure}
  \hspace{1.1cm}
   \begin{subfigure}{8cm}
    \centering\includegraphics[width=8cm]{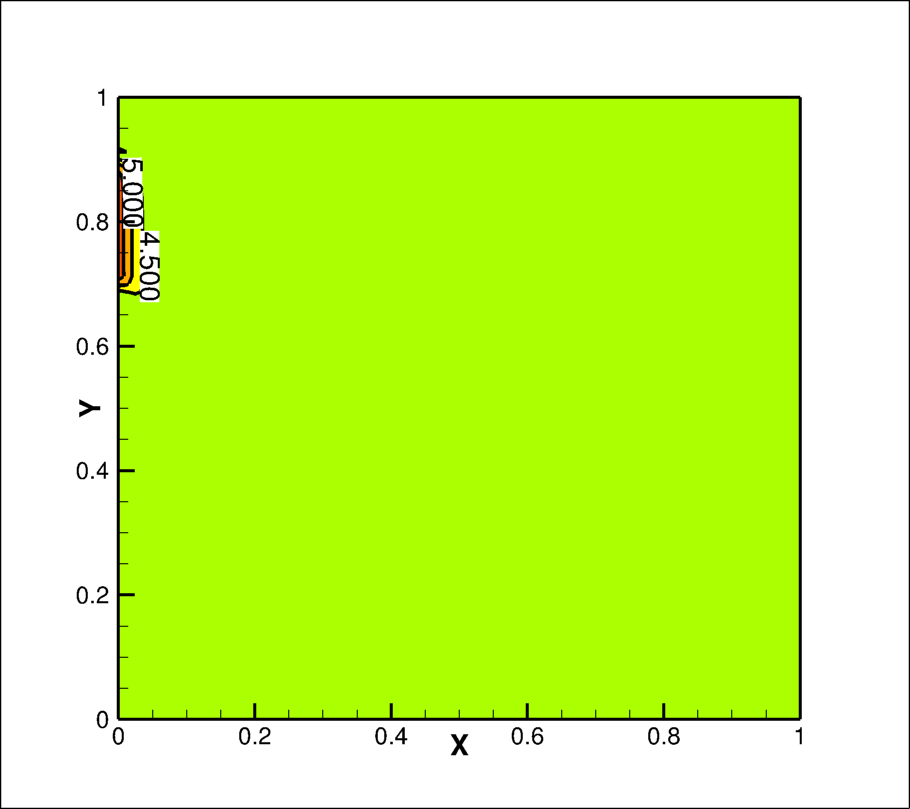}
    \caption{}
    \label{G_C_N50}
  \end{subfigure}
  \caption{The progression of collimated beam in (a) non-participating medium $Pl=0$; participating medium for (b) $Pl=1$ (c) $Pl=10$ and (d) $Pl=50$, for the irradiation value of 1000 $W/m^2$ applied on the semitransparent wall for case A}
\label{G_collimated}
\end{figure} 

A semitransparent window of width ratio 0.2 is created at the height ratio 0.8 on the left vertical wall. A collimated irradiation of value 1000 $W/m^2$ is applied on this semitransparent window at an angle of $135^0$. The dynamics of fluid flow and heat transfer are studied below

The collimated beam progression into the cavity in the direction of $135^0$ from the semitransparent window can be best represented by irradiation contours. The irradiation contours inside the cavity for Planck numbers 0, 1, 10 and 50 are shown in Fig. \ref{G_collimated} (a), (b), (c) and (d), respectively. The Planck number ($Pl=0$) corresponds to transparent medium, i.e., neither absorption nor emission by fluid therefore, the irradiation strength remains constant till it reaches to the bottom wall and spreads equal to a window width ratio 0.2 and strikes on bottom wall at a non-dimensional distance 0.7 from the left corner of the cavity. Whereas, the energy of collimated beam reduces along the line of progression in a participating medium for non-zero Planck numbers of the medium. The collimated irradiation does not reach to the bottom wall for Pl=10 as can be seen from the fig \ref{G_C_N10}. Moreover, the collimated beam energy gets absorbed near to the window for high optical thickness Pl=50 (see fig. \ref{G_C_N50}).

\subsubsection{Characteristics of Stream Function and Temperature Field}
\begin{figure}[!b]
\begin{subfigure}{8cm}
    \centering\includegraphics[width=8cm]{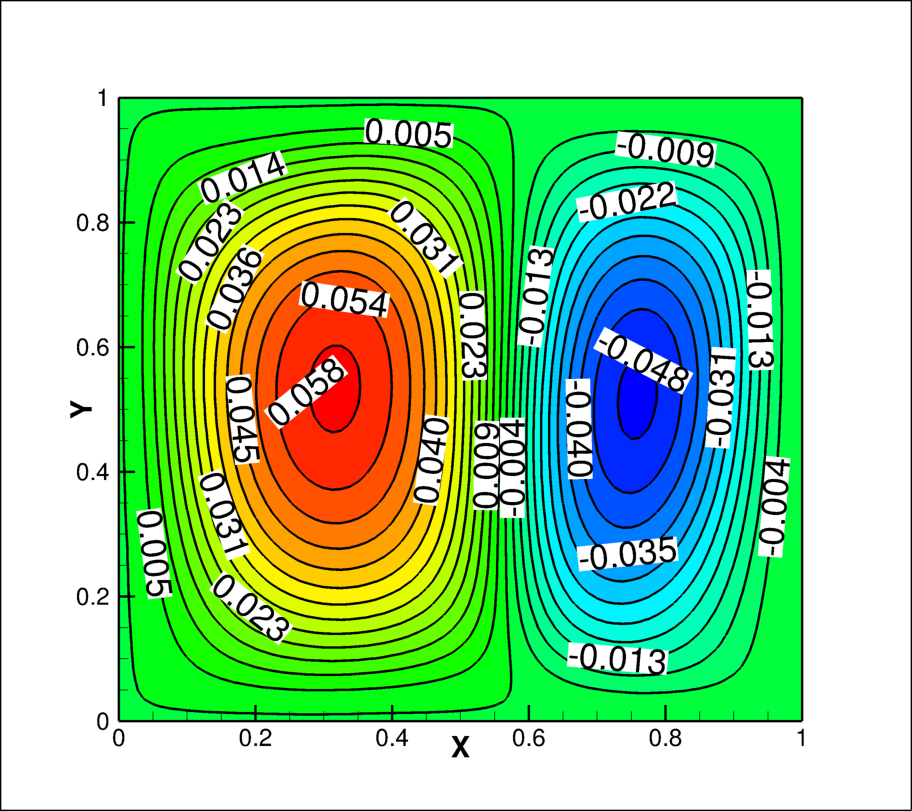}
    \caption{}
    \label{G_SF_N0}
  \end{subfigure}
   \begin{subfigure}{8cm}
    \centering\includegraphics[width=8cm]{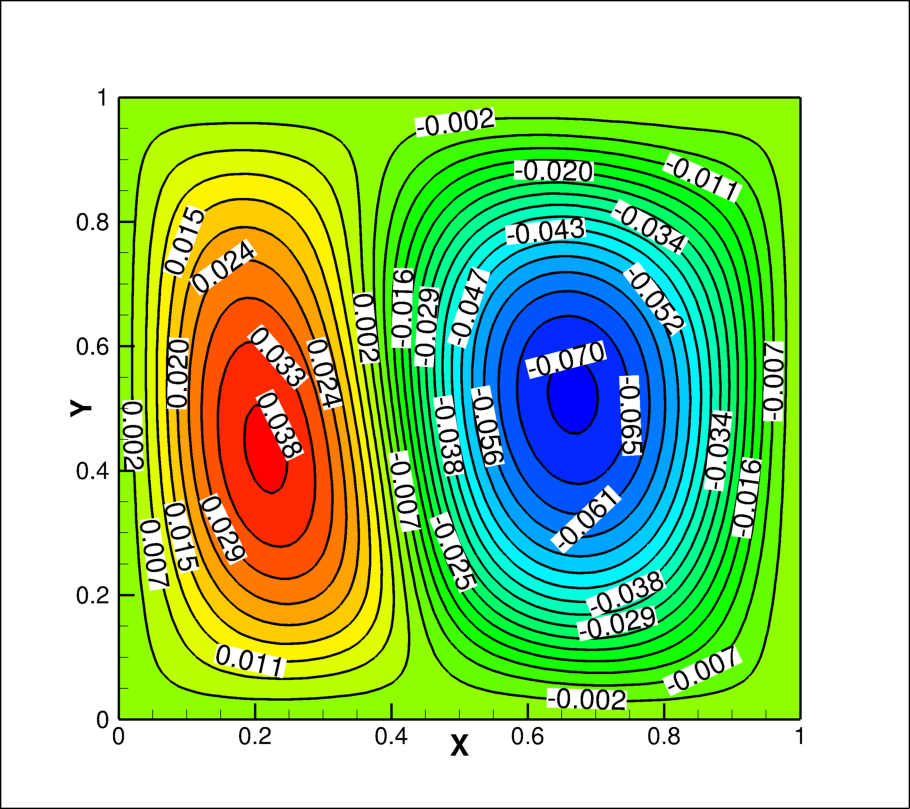}
    \caption{}
    \label{G_SF_N1}
  \end{subfigure}
  \begin{subfigure}{8cm}
    \centering\includegraphics[width=8cm]{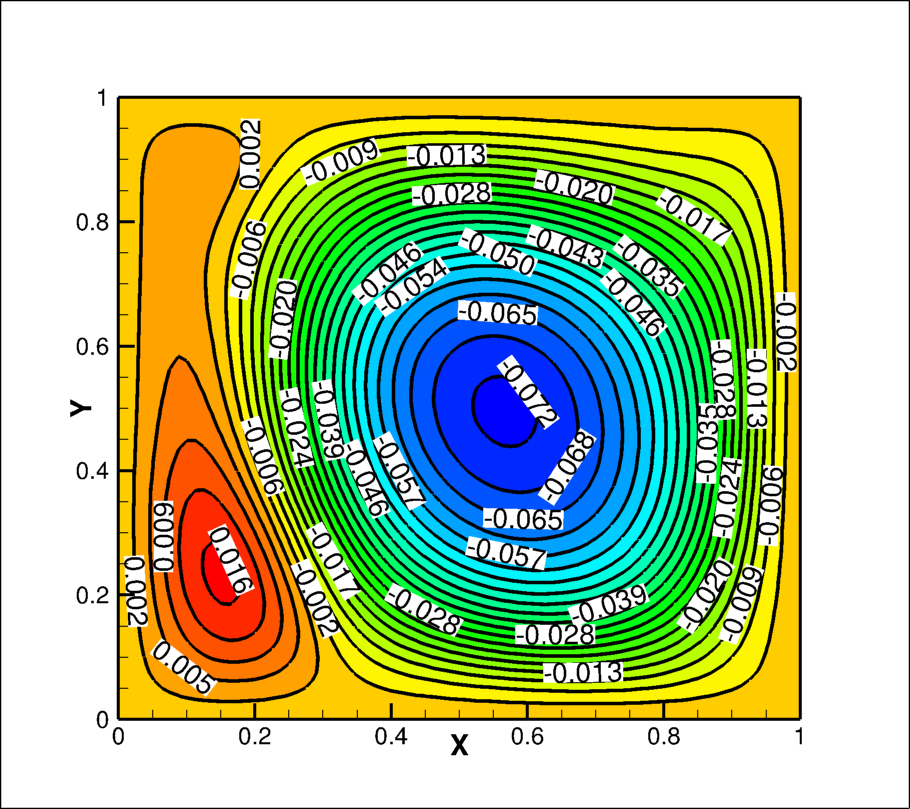}
    \caption{}
    \label{G_SF_N10}
  \end{subfigure}
  \hspace{1.1cm}
   \begin{subfigure}{8cm}
    \centering\includegraphics[width=8cm]{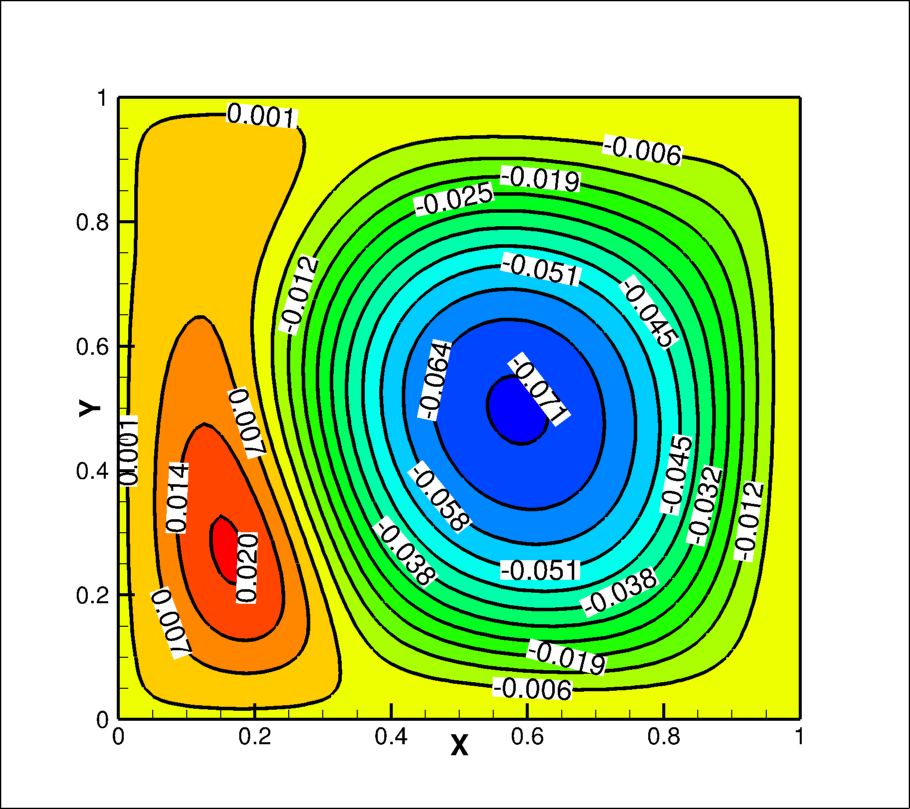}
    \caption{}
    \label{G_SF_N50}
  \end{subfigure}
  \caption{The contours of the non-dimensional stream function for (a) $Pl=0$  (b) $Pl=1$ (c) $Pl=10$ and (d) $Pl=50$ for case A}
\label{G_SF_AR1}
\end{figure} 

The effect of collimated beam irradiation on the stream function for Planck numbers 0, 1, 10 and 50 are shown in fig. \ref{G_SF_AR1}(a)-(d) respectively. Figure \ref{G_SF_N0} depicts stream function contour for the transparent media, where two asymmetrical vortices can be observed. The left vortex is larger in size than the right vortex. As the medium behaves transparent for radiation transfer, all collimated beam energy strikes on the bottom wall at a non-dimensional distance 0.7 from the left corner having non-dimensional spread 0.2, thus transfer the most of the energy to the bottom wall. There is enhancement in the buoyancy force in upward direction at this location (i.e., over the spread of collimated beam on the bottom wall), this intern increases the force in upward direction in the right vortex by making right vortex thinner and rest space is occupied by the left vortex. The reverse trend is observed for the participating media for Planck numbers ($Pl= 1, 10$ and $50$), where left vortex is smaller in size than right vortex. This is due to the fact that collimated beam is travelling through the left vortex which absorbs the radiation energy and creates local heating of fluid, this enhances local upward buoyancy force in the left vortex, whereas some energy is also being transferred to right vortex through absorption in the right vortex and also absorption by the bottom. The energy absorbed by left vortex may be higher due to large distance traveled by collimated beam in the left vortex, this causes decrease in the size of the left vortex and increases in size of the right vortex. The size of left vortex keeps on decreasing till Planck number 10, afterwards its size increases. Also, the flow rate in left vortex also keeps on decreasing till $Pl=10$ and then increases. The reverse trend is found in the right vortex.

\begin{figure}[!t]
\begin{subfigure}{8cm}
    \centering\includegraphics[width=8cm]{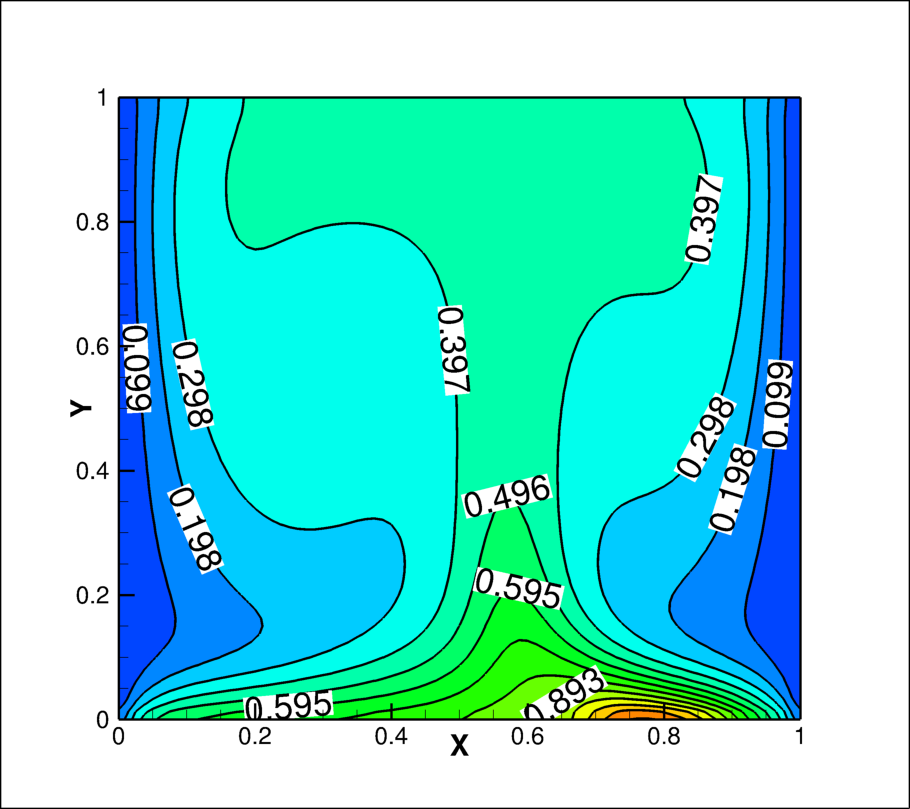}
    \caption{}
    \label{G_T_N0}
  \end{subfigure}
   \begin{subfigure}{8cm}
    \centering\includegraphics[width=8cm]{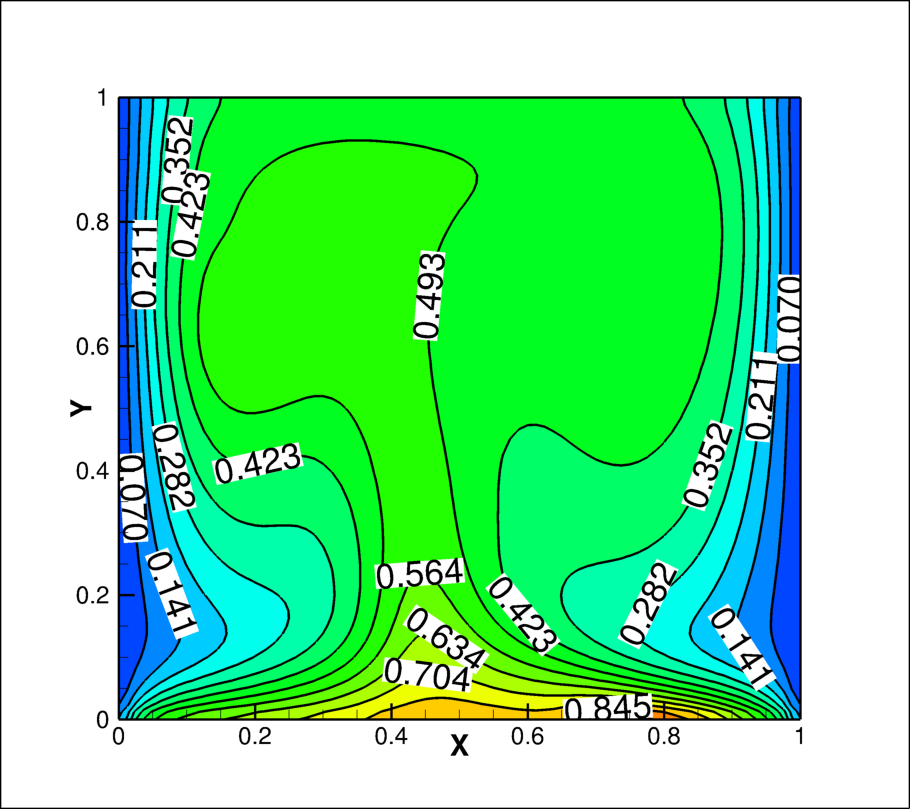}
    \caption{}
    \label{G_T_N1}
  \end{subfigure}
  \begin{subfigure}{8cm}
    \centering\includegraphics[width=8cm]{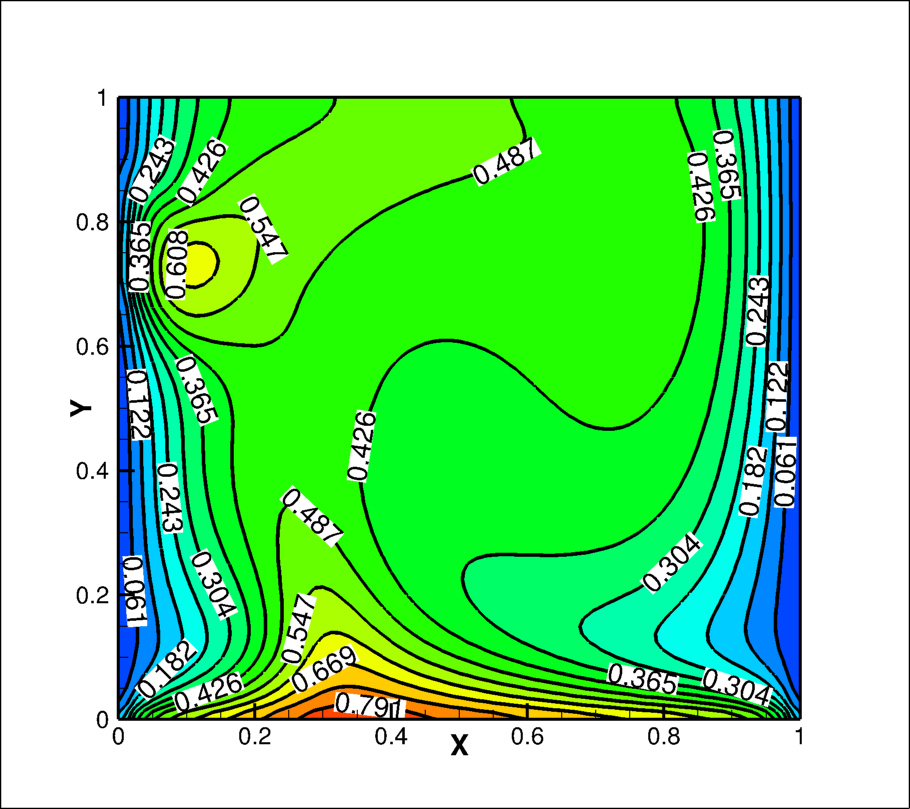}
    \caption{}
    \label{G_T_N10}
  \end{subfigure}
  \hspace{1.1cm}
   \begin{subfigure}{8cm}
    \centering\includegraphics[width=8cm]{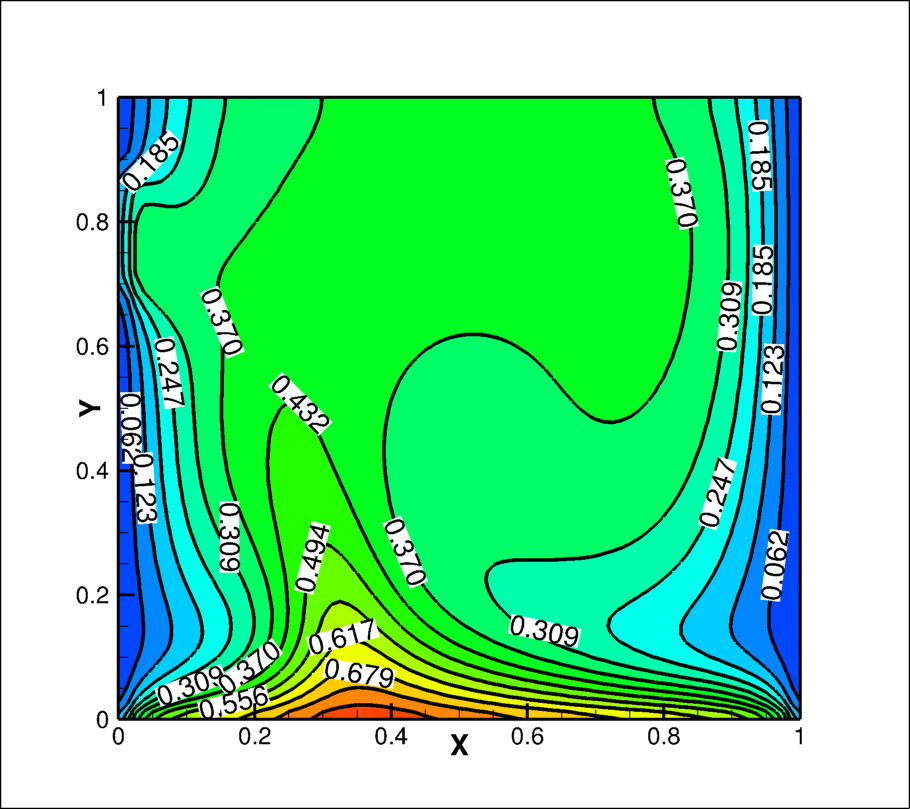}
    \caption{}
    \label{G_T_N50}
  \end{subfigure}
  \caption{The non-dimensional temperature contours for (a) $Pl=0$  (b) $Pl=1$ (c) $Pl=10$ and (d) $Pl=50$ for case A}
\label{G_T_AR1}
\end{figure} 

The  effect of collimated beam radiation on the temperature field inside the cavity for the Planck number 0, 1, 10 and 50 are shown in figs \ref{G_T_AR1}(a), (b), (c) and (d) respectively. The symmetrical isotherm lines about the mid vertical line of the cavity \cite{chanakya2020effects} becomes asymmetrical with inclusion of collimated beam and these lines tilt either right or left to the vertical line depending upon the medium behaviour for the radiation energy. The isothermal lines are bent towards left for the participating medium as left vortex is smaller in size (fig \ref{G_T_AR1}). The clustering of isotherm lines appears at the strike zone at the bottom wall for non-participating medium, whereas almost uniform temperature is spread in the core near to the top adiabatic wall. Furthermore, the isotherm lines are more closely placed at the bottom and on left wall with increase in Planck number of medium and also localized heating of the fluid is observed near to the semitransparent wall for the case Pl=10 (see fig \ref{G_T_N10}), this may be due to most of the collimated energy is getting absorbed within few distance from the semitransparent wall (see fig \ref{G_C_N10}), whereas the effect of localized heating is limited to semitransparent wall for the case of N=50 (see fig \ref{G_T_N50}), this is because of almost all the collimated irradiation is absorbed near to semitransparent wall. The wall is being isothermal, the energy is transferred out from the cavity. The maximum non-dimensional temperature inside the cavity is on the bottom wall but at different locations. It is on the strike zone for Planck number 0 and 1, and at the junction point of two vortices for Planck number 10 and 50.

\subsubsection{Velocity and Temperature Variations}

\begin{figure}[!b]
	\centering
	\includegraphics[width=80mm,scale=1]{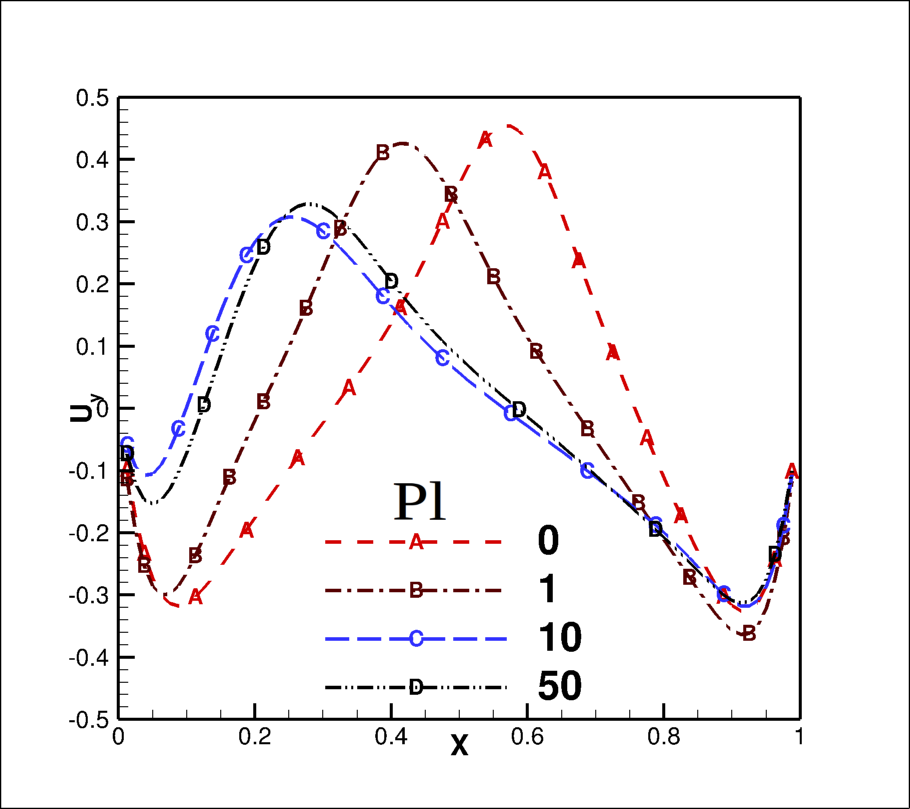}
	\caption{The variation of the non-dimensional vertical velocity along the horizontal line at the mid height of the cavity for various Planck numbers for case A}
	\label{Mid_vet_vel_A}
\end{figure}

The variations of non-dimensional vertical velocity in horizontal-direction at mid height of the cavity for the  Planck number 0, 1, 10 and 50 are depicted in fig \ref{Mid_vet_vel_A}. The vertical velocity is in the downward direction near to both the cold walls and reaches to maximum at same distance of 0.1 from the right vertical walls for Planck numbers. However, this distance 0.075 for Planck number 0 and 1 and 0.05 for Planck number 10 and 50 from the left wall. This maximum vertical velocity in the downward direction near to right vertical walls is 0.33 for all Planck numbers except for Planck number 1 where the value is 0.39. Furthermore, the maximum non-dimensional vertical velocity in downward direction near to the left walls for the Planck number 0, 1, 10 and 50 are 0.32, 0.3, 0.15 and 0.1, respectively. The vertical velocity in downward directions keep on decreasing and reach to zero at centre points of each vortex. The maximum vertical velocity in upward directions achieved at the junction of two vortices and their non-dimensional values are 0.43, 0.4, 0.32 and 0.3 achieved at distance 0.61, 0.4, 0.3 and 0.22 from left wall for Planck number 0, 1, 10 and 50, respectively. 

\begin{figure}[!t]
\begin{subfigure}{8cm}
    \centering
    \includegraphics[width=8cm,scale=1]{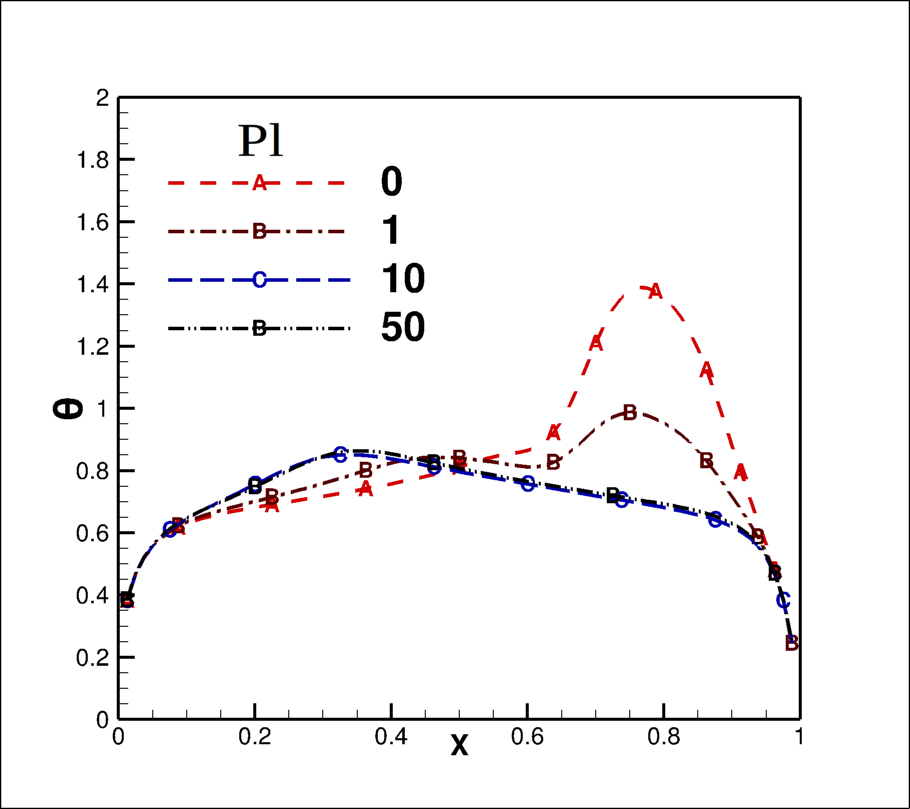}
    \caption{}
    \label{bot_temp_A}
  \end{subfigure}
   \begin{subfigure}{8cm}
    \centering
    \includegraphics[width=8cm,scale=1]{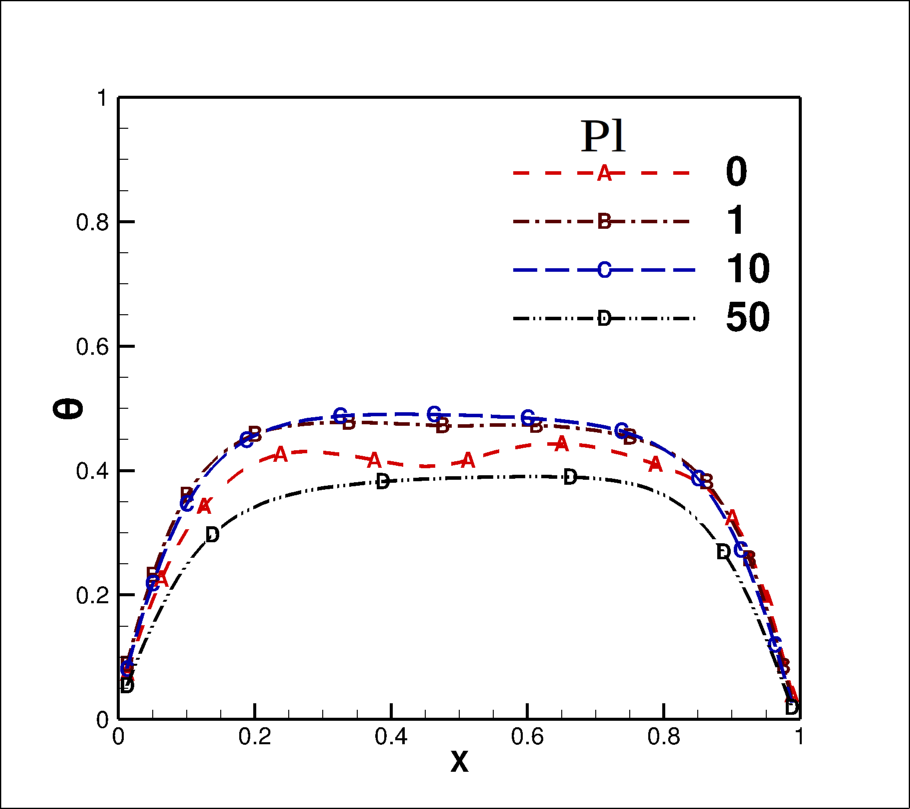}
    \caption{}
    \label{top_temp_A}
  \end{subfigure}
  \hspace{15cm}
   \caption{The variation of the non-dimensional temperature on (a) bottom wall (b) top wall for various Planck numbers case A}
\label{Nu_bot_AR1}
\end{figure} 

The non-dimensional temperature on the bottom wall increases rapidly from the left end till the distance 0.1 afterwards its rate of increase is slow till the strike length of collimated beam (fig \ref{bot_temp_A}). Afterwards, there is sudden raise in temperature at the strike zone and reaches to a maximum value of non-dimensional temperature 1.4 then starts decreasing till the right wall $Pl=0$. For the case $Pl=1$, two maximas in the temperature curve can be observed that corresponds to the strike zone of the collimated beam and stagnation point developed at the junction of two vortices. Nevertheless, the location of the global maxima corresponds to strike zone of collimated beam that remains fixed, it is the highest for the radiatively transparent fluid and keeps on decreasing for with increasing Planck number of the medium. Thehigh temperature zone due to collimated beam strike cannot be seen for Planck number values of 10 and 50, due to absorption of radiative energy within the fluid before it reaches to the bottom wall. The maximum non-dimensional temperature is found at the junction of two vortices  (i.e, the stagnation point) for Pl=10 and 50. The non-dimensional temperature on the top wall (fig \ref{top_temp_A}) increases from both the ends upto distance of 0.2 and remain almost constant at the middle portion of the curve for $Pl=1 , 10$ and 50. However, a little decrease in temperature is seen at the middle for the curve for $Pl=0$. There is no major difference in the temperature profile is observed  for Pl=1 and 10. The maximum non-dimensional value for temperature 0.41. The average temperature on the top wall increase with Planck number till 10, then it decreases for $Pl=50$.

\subsubsection{Variation of Nusselt Number}

\begin{figure}[!t]
\begin{subfigure}{8cm}
    \centering\includegraphics[width=8cm]{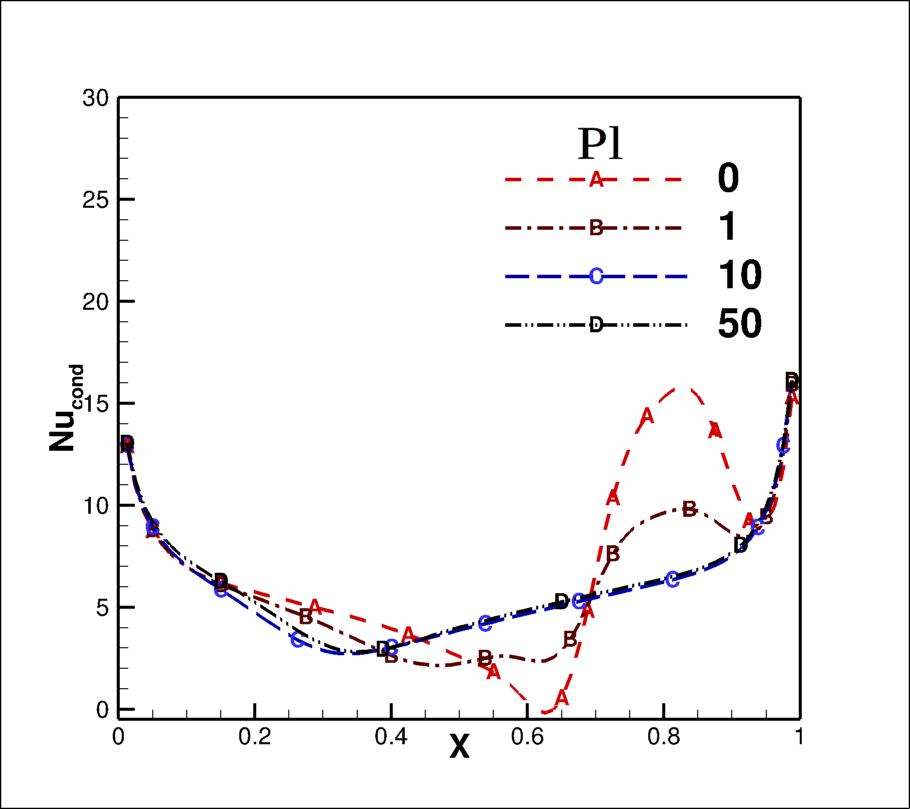}
    \caption{}
    \label{Nu_cond_bot_A}
  \end{subfigure}
   \begin{subfigure}{8cm}
    \centering\includegraphics[width=8cm]{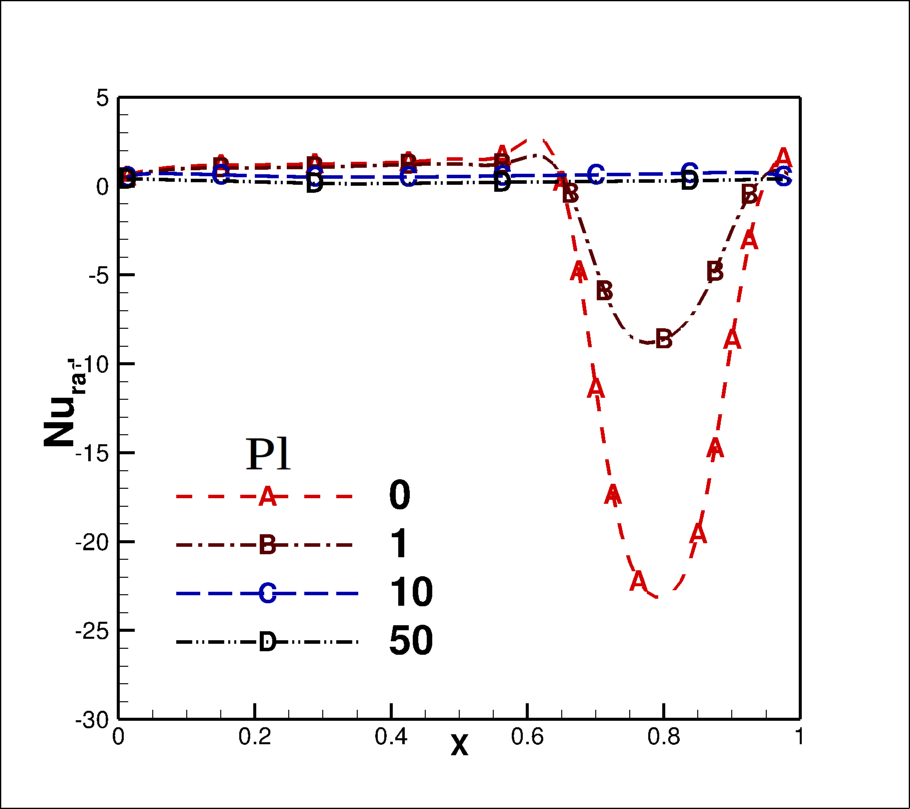}
    \caption{}
    \label{Nu_rad_bot_A}
  \end{subfigure}
  \hspace{15cm}
  \begin{subfigure}{17cm}
     \centering\includegraphics[width=8cm]{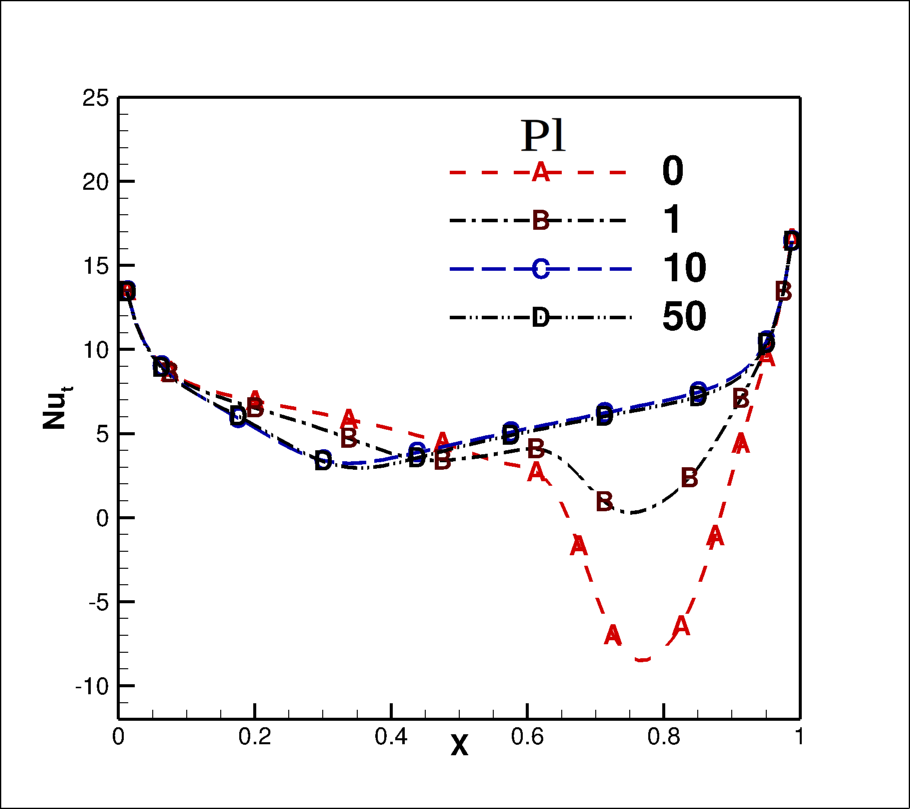}
    \caption{}
    \label{Nu_tot_bot_A}
  \end{subfigure}
   \caption{The variation of (a) conduction (b) radiation and (c) total Nusselt numbers for different Planck numbers on the bottom wall for case A}
\label{Nu_bot_AR1}
\end{figure} 

The conduction ($Nu_{cond}$), radiation ($Nu_{rad}$) and total ($Nu_{tot}$) Nusselt numbers variation on the bottom wall are presented in fig \ref{Nu_bot_AR1} for various values of Planck numbers. The variation of conduction Nusselt number on both ends of the bottom wall are similar for all the Planck numbers. The behaviour of the conduction Nusselt number graph is same for all Planck numbers of the medium till non-dimensional distance of 0.15 from the left corner and then slowly decreases to a minimum value of almost zero at a non-dimensional distance of 0.61 from left corner $Pl=0$. All of a sudden, it increases to 16 at the strike length of collimated beam, it further decreases to 9 than starts increasing and reaches to a maximum value of 17 on the right side of isothermal wall for the case of transparent medium. The lowest conduction Nusselt is obtained at distance of 0.4 and remains constant till the strike point of collimated beam, then increases to value of 8 for the case of $Pl=1$, further its behaviour is similar to the transparent medium case. Whereas, the conduction Nusselt number behaviour remains unaffected by the collimated beam for Planck numbers 10 and 50. However, $Nu_{rad}$ remains constant for the values of $Pl=0$ and $1$ till a distance of 0.6 on the bottom wall, and an inverted cone having a peak value of 22 and 7, are observed for the $Pl = 0$ and 1 respectively, whereas, the radiation Nusselt number remains constant over the entire length and its value is almost zero over the entire length for the Pl=10 and 50.

The total Nusselt number which is a linear combination of conduction and radiation Nusselt numbers is dominated by conduction Nusselt number in the most of the length expect of the bottom wall the length over which collimated beam strikes (fig \ref{Nu_tot_bot_A}). The total Nusselt number at the beam strike zone is dominated by the radiation Nusselt number in the curve. Whereas, no peak appears in the total Nusselt number for the Planck number 10 and 50 which indicates that the collimated beam energy gets absorbed within the fluid before reaching to the bottom wall.

\begin{figure}[!t]
\begin{subfigure}{8cm}
    \centering\includegraphics[width=8cm]{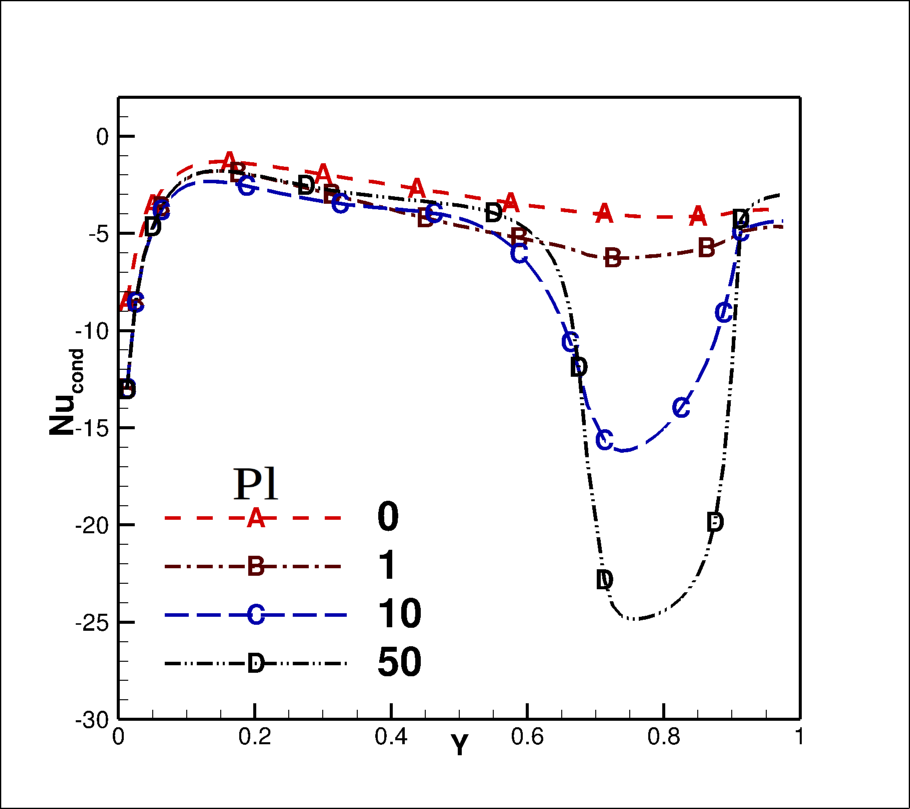}
    \caption{}
    \label{Nu_cond_left_A}
  \end{subfigure}
   \begin{subfigure}{8cm}
    \centering\includegraphics[width=8cm]{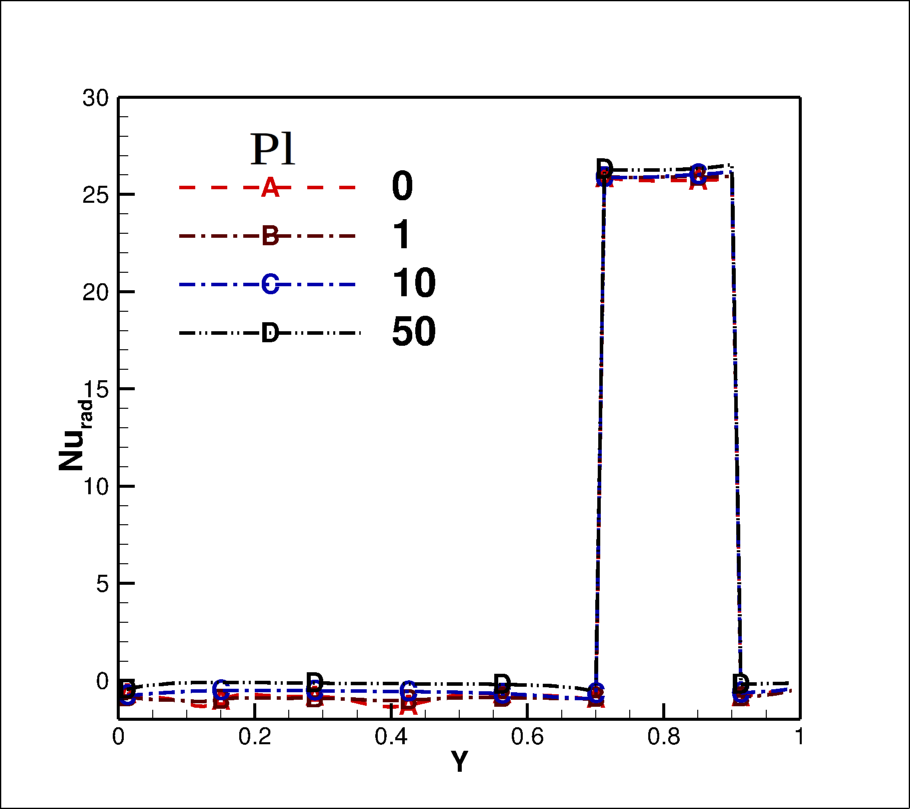}
    \caption{}
    \label{Nu_rad_left_A}
  \end{subfigure}
  \hspace{15cm}
    \begin{subfigure}{17cm}
     \centering\includegraphics[width=8cm]{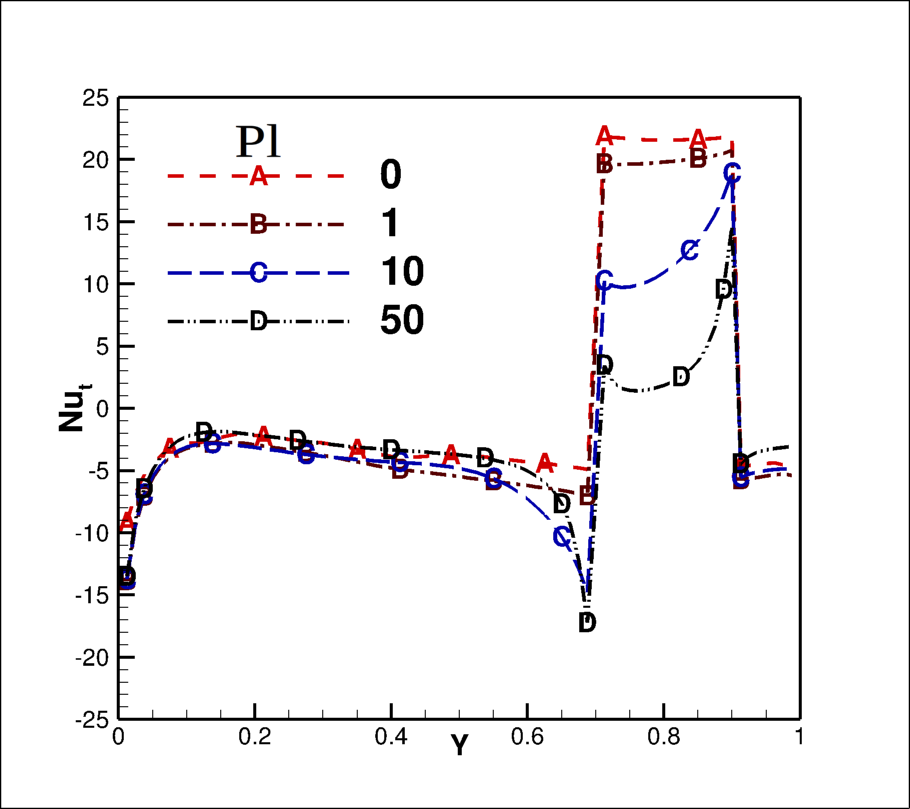}
    \caption{}
    \label{Nu_tot_left_A}
  \end{subfigure}
   \caption{The variation of (a) conduction (b) radiation and (c) total Nusselt numbers for different Planck numbers on the left wall for case A}
\label{Nu_left_AR1}
\end{figure} 

The variation of conduction, radiation and total Nusselt numbers on the left isothermal wall which also includes the semitransparent window for the Pl=0, 1, 10 and 50 are depicted in Fig \ref{Nu_left_AR1} (a), (b) and (c), respectively. There is sudden decrease in conduction Nusselt number over few height from the bottom and reaches to a minimum value of 1 and it almost remains constant over the rest height of the left wall for transparent medium cases. Nevertheless small increment happens on the semitransparent wall for $Pl=1$. However, for $Pl=10$ and 50, there is sudden increment in the conduction Nusselt number on the semitransparent window for $Pl=10$ and 50. This is mostly in the negative direction which reveals that the energy leaves from the wall from the conduction mode of the heat transfer. The maximum conduction Nusselt number value is 24 and found on the semitransparent window for the $Pl=50$. This value decreases to 3 after strike zone and remains constant over the rest height of the vertical wall. The maximum conduction Nusselt number is 15 on the semitransparent window for $Pl=10$. The radiation Nusselt number is found to be constant over the height of the left wall expect at the semitransparent wall because of irradiation G=1000 $W/m^2$ is applied at the semitransparent wall (fig \ref{Nu_rad_left_A}). It can also be noticed that the radiative Nusselt number is negative over entire vertical length expect at the semitransparent window where the radiative Nusselt number is positive, this reveals that the radiative heat flux is coming inside the cavity. The total Nusselt number on the left wall is negative Nusselt number over most of the height indicates that energy is leaving from the domain from the other height except length of the semitransparent window, however, the total Nusselt number is positive on the semitransparent window indicates that the net energy is entering into the domain from this portion of the left wall.

\begin{figure}[!t]
\begin{subfigure}{8cm}
    \centering\includegraphics[width=8cm]{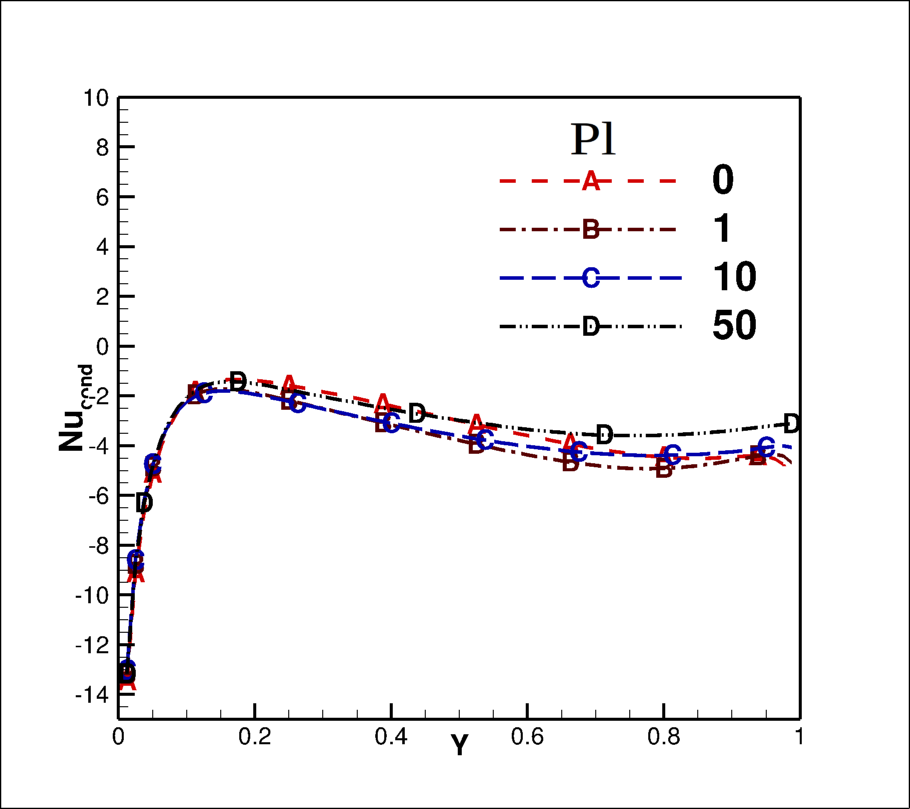}
    \caption{}
    \label{Nu_cond_right_A}
  \end{subfigure}
   \begin{subfigure}{8cm}
    \centering\includegraphics[width=8cm]{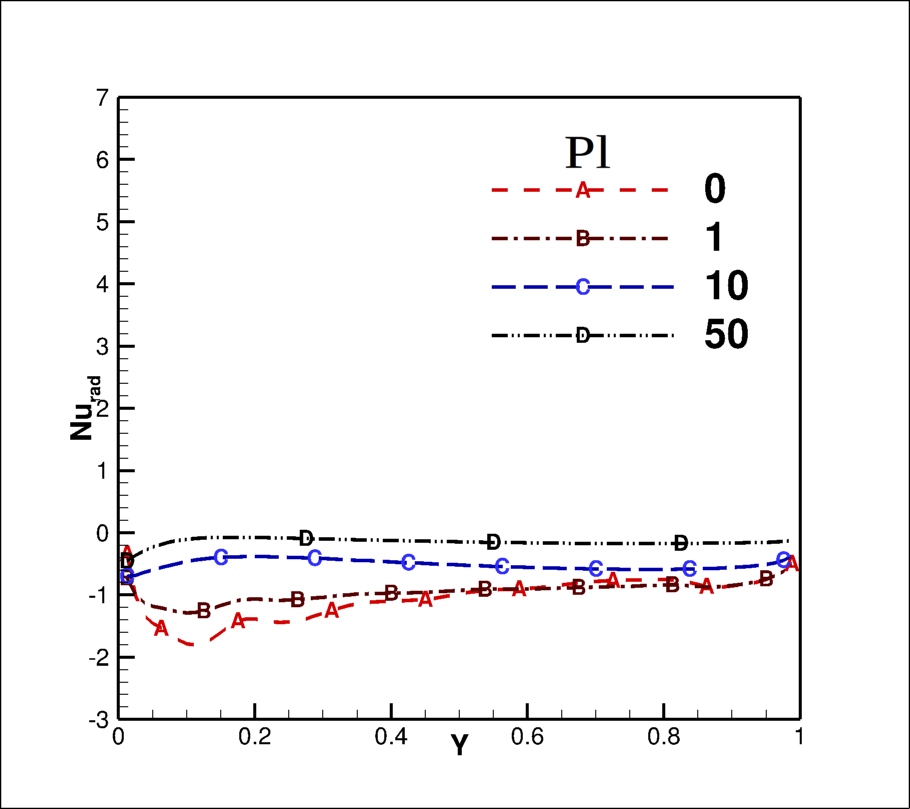}
    \caption{}
    \label{Nu_rad_right_A}
  \end{subfigure}
  \hspace{15cm}
  \begin{subfigure}{17cm}
     \centering\includegraphics[width=8cm]{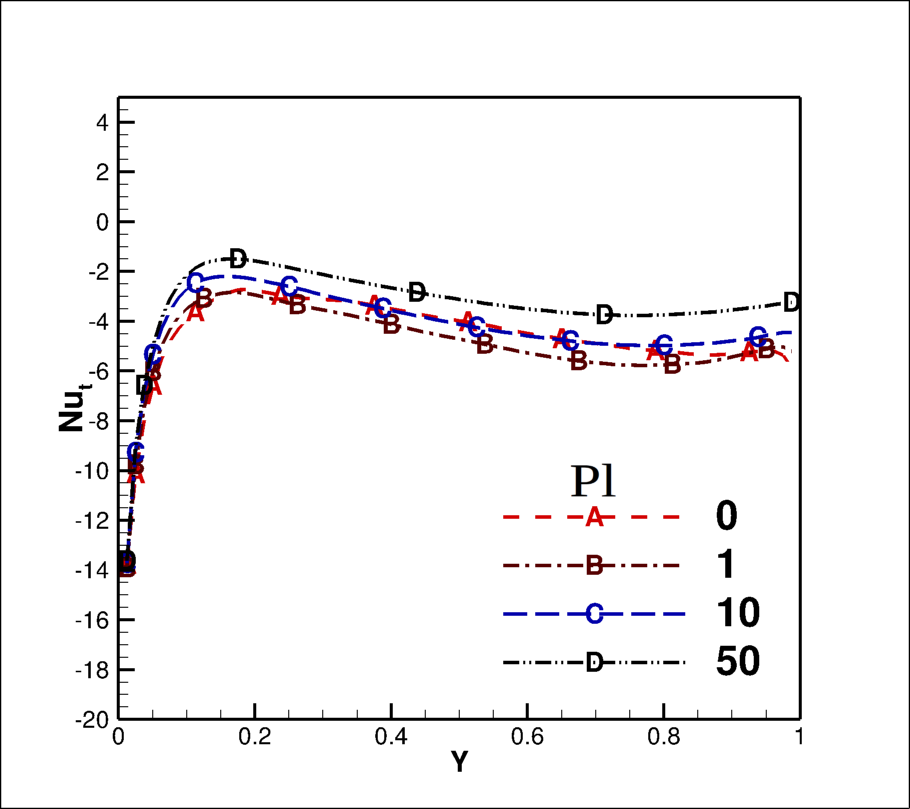}
    \caption{}
    \label{Nu_tot_right_A}
  \end{subfigure}
   \caption{The variation of (a) conduction (b) radiation and (c) total Nusselt numbers for different Planck numbers on the right wall for case A}
\label{Nu_right_AR1}
\end{figure} 

The variation of conduction Nusselt number on the right wall  (fig \ref{Nu_cond_right_A}) is similar to the left wall except the right wall does not contain the semitransparent window, thus the phenomenon happens on the semitransparent wall does not appear on the right wall. There is no major change in the conduction Nusselt number with the  Planck numbers of the medium. Nevertheless, the conduction Nusselt number increases little for $Pl=1$ compared to $Pl=0$, then decreases for $Pl=10$ and $Pl=50$. The radiation Nusselt number is almost constant over the whole height of the right wall and decreases for Planck number $Pl=0$ to $Pl=50$. The radiation Nusselt number is almost zero for $Pl=0$, on the right wall. The total Nusselt number is similar to conduction Nusselt number curve, but the difference increases from $Pl=1$ to $Pl=50$.

\subsection{Progression of Collimated Beam in other Aspect Ratio of Semitransparent Window}

\begin{figure}[!b]
\begin{subfigure}{8cm}
    \centering\includegraphics[width=8cm]{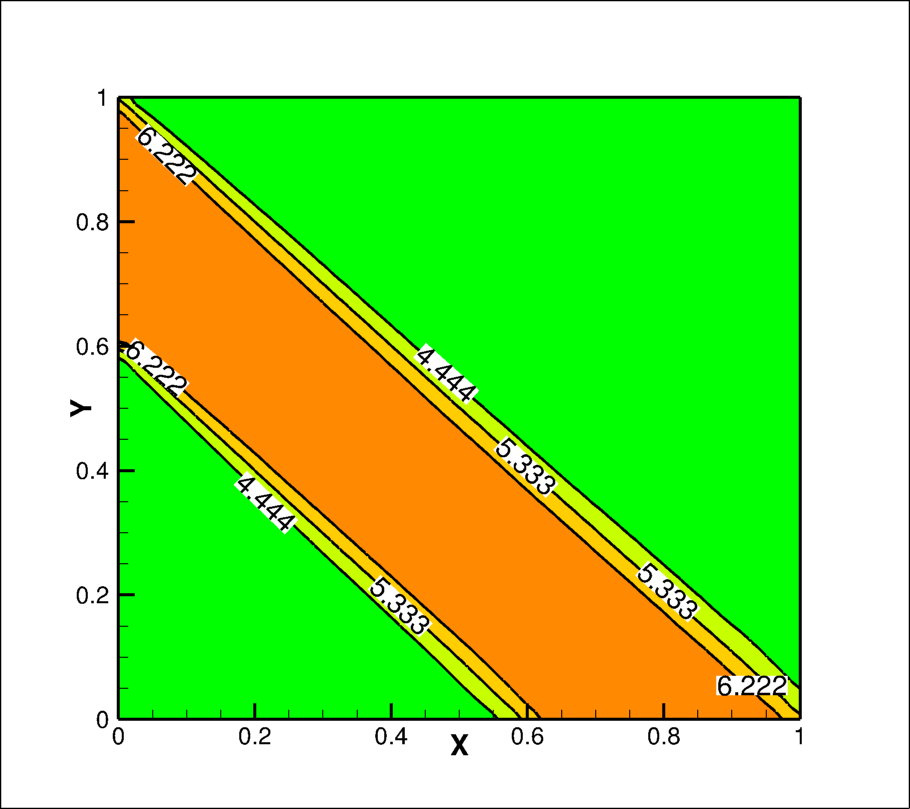}
    \caption{}
    \label{G_As_N0}
  \end{subfigure}
   \begin{subfigure}{8cm}
    \centering\includegraphics[width=8cm]{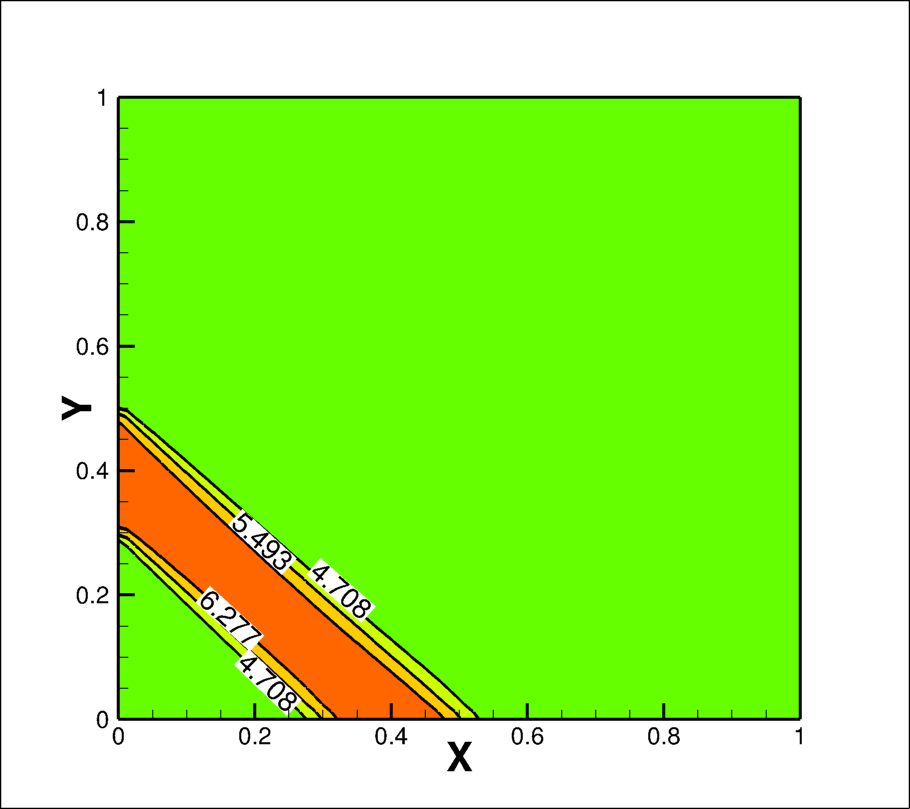}
    \caption{}
    \label{G_As_N1}
  \end{subfigure}
  \begin{subfigure}{18cm}
    \centering\includegraphics[width=8cm]{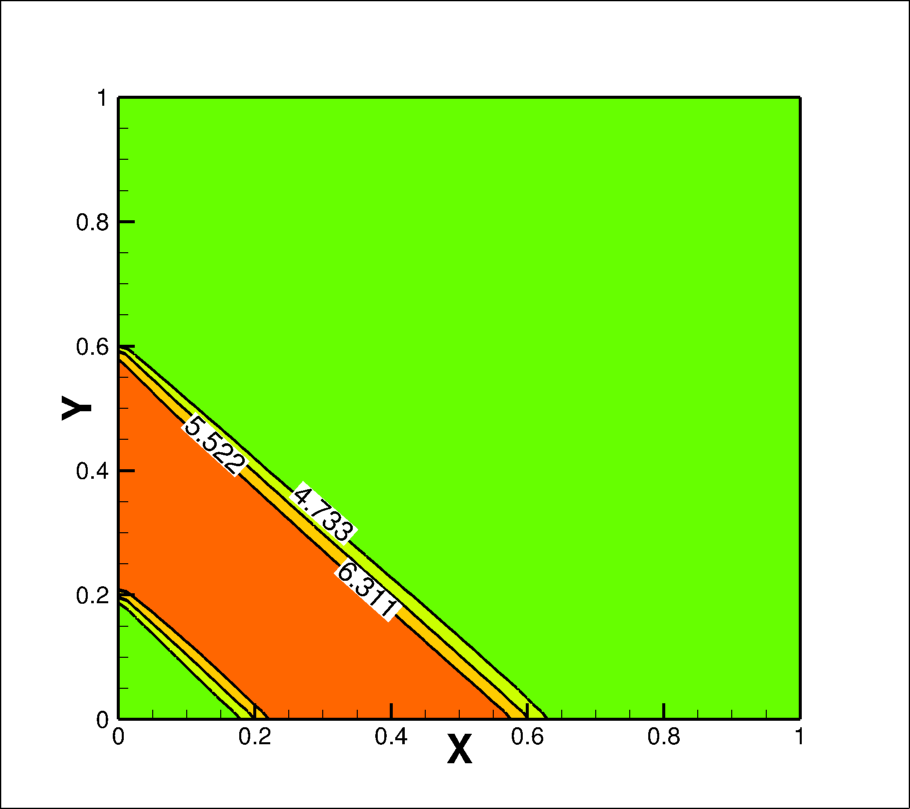}
    \caption{}
    \label{G_as_N10}
  \end{subfigure}
  \hspace{1.1cm}
  \caption{The progression of a collimated beam in a non-participating medium for (a) case B: $h_r=0.8$ and $w_r=0.4$ (b) case C: $h_r=0.4$ and $w_r=0.2$ and (c) case D: $h_r=0.4$ and $w_r=0.4$ for the collimated irradiation value of 1000 $W/m^2$ applied on the semitransparent wall at an angle of $135^0$ }
\label{G_Aspect_collimated}
\end{figure} 

The collimated irradiation contours for the aspect ratios of case B, C, and D are depicted in figs. \ref{G_Aspect_collimated}(a), (b) and (c) respectively. The purpose of present graph is to show the position of the semitransparent window and its width for different cases. The transparent medium has been selected to present the collimated beam contours, thus collimated irradiation contours remain same through out the progression of the beam. However, collimated beam contours will be different for different Planck numbers of the medium inside the cavity as shown fig \ref{G_collimated}; however they are not presented here for the brevity. The effects of these aspect ratios and Planck numbers of the medium on the fluid flow and the heat transfer will be presented in the subsequent sections:

\subsection{Characteristics of the Stream Function and the Temperature Field}
\subsubsection{Case B: $h_r$=0.8 $w_r$=0.4}
\begin{figure}[!b]
\begin{subfigure}{8cm}
    \centering\includegraphics[width=8cm]{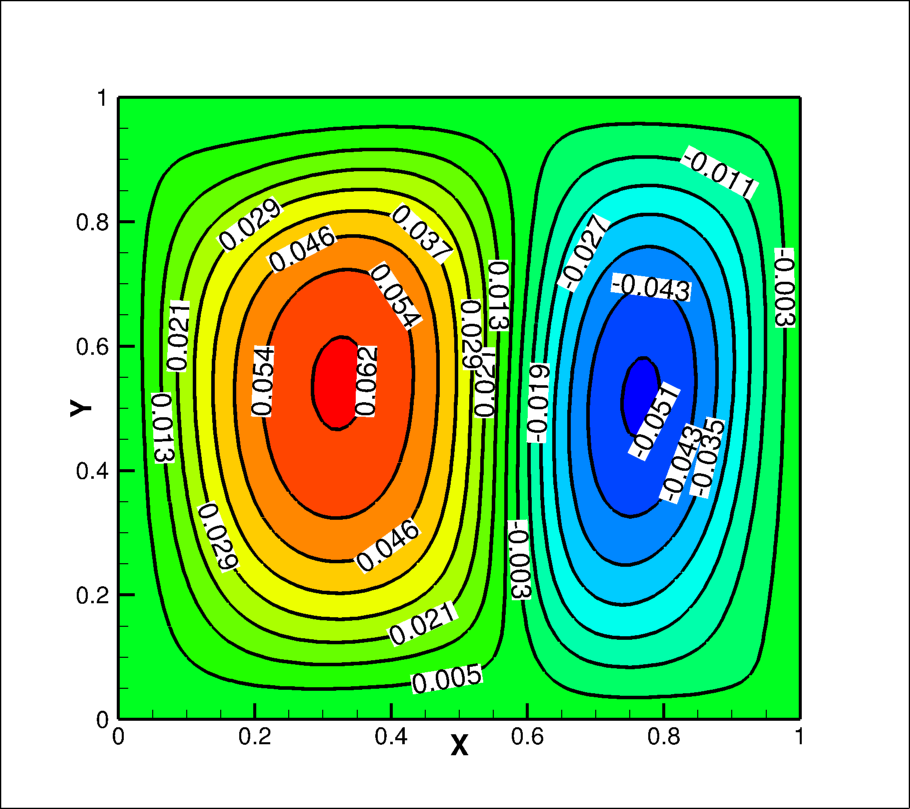}
    \caption{}
    \label{B_G_SF_N0}
  \end{subfigure}
   \begin{subfigure}{8cm}
    \centering\includegraphics[width=8cm]{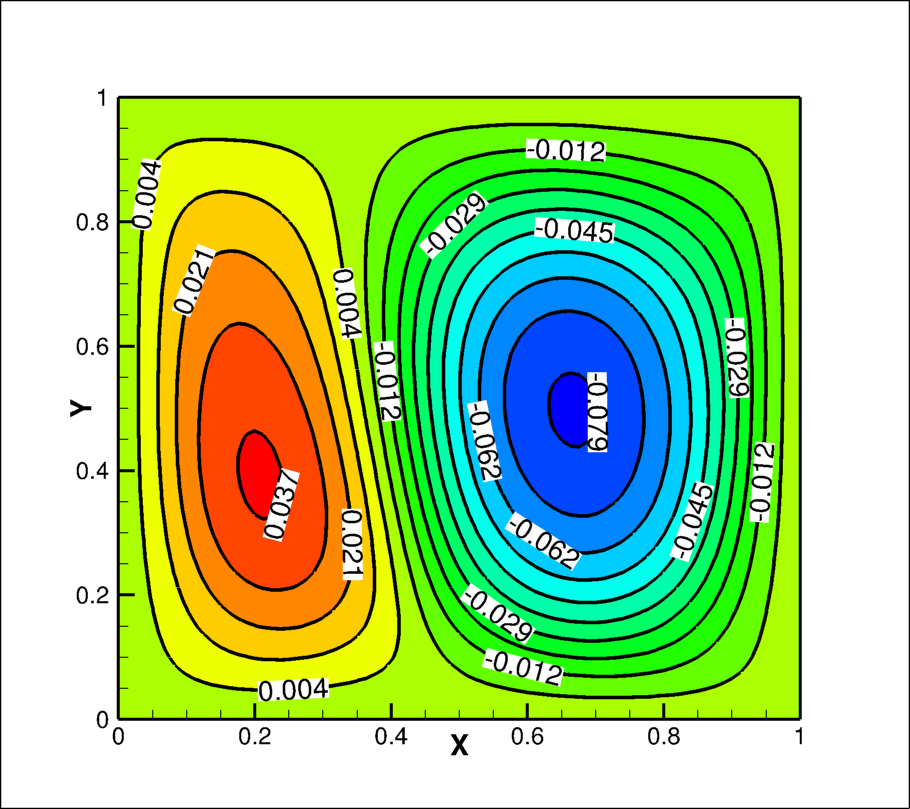}
    \caption{}
    \label{B_G_SF_N1}
  \end{subfigure}
  \begin{subfigure}{8cm}
    \centering\includegraphics[width=8cm]{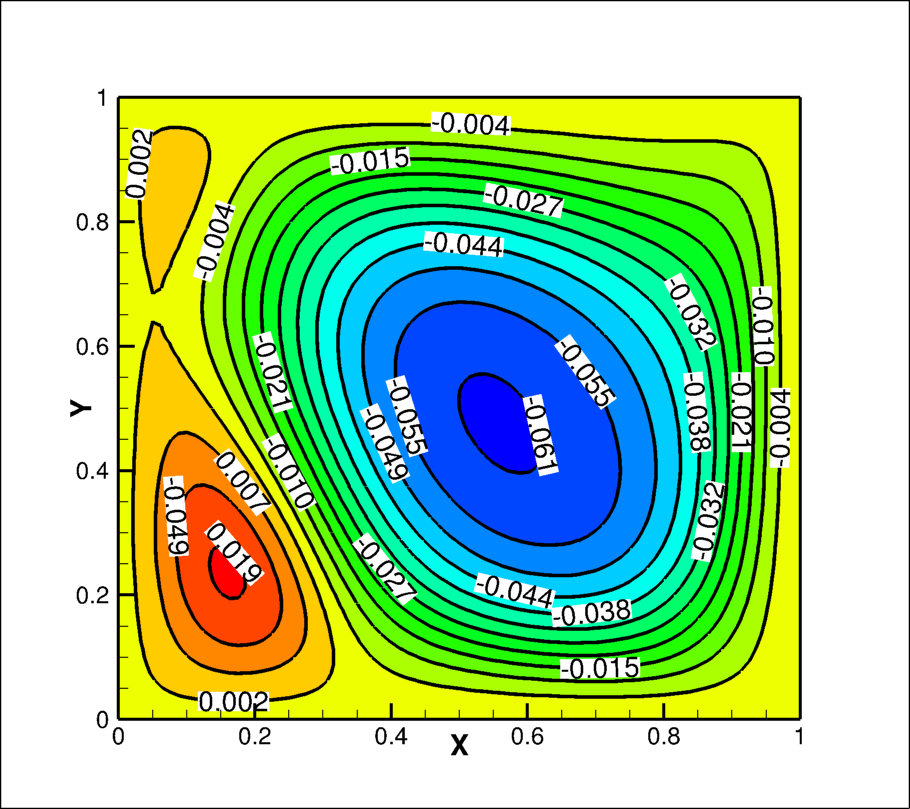}
    \caption{}
    \label{B_G_SF_N10}
  \end{subfigure}
  \hspace{1.1cm}
   \begin{subfigure}{8cm}
    \centering\includegraphics[width=8cm]{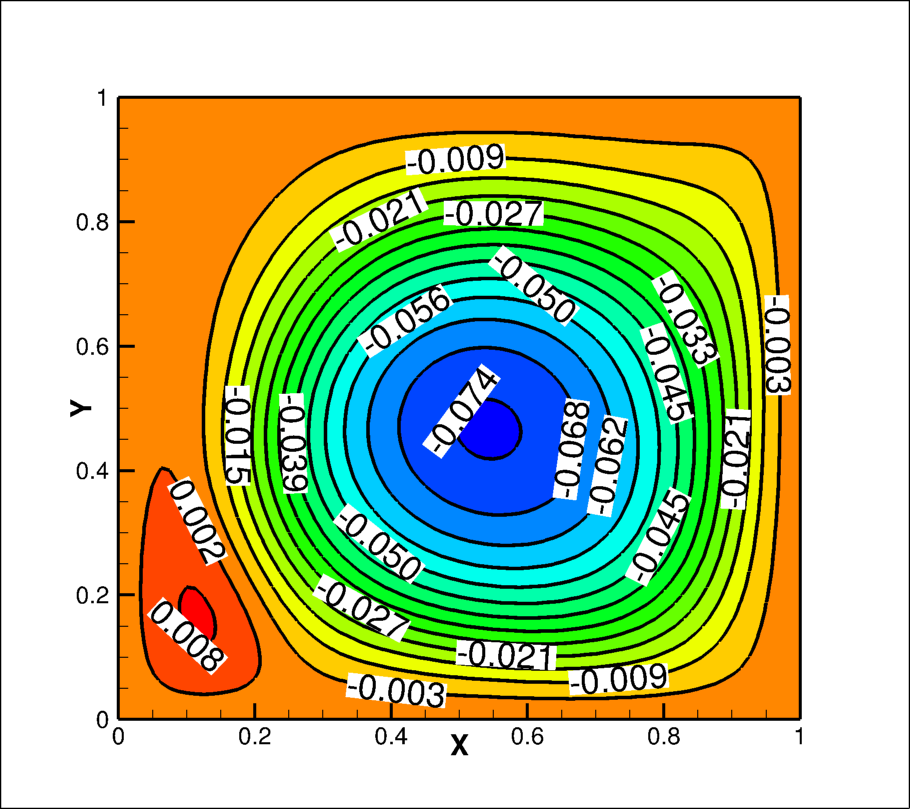}
    \caption{}
    \label{B_G_SF_N50}
  \end{subfigure}
  \caption{The contours of the non-dimensional stream function for (a) $Pl=0$  (b) $Pl=1$ (c) $Pl=10$ and (d) $Pl=50$ for case B}
\label{G_SF_AR2}
\end{figure}

The effect of the collimated beam for semitransparent window's aspect ratio $h_{r}$=0.8 and $w_{r}$=0.4  on the stream function for the range of Planck numbers 0--50 are depicted in fig \ref{G_SF_AR2}. The dynamics of two vortices are almost similar to case A (fig \ref{G_SF_AR1}), except little increase in the stream function i.e., increase in the flow rate in respective vortices for Planck numbers 0 and 1, however, on further increase of the Planck number ($Pl=10$) of the medium, the left vortex breaks into two part-upper left vortex and lower left vortex. The flow rate in lower left vortex is higher than the upper left vortex. An interesting fact to notice that the flow rate in right vortex also decreases. This upper left vortex disappears for Planck number $Pl=50$ and flow rate in lower left vortex also decreases. Nevertheless, the flow rate in the right vortex has now increased.

\begin{figure}[!t]
\begin{subfigure}{8cm}
    \centering\includegraphics[width=8cm]{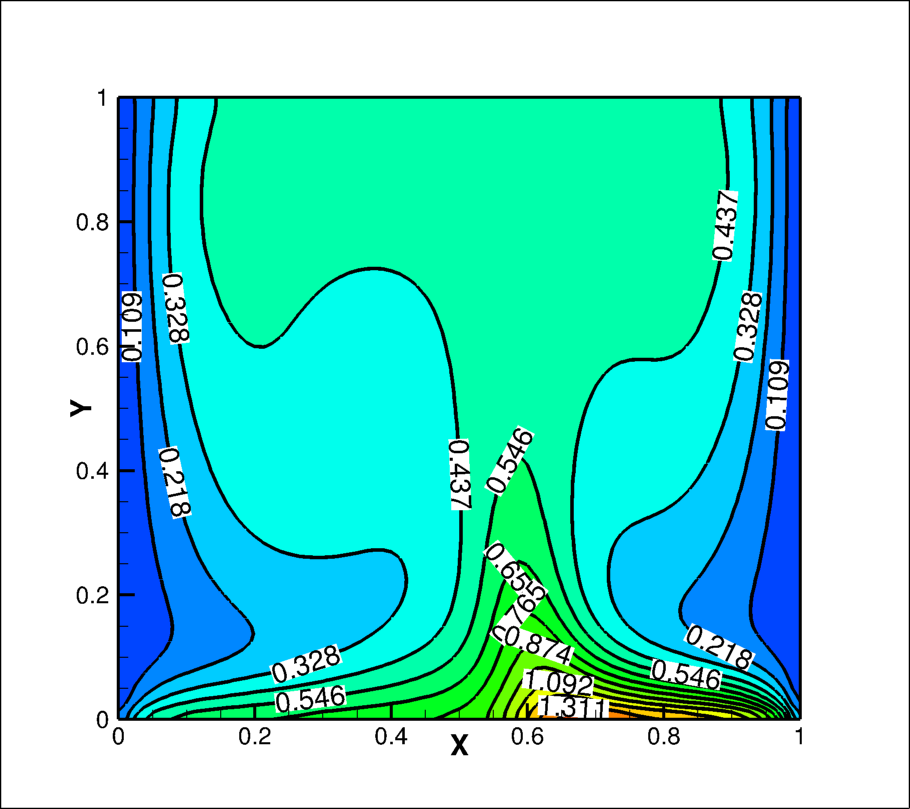}
    \caption{}
    \label{B_G_T_N0}
  \end{subfigure}
   \begin{subfigure}{8cm}
    \centering\includegraphics[width=8cm]{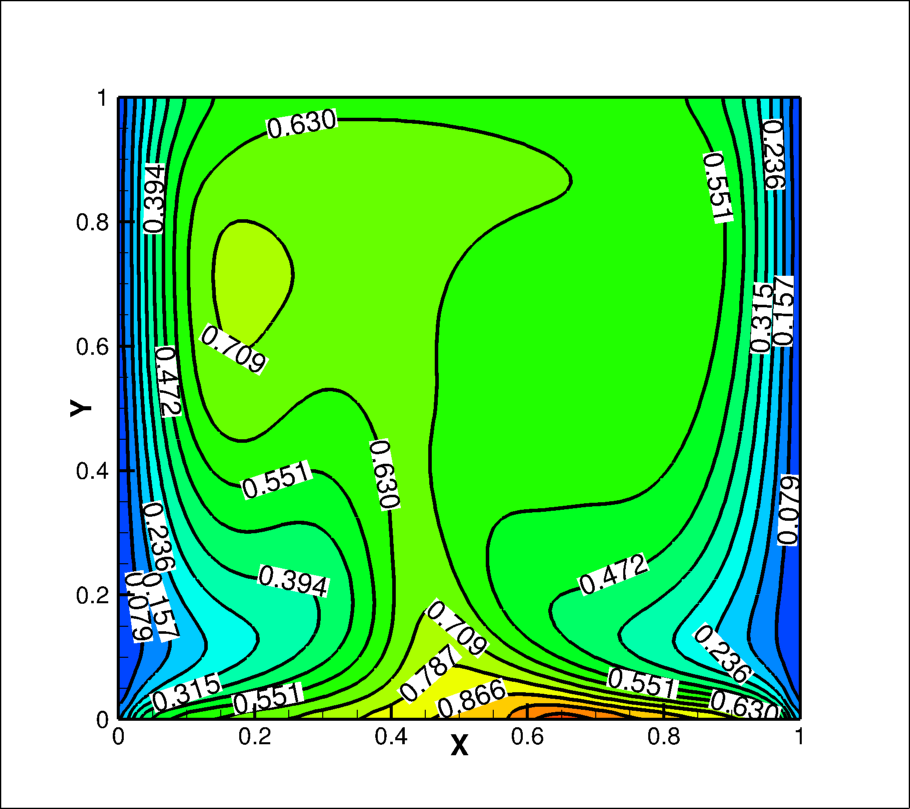}
    \caption{}
    \label{B_G_T_N1}
  \end{subfigure}
  \begin{subfigure}{8cm}
    \centering\includegraphics[width=8cm]{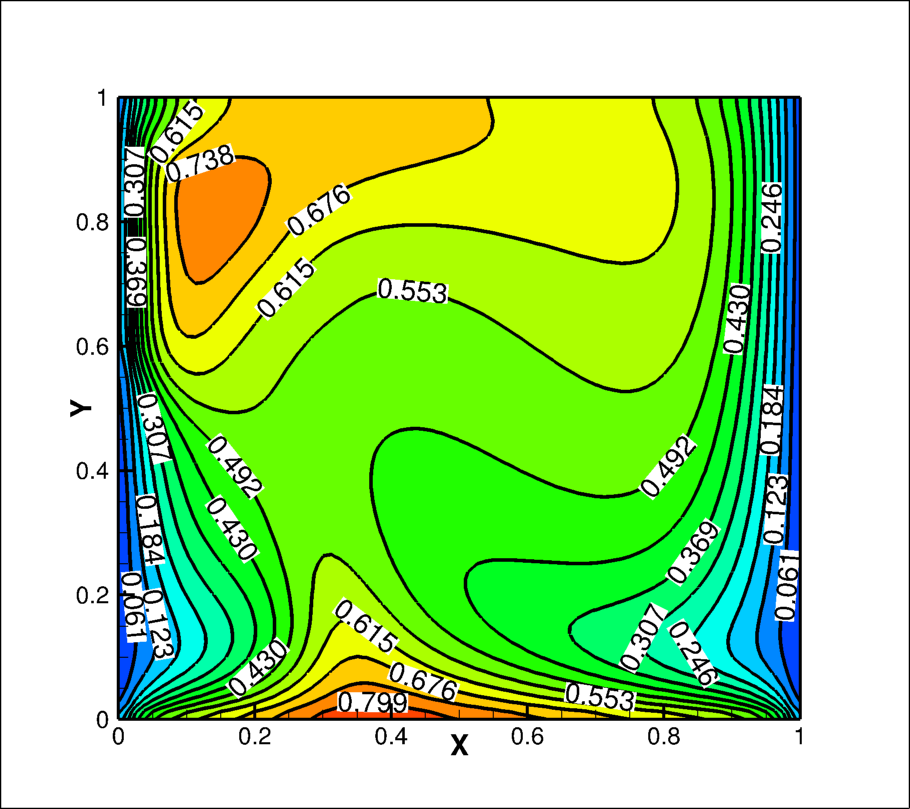}
    \caption{}
    \label{B_G_T_N10}
  \end{subfigure}
  \hspace{1.1cm}
   \begin{subfigure}{8cm}
    \centering\includegraphics[width=8cm]{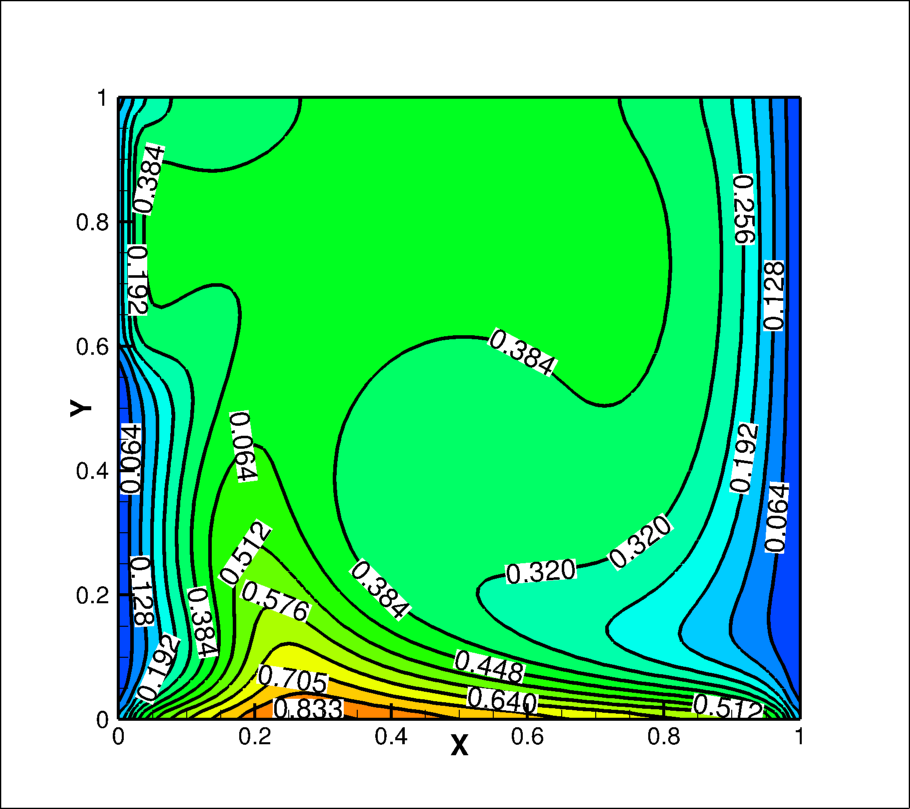}
    \caption{}
    \label{B_G_T_N50}
  \end{subfigure}
  \caption{The contours of the non-dimensional temperature for (a) $Pl=0$  (b) $Pl=1$ (c) $Pl=10$ and (d) $Pl=50$ for case B}
\label{G_T_AR2}
\end{figure} 

Figure \ref{G_T_AR2} depicts the non-dimensional temperature contours for the range of Planck number Pl=0--50. The qualitative behaviour for the case $Pl=0$ (fig \ref{B_G_T_N0}), is same as explained for the fig \ref{G_T_N0}. Moreover, the local heating of the fluid happens near to the semitransparent wall for $Pl=1$, (fig \ref{B_G_T_N1}), this local heating further shifts near to the semitransparent wall for $Pl=10$, also, the clustering of the isothermal lines also increased. Further, more temperature variations are seen in the upper part of the cavity. The local heating of the fluid does not happen for $Pl=50$ as the maximum energy of the collimated beam gets absorbed near to the semitransparent wall as can be envisaged from fig \ref{G_C_N50} and transferred out of the cavity as semitransparent wall is being isothermal. The clustering of isotherm lines near to the semitransparent window have also decreased. The location of the maximum temperature is at the beam strike zone at the bottom of the cavity for $Pl=0$ and 1, however it is at location of two vortices (at the bottom wall) for $Pl=10$ and 50.

\subsubsection{Case C: $h_r$=0.4 $w_r$=0.2}
\begin{figure}[!b]
\begin{subfigure}{8cm}
    \centering\includegraphics[width=8cm]{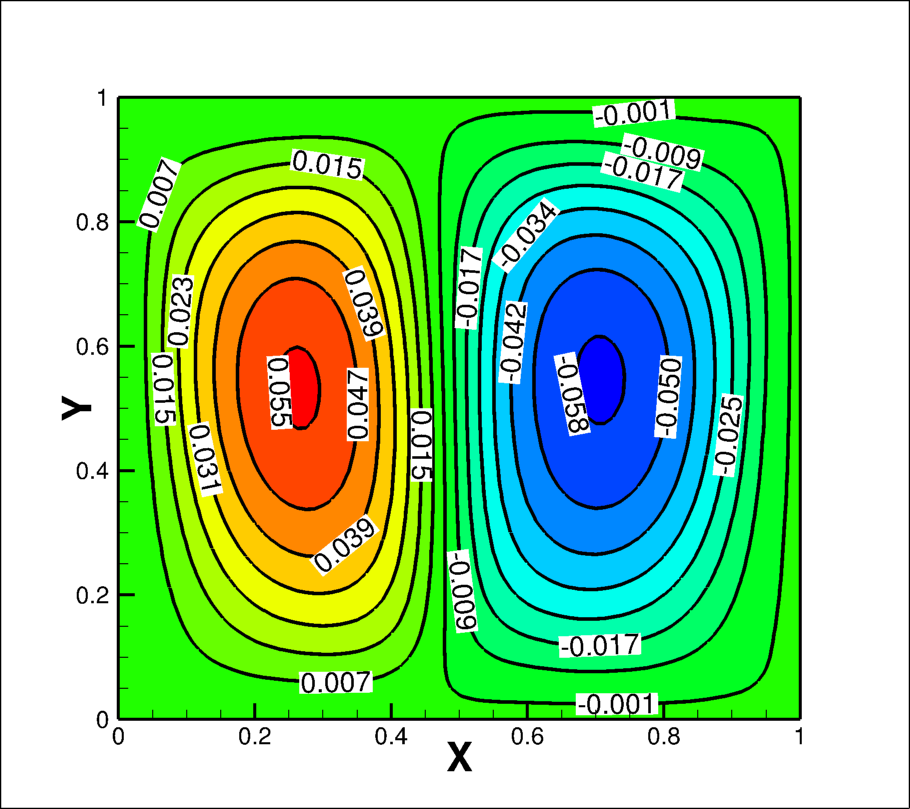}
    \caption{}
    \label{C_G_SF_N0}
  \end{subfigure}
   \begin{subfigure}{8cm}
    \centering\includegraphics[width=8cm]{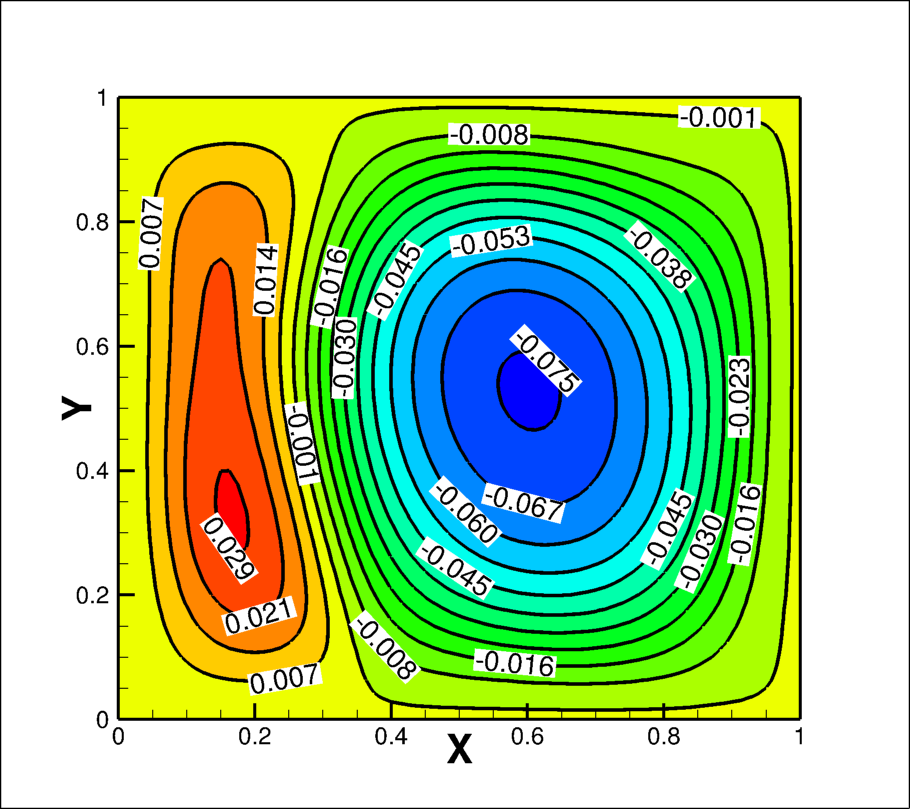}
    \caption{}
    \label{C_G_SF_N1}
  \end{subfigure}
  \begin{subfigure}{8cm}
    \centering\includegraphics[width=8cm]{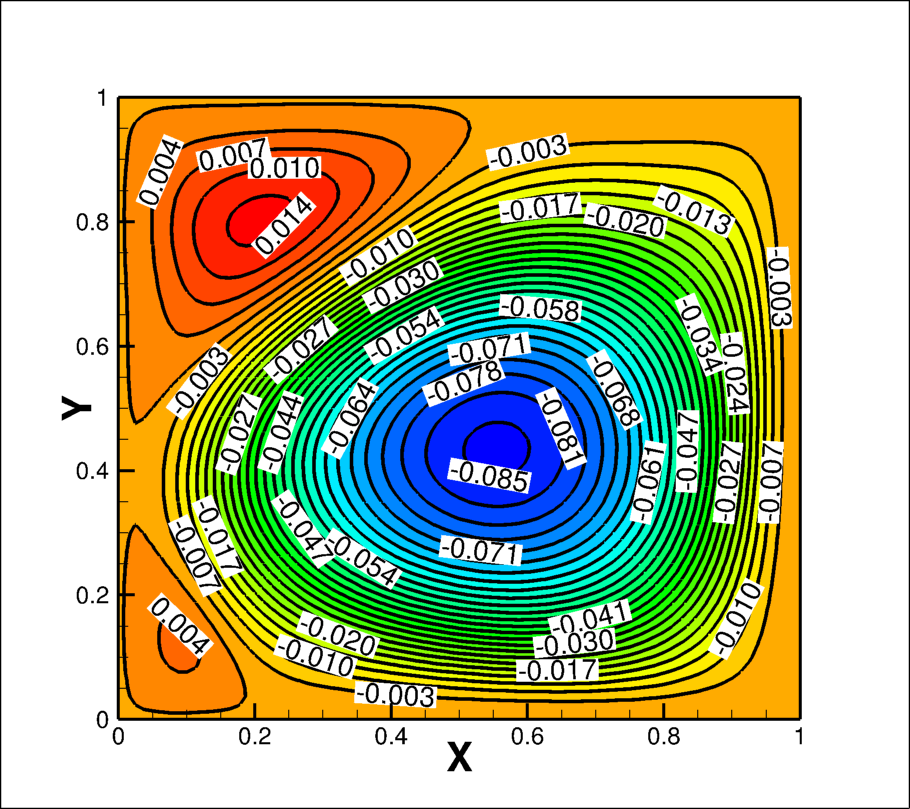}
    \caption{}
    \label{C_G_SF_N10}
  \end{subfigure}
  \hspace{1.1cm}
   \begin{subfigure}{8cm}
    \centering\includegraphics[width=8cm]{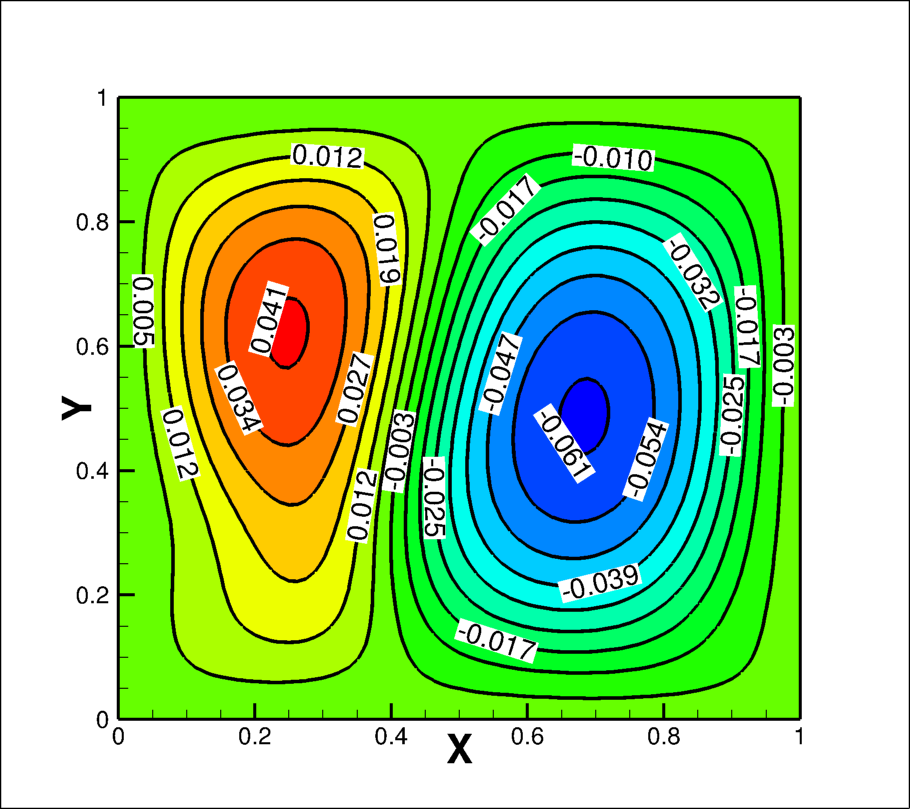}
    \caption{}
    \label{C_G_SF_N50}
  \end{subfigure}
  \caption{The contours of the non-dimensional stream function for (a) $Pl=0$  (b) $Pl=1$ (c) $Pl=10$ and (d) $Pl=50$ for case C}
\label{G_SF_AR3}
\end{figure} 

Unlike to cases A and B for the transparent medium ($Pl=0$) the left vortex is little smaller in size to the right side vortex (compare  (\ref{G_SF_N0}), (\ref{B_G_SF_N0}) to (\ref{C_G_SF_N0})). This is mainly due to the fact that the collimated beam incidence takes place between the non-dimensional length 0.3 to 0.5 on the bottom wall. This incidence length is below to the left vortex, thus higher buoyancy causes the reduction of size of left vortex. One interesting fact to notice that fluid velocity has almost $90^0$ turn in the right vortex at end point of collimated strike. Also, the flow rate in the right vortex is higher than the left vortex unlike to cases A and B for transparent medium with increase of Planck number of the medium, the flow rate in the right vortex keeps on increasing till $Pl=10$. The size of left vortex also keeps on decreasing with Planck number of the medium and the left vortex breaks into two for $Pl=10$. However, total a different situation appears for these two vortices for $Pl=50$. The left vortex has grown and right vortex reduced in the size. The flow rate in left vortex has also increased whereas it is reduced in the right vortex.

\begin{figure}[!t]
\begin{subfigure}{8cm}
    \centering\includegraphics[width=8cm]{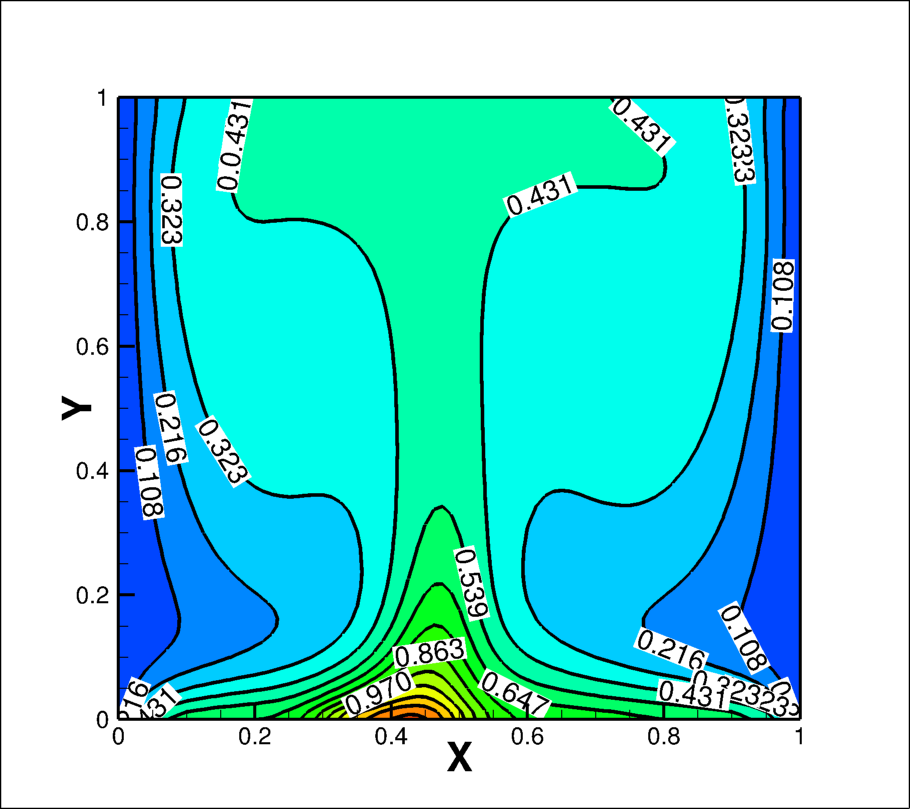}
    \caption{}
    \label{C_G_T_N0}
  \end{subfigure}
   \begin{subfigure}{8cm}
    \centering\includegraphics[width=8cm]{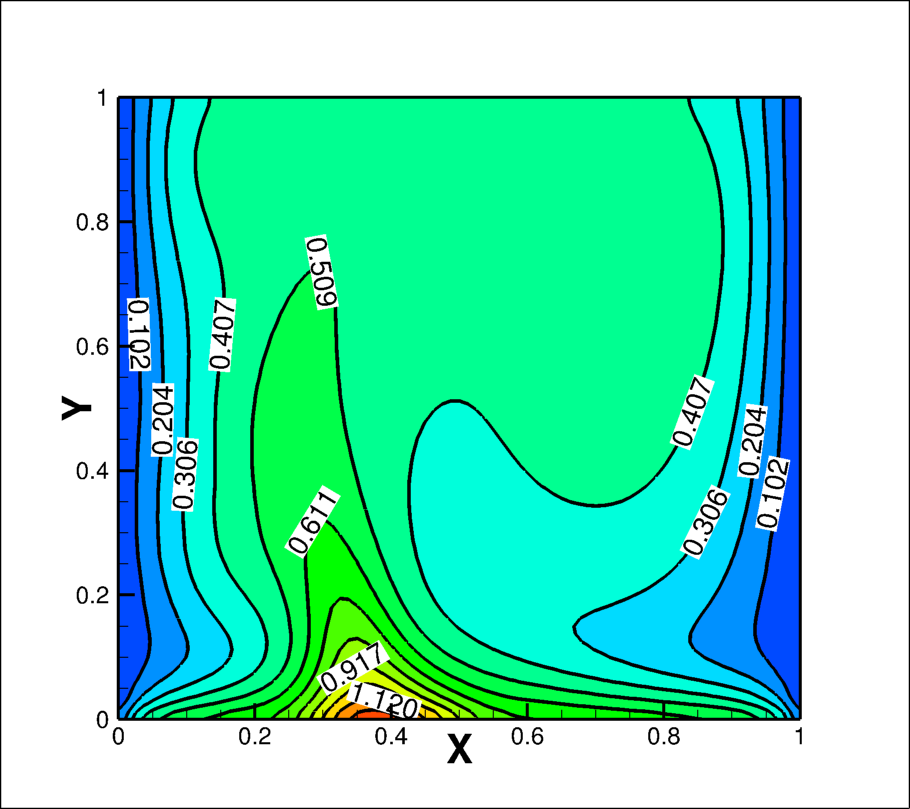}
    \caption{}
    \label{C_G_T_N1}
  \end{subfigure}
  \begin{subfigure}{8cm}
    \centering\includegraphics[width=8cm]{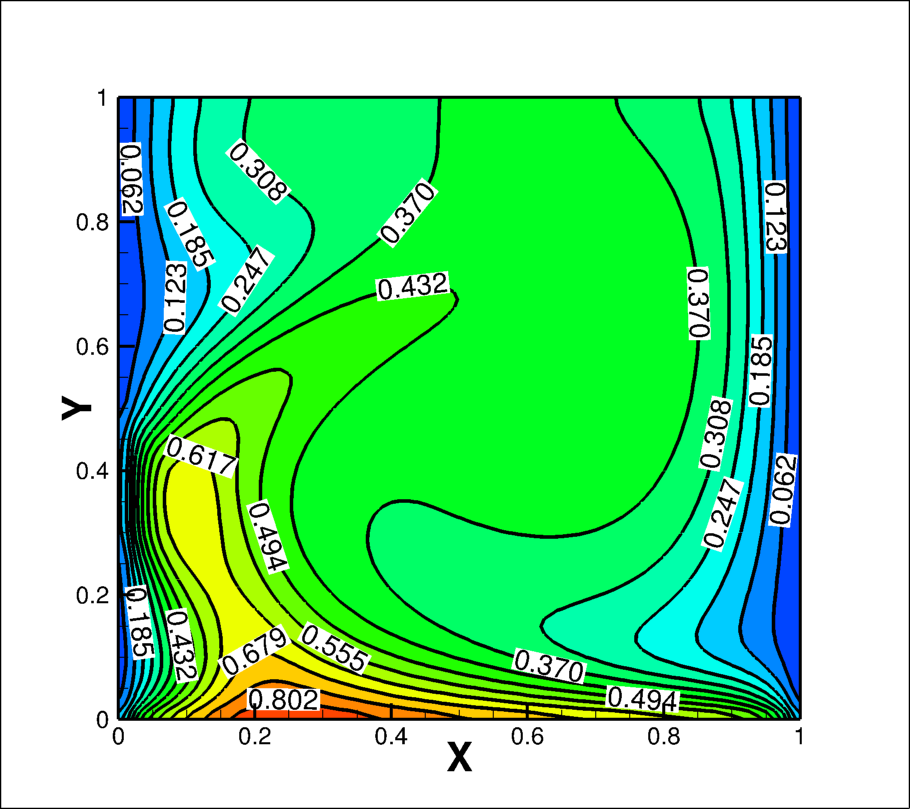}
    \caption{}
    \label{C_G_T_N10}
  \end{subfigure}
  \hspace{1.1cm}
   \begin{subfigure}{8cm}
    \centering\includegraphics[width=8cm]{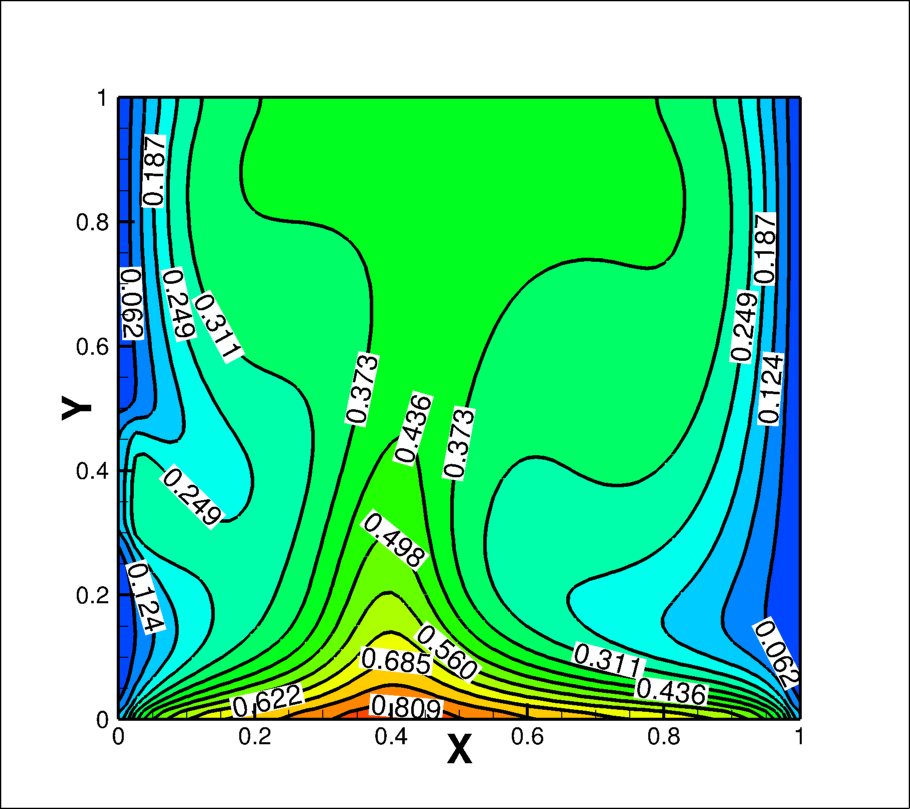}
    \caption{}
    \label{C_G_T_N50}
  \end{subfigure}
  \caption{The contours of the non-dimensional temperature for (a) $Pl=0$  (b) $Pl=1$ (c) $Pl=10$ and (d) $Pl=50$ for case C}
\label{G_T_AR3}
\end{figure} 

The plume is arising from the collimated incidence length and it is almost vertical from the transparent medium ($Pl=0$) (see fig \ref{C_G_T_N0}), this plume is bent towards left for Planck number ($Pl=1$)(see fig \ref{C_G_T_N1}) case. It gets totally bent and nearly touches the left isothermal wall for Planck number $Pl=10$ (see fig \ref{C_G_T_N10}). On contrary to this, the plume is bent toward right for $Pl=50$ (see fig \ref{C_G_T_N50}). The isotherm lines are also clustered and parallel to the semitransparent window for Planck number $Pl=50$. In this case also, the maximum temperature rise can be see on the bottom wall like cases A and B and is at the point of incident of the collimated beam for Planck number 0 and 1 and at the point of plume rise for Planck number 10 and 50.

\subsection{Case D: $h_r$=0.4 $w_r$=0.4}

\begin{figure}[!b]
\begin{subfigure}{8cm}
    \centering\includegraphics[width=8cm]{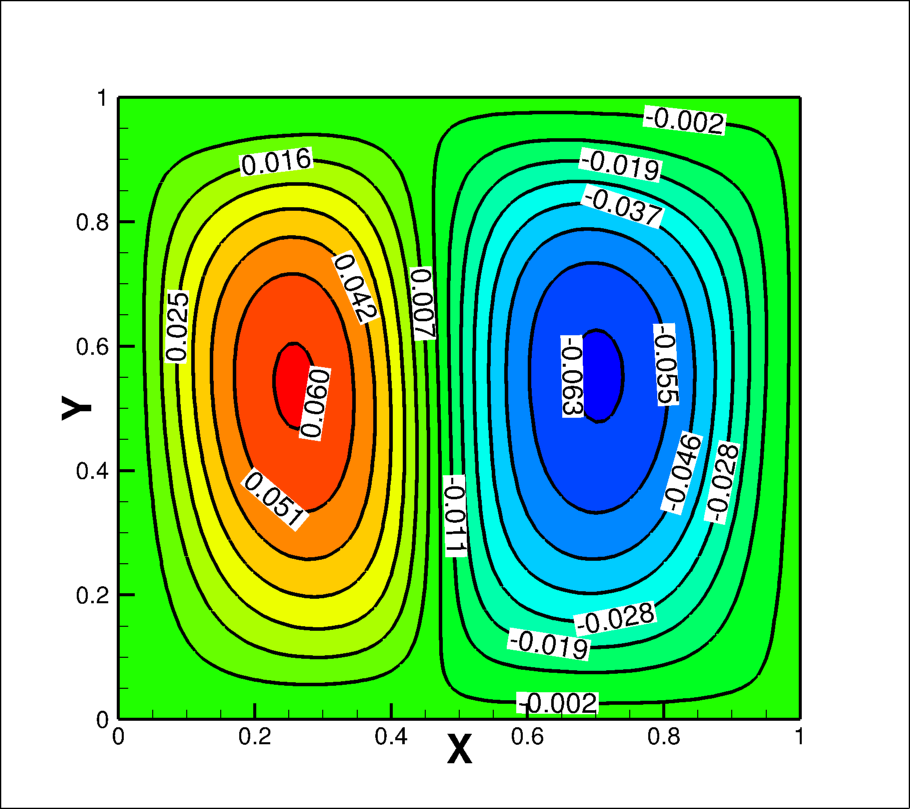}
    \caption{}
    \label{D_G_SF_N0}
  \end{subfigure}
   \begin{subfigure}{8cm}
    \centering\includegraphics[width=8cm]{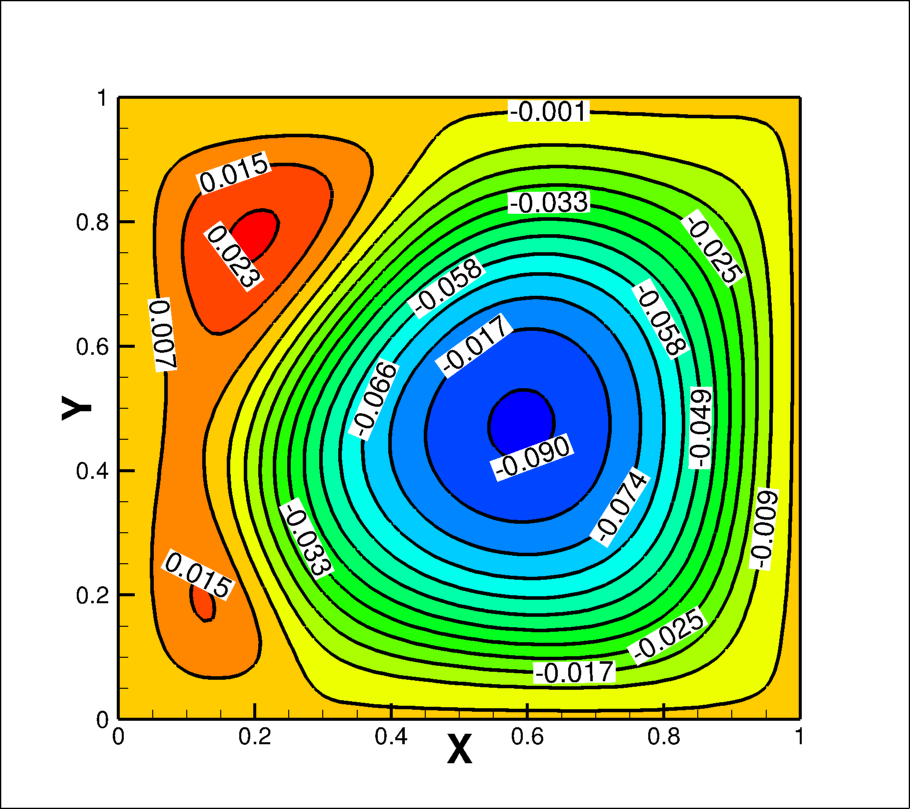}
    \caption{}
    \label{D_G_SF_N1}
  \end{subfigure}
  \begin{subfigure}{8cm}
    \centering\includegraphics[width=8cm]{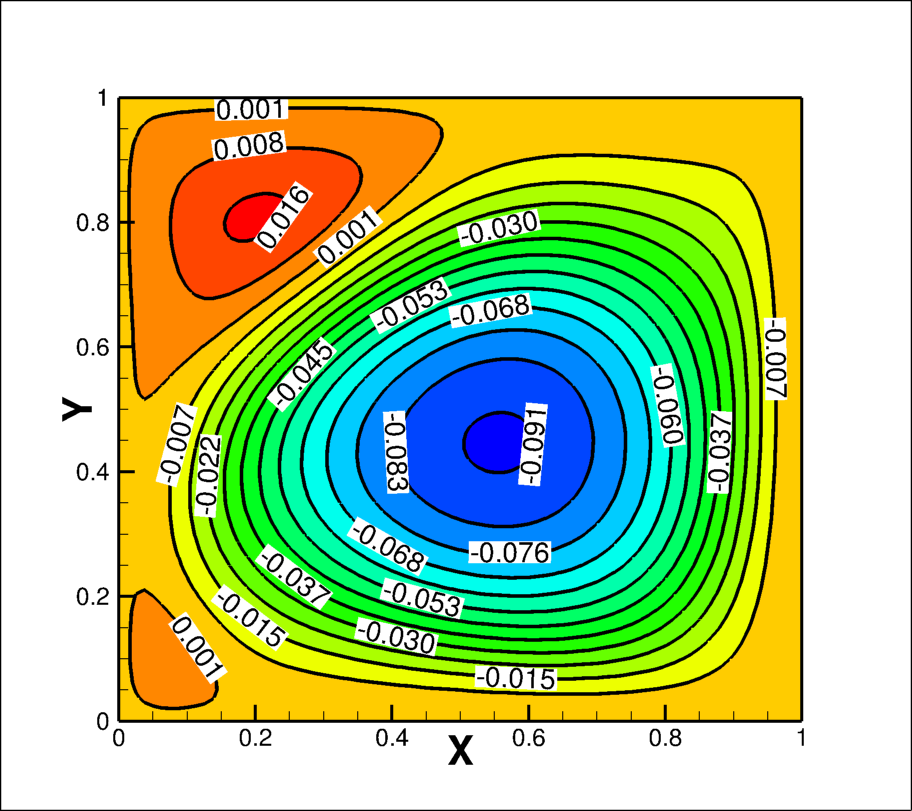}
    \caption{}
    \label{D_G_SF_N10}
  \end{subfigure}
  \hspace{1.1cm}
   \begin{subfigure}{8cm}
    \centering\includegraphics[width=8cm]{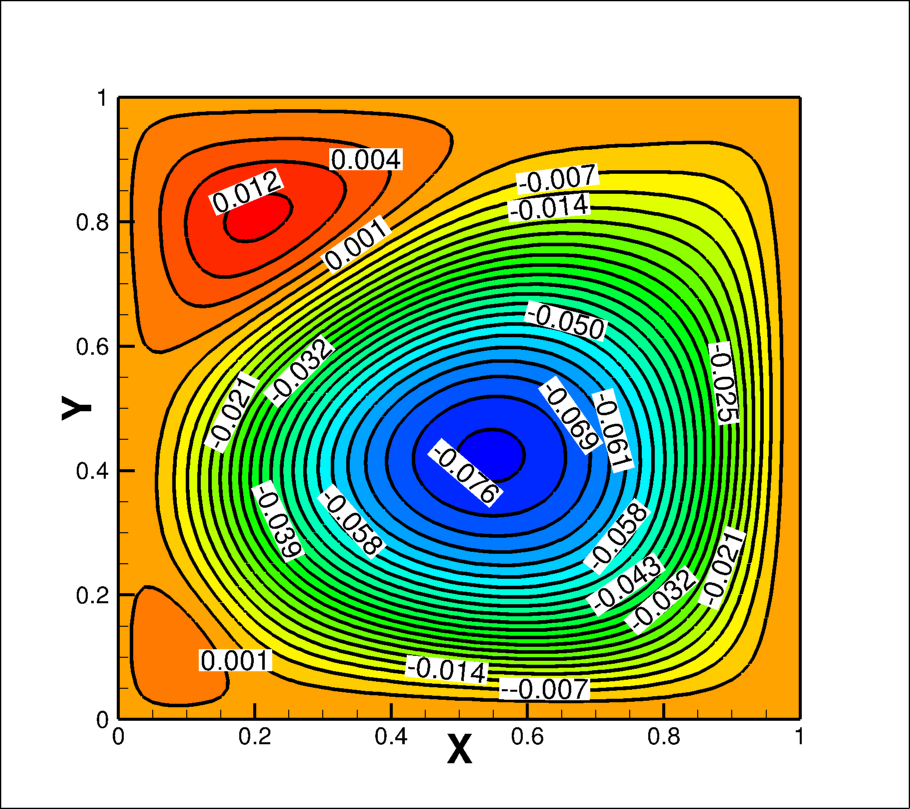}
    \caption{}
    \label{D_G_SF_N50}
  \end{subfigure}
  \caption{The Contours of the non-dimensional stream function for (a) $Pl=0$  (b) $Pl=1$ (c) $Pl=10$ and (d) $Pl=50$ for case D}
\label{G_SF_AR4}
\end{figure} 

There is no major change occurs in the fluid flow behaviour with the increase of window width ratio for the transparent medium case (fig \ref{G_SF_AR4}). Nevertheless, the flow rate in both the vortices increases (compare fig \ref{C_G_SF_N0} and \ref{D_G_SF_N0}). However, the size of the left vortex reduces drastically for $Pl=1$ and breaks into two parts-upper left vortex and lower left vortex but both are connected. Moreover, they are disconnected for $Pl=1$. The lower left vortex has lower flow rate compared to upper left vortex, also the flow rate in right vortex increases. On further increase of the optical thickness of the medium (Pl=50), the flow rates in the right vortex and upper left vortex, decrease and there is no change in lower left vortex.

\begin{figure}[!t]
\begin{subfigure}{8cm}
    \centering\includegraphics[width=8cm]{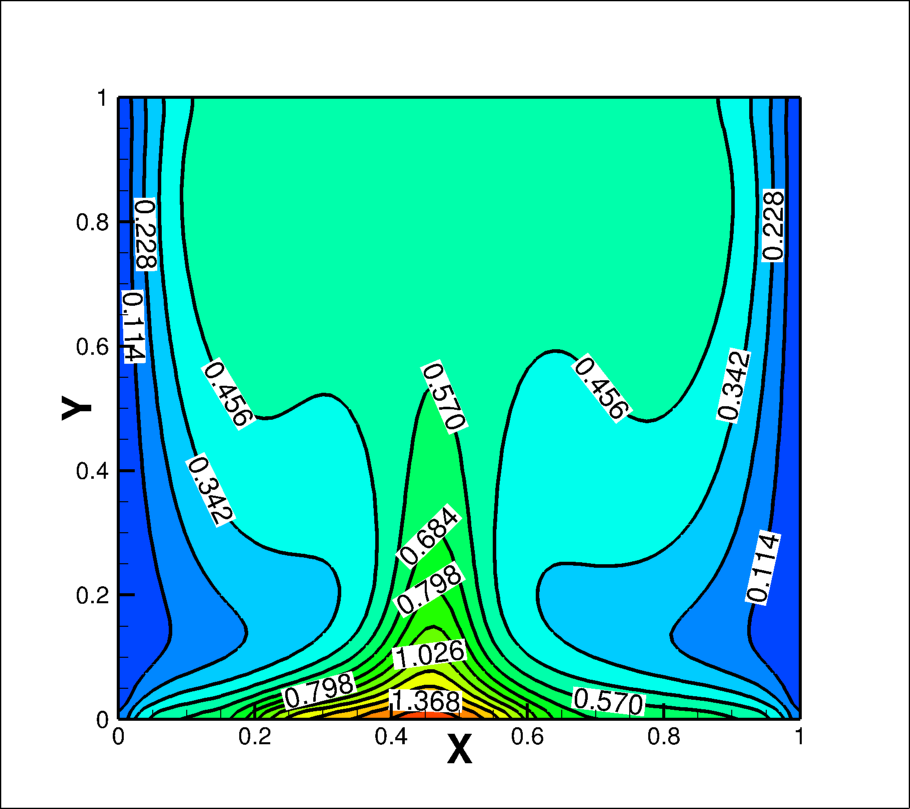}
    \caption{}
    \label{D_G_T_N0}
  \end{subfigure}
   \begin{subfigure}{8cm}
    \centering\includegraphics[width=8cm]{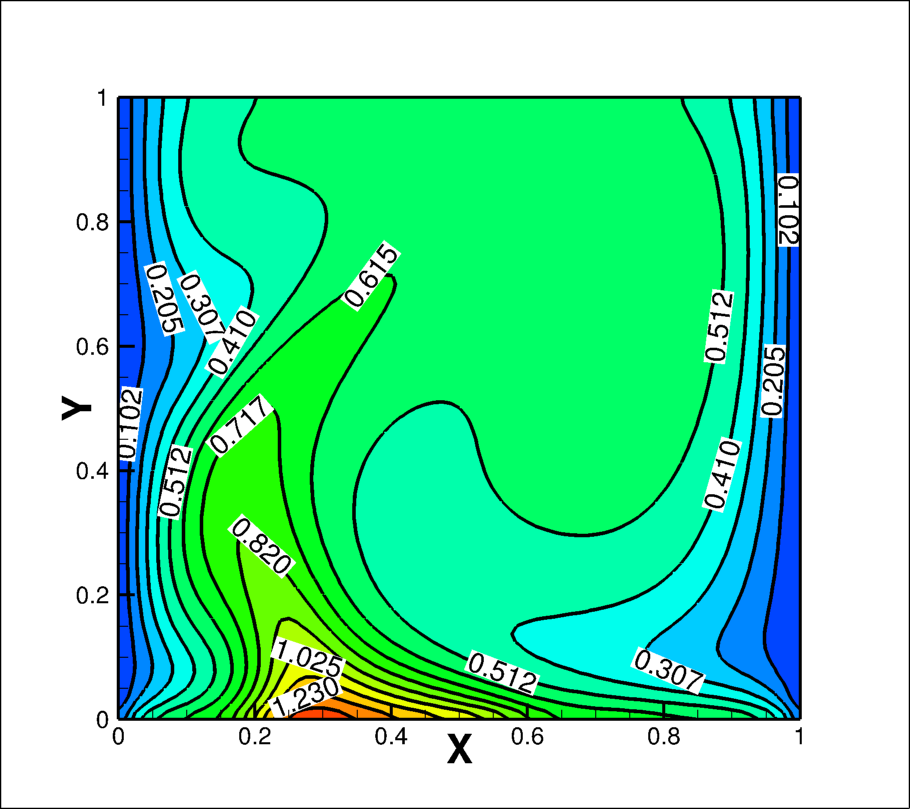}
    \caption{}
    \label{D_G_T_N1}
  \end{subfigure}
  \begin{subfigure}{8cm}
    \centering\includegraphics[width=8cm]{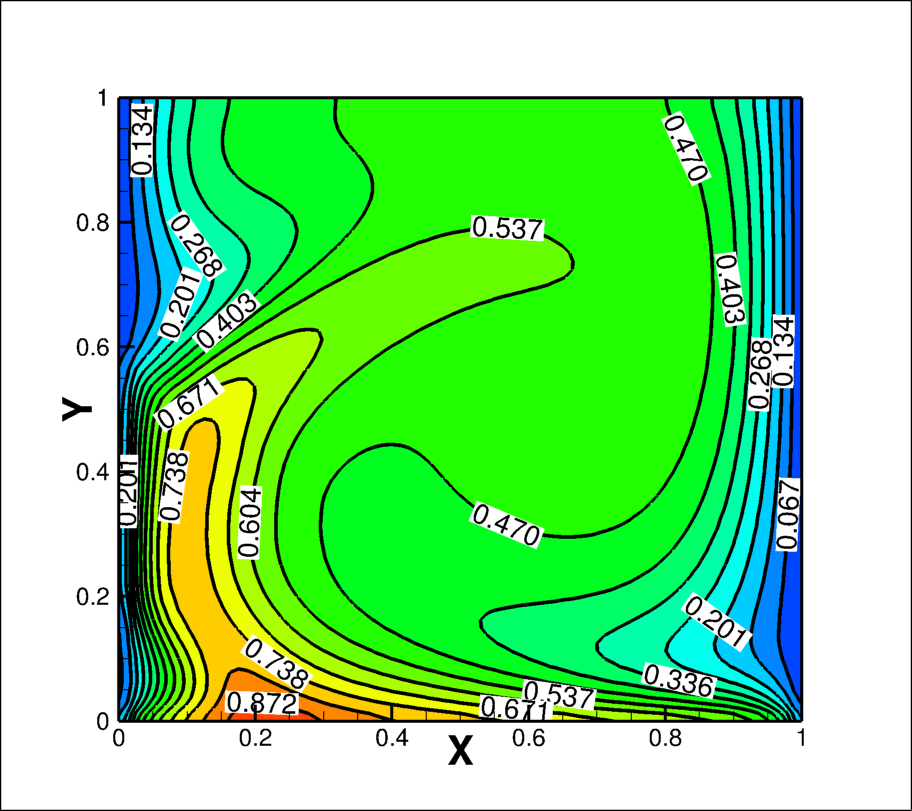}
    \caption{}
    \label{D_G_T_N10}
  \end{subfigure}
  \hspace{1.1cm}
   \begin{subfigure}{8cm}
    \centering\includegraphics[width=8cm]{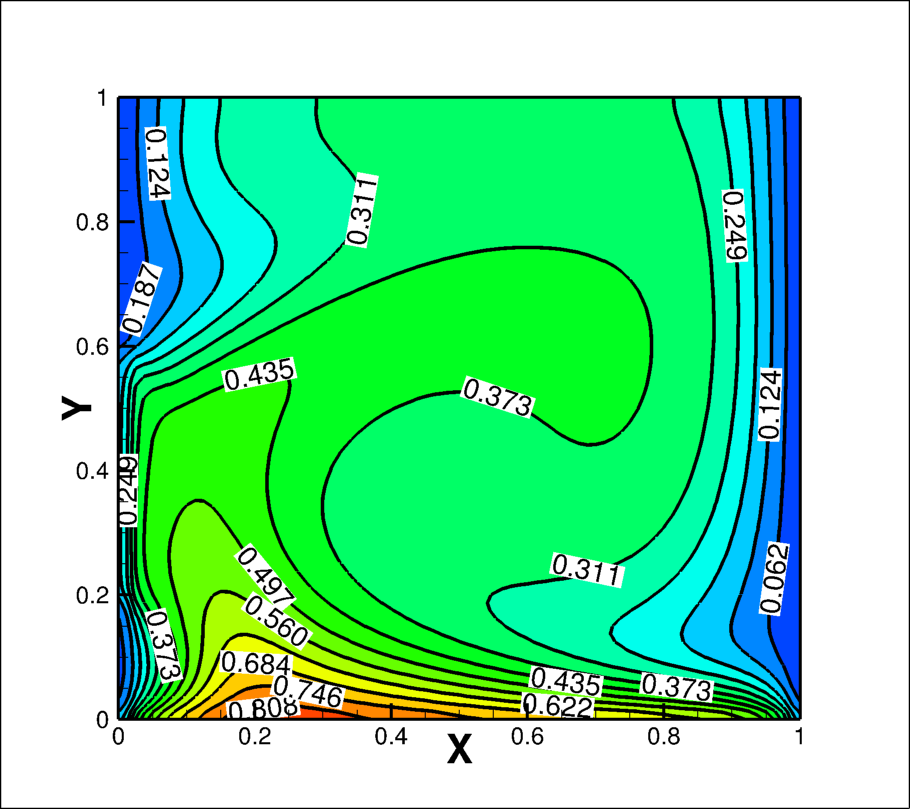}
    \caption{}
    \label{D_G_T_N50}
  \end{subfigure}
  \caption{The contours of the non-dimensional temperature for (a) $Pl=0$  (b) $Pl=1$ (c) $Pl=10$ and (d) $Pl=50$ for case D}
\label{G_T_AR4}
\end{figure} 

Similarly, no major change in the temperature contours (fig \ref{G_T_AR4}) appear for transparent and the participating medium cases except the plume is near to the left wall for Pl=50 case (fig \ref{D_G_T_N10}) where isothermal lines are still bent towards left. The isotherm lines are parallel and closely packed near to the semitransparent window. 

\subsection{Variation of Non-dimensional Temperature for Different Aspect Ratios}

\begin{figure}[!t]
\begin{subfigure}{8cm}
    \centering\includegraphics[width=8cm]{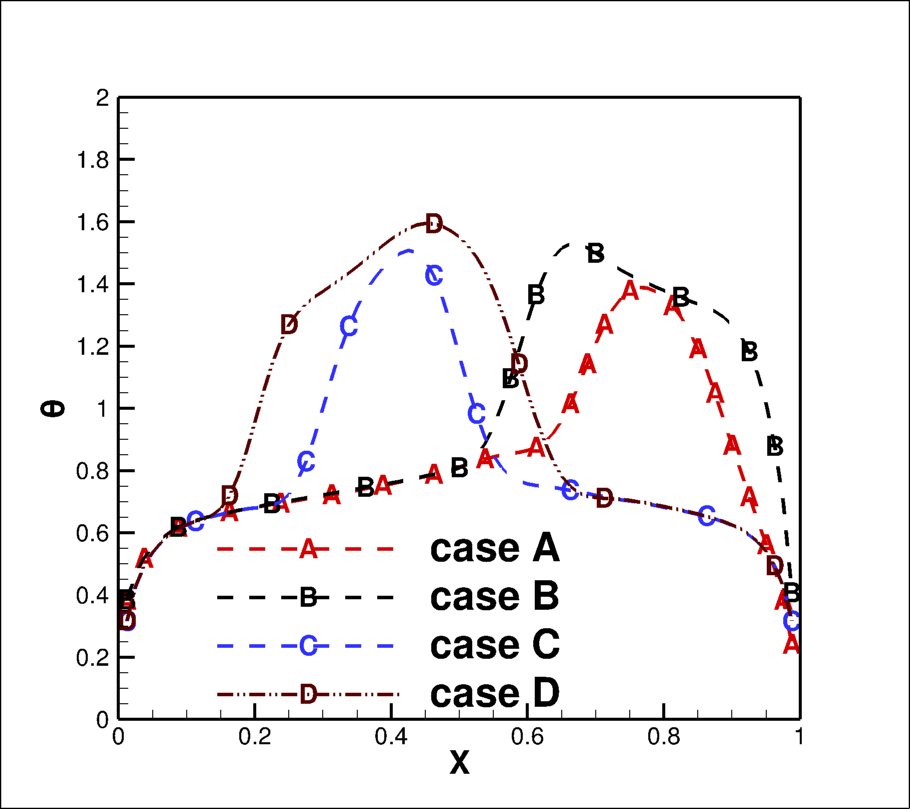}
    \caption{}
    \label{AsA_N0}
  \end{subfigure}
   \begin{subfigure}{8cm}
    \centering\includegraphics[width=8cm]{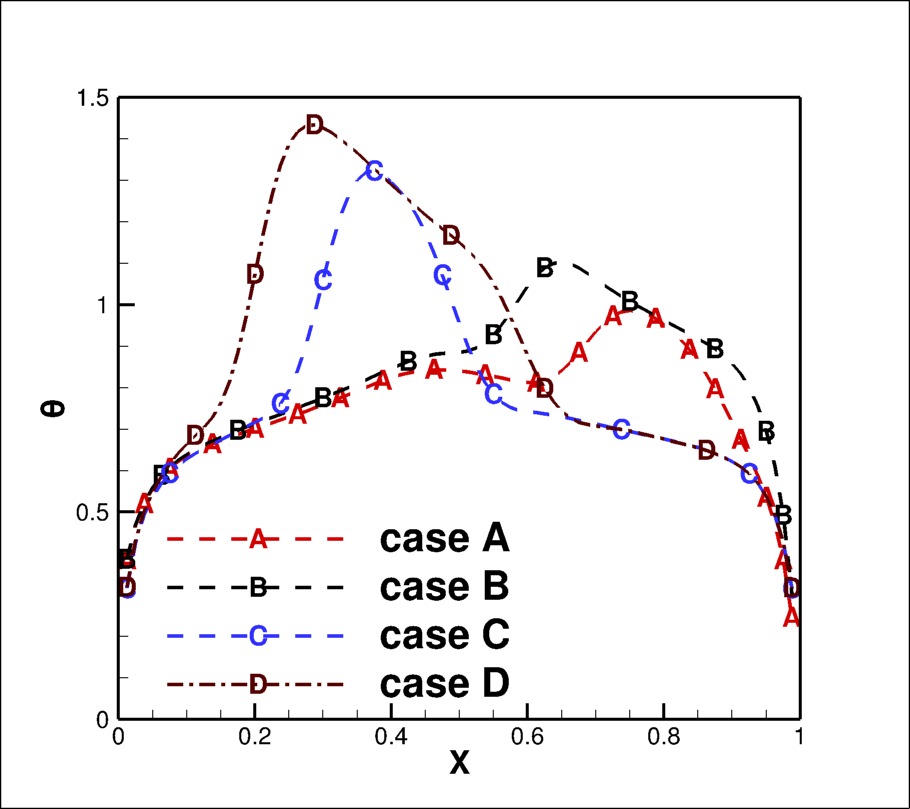}
    \caption{}
    \label{AsB_N1}
  \end{subfigure}
  \begin{subfigure}{8cm}
    \centering\includegraphics[width=8cm]{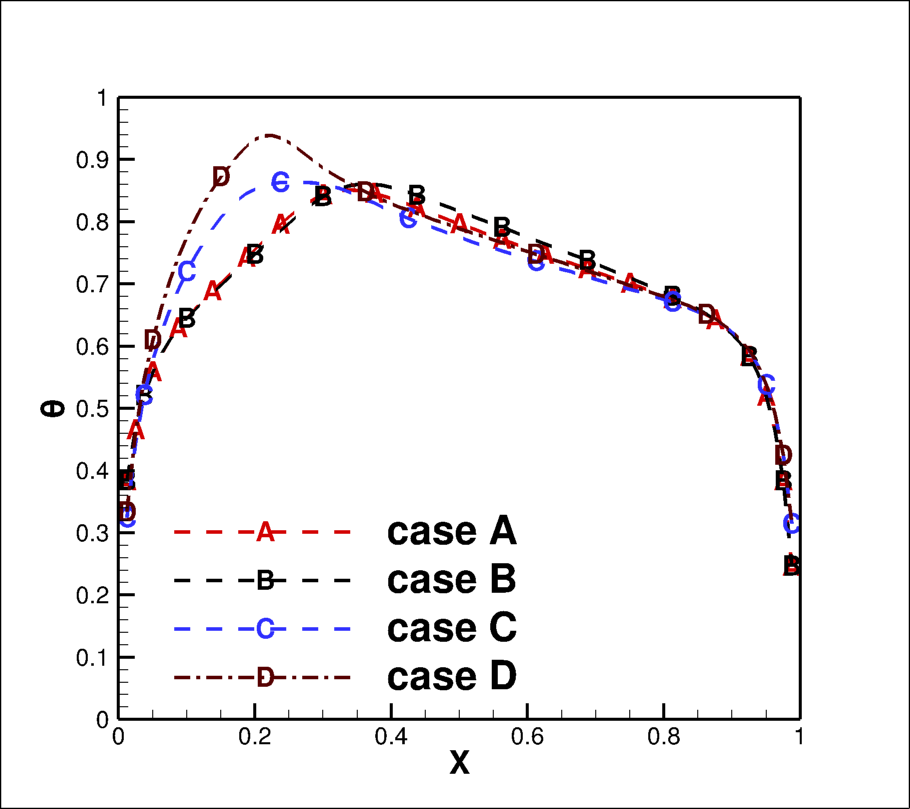}
    \caption{}
    \label{AsC_N10}
  \end{subfigure}
  \hspace{1.1cm}
   \begin{subfigure}{8cm}
    \centering\includegraphics[width=8cm]{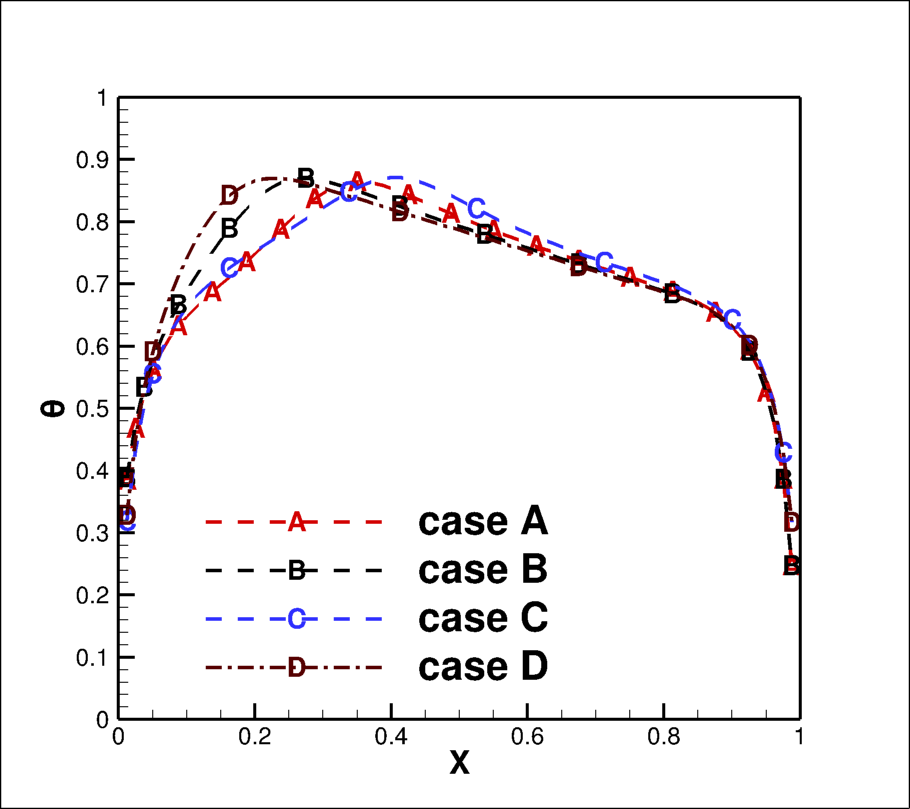}
    \caption{}
    \label{AsD_N50}
  \end{subfigure}
  \caption{The variations of the non-dimensional temperature on the bottom wall for (a) $Pl=0$  (b) $Pl=1$ (c) $Pl=10$ and (d) $Pl=50$}
\label{AsN_AR2}
\end{figure} 

Figure \ref{AsN_AR2} depicts the non-dimensional temperature variation on the bottom wall for the different cases, also for the range of Planck numbers of the medium $Pl=0-50$. The peak of the temperature profile shifts to left from case A to C, for Planck numbers 0 and 1, Whereas temperature profile almost remains same for case A and B for Planck number 10. Furthermore, no ordered way shifting of temperature peak appears for the different cases for Planck number 50. The peak values of non-dimensional temperature are 1.3, 1.5, 1.45 and 1.6 for cases A, B, C and D, respectively and appears at respective non-dimensional distance 0.8, 0.7, 0.4 and 0.45 from the left wall for Planck number 0. Similarly, the peak values are 0.9, 1.1, 1.3 and 1.4 for cases A, B, C and D respectively at respective non-dimensional distance 0.75, 0.65, 0.4 and 0.35 from the left wall for Planck number 1. Moreover these values for Planck number 10 are 0.7 for cases A, B and C and for case D it is 0.95, whereas the maximum value is 0.85 for $Pl=50$ for all the cases.

\begin{table}[!b]
\caption{The non-dimensional maximum stream function values for various cases over the range of Planck numbers}
\label{SF_As_table}
\begin{tabular}{|l|c|l|c|c|c|c|c|c|c|c|c|c|}
\hline
\multirow{2}{*}{Pl} 
& \multicolumn{3}{c|}{case A}                 & \multicolumn{3}{c|}{case B}              & \multicolumn{3}{c|}{case C}              & \multicolumn{3}{c|}{case D}              \\ \cline{2-13} 
& \multicolumn{2}{c|}{Left}  & Right     & \multicolumn{2}{c|}{Left}  & Right  & \multicolumn{2}{c|}{Left}  & Right  & \multicolumn{2}{c|}{Left}  & Right  \\ \hline
0                   & \multicolumn{2}{c|}{0.052} & -0.043    & \multicolumn{2}{c|}{0.062} & -0.051 & \multicolumn{2}{c|}{0.055} & -0.058 & \multicolumn{2}{c|}{0.060} & -0.063 \\ \hline
1                   & \multicolumn{2}{c|}{0.032} & -0.064    & \multicolumn{2}{c|}{0.037} & -0.079 & \multicolumn{2}{c|}{0.029} & -0.075 & 0.015(b)     & 0.023(t)    & -0.090 \\ \hline
10                  & \multicolumn{2}{c|}{0.013} & -0.068    & 0.019(b)     & 0.002(t)    & -0.061 & 0.004(b)     & 0.014(t)    & -0.085 & 0.001(b)     & 0.016(t)    & -0.091 \\ \hline
50                  & \multicolumn{2}{c|}{0.020} & -0.071    & \multicolumn{2}{c|}{0.008} & -0.071 & \multicolumn{2}{c|}{0.041} & -0.061 & 0.001(b)     & 0.012(t)    & -0.076 \\ \hline
\end{tabular}
\end{table}

The maximum non-dimensional stream function values in the vortices for Planck numbers $Pl=0$ to $Pl=50$ and all cases have been presented in Table \ref{SF_As_table}. As the left vortex breaks into two parts, thus the stream function values of these two vortex have been shown in table for left vortex by bifurcating into two parts. The first bifurcated part shows stream function of upper left vortex and second bifurcated part shows the stream function of the lower left vortex. There is no monotone order for increasing or decreasing. The maximum and minimum values of stream function are found right vortex for Planck number 10 for case D, and left vortex for Planck number 0 and case B, respectively.

\begin{table}[!b]
\centering
\caption{The maximum non-dimensional temperature inside the cavity for different cases over a range of Planck numbers}
\label{MaxT_As}
\begin{tabular}{|l|l|l|l|l|l|l|l|l|}
\hline
\multirow{1}{*}{Pl} 
& \multicolumn{2}{c|}{case A} & \multicolumn{2}{c|}{case B} & \multicolumn{2}{c|}{case C} & \multicolumn{2}{c|}{case D} \\ \hline
0 & \multicolumn{2}{l|}{1.388} & \multicolumn{2}{l|}{1.529} & \multicolumn{2}{l|}{1.509} & \multicolumn{2}{l|}{1.596} \\ \hline
1 & \multicolumn{2}{l|}{0.986} & \multicolumn{2}{l|}{1.102} & \multicolumn{2}{l|}{1.324} & \multicolumn{2}{l|}{1.434} \\ \hline
10 & \multicolumn{2}{l|}{0.851} & \multicolumn{2}{l|}{0.860} & \multicolumn{2}{l|}{0.863} & \multicolumn{2}{l|}{0.939} \\ \hline
50 & \multicolumn{2}{l|}{0.864} & \multicolumn{2}{l|}{0.869} & \multicolumn{2}{l|}{0.872} & \multicolumn{2}{l|}{0.870} \\ \hline
\end{tabular}
\end{table}
Similarly, the maximum non-dimensional temperature inside the cavity is shown in Table \ref{MaxT_As}. In few scenarios, the maximum non-dimensional temperature has increased beyond one. The maximum non-dimensional temperature is found for case D and Planck number $Pl=0$, and the minimum non-dimensional temperature is found for case A and Planck number $Pl=10$.

\subsubsection{Variation of Nusselt Number for different Aspect Ratios Cases}

\begin{figure}[!t]
\begin{subfigure}{8cm}
    \centering\includegraphics[width=8cm]{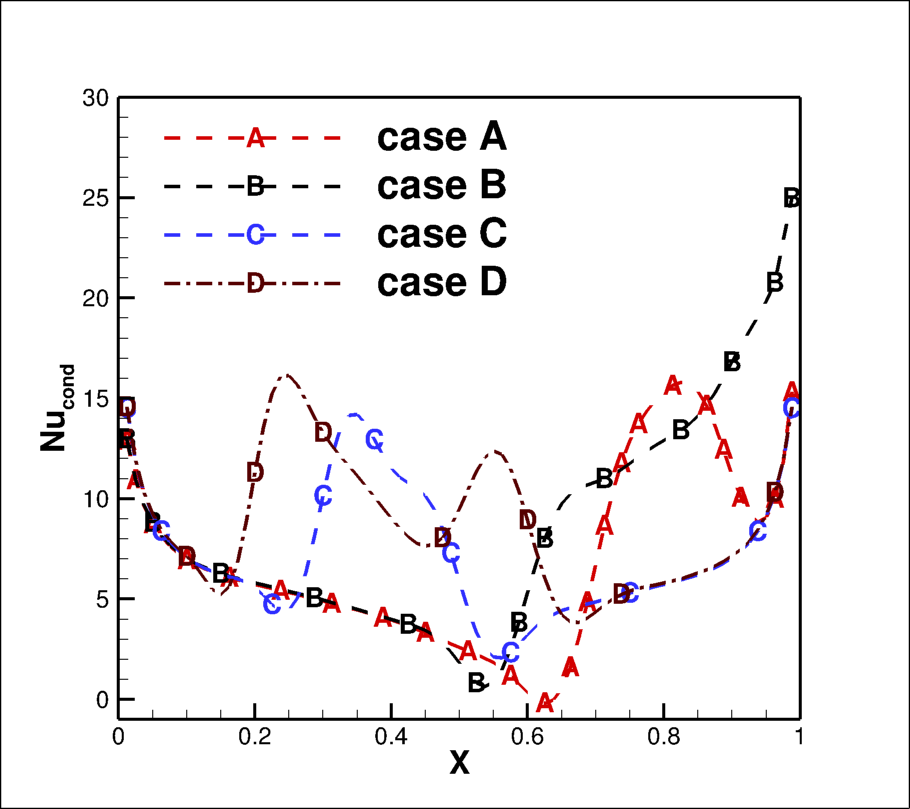}
    \caption{}
    \label{AsA_cond_NuN0}
  \end{subfigure}
   \begin{subfigure}{8cm}
    \centering\includegraphics[width=8cm]{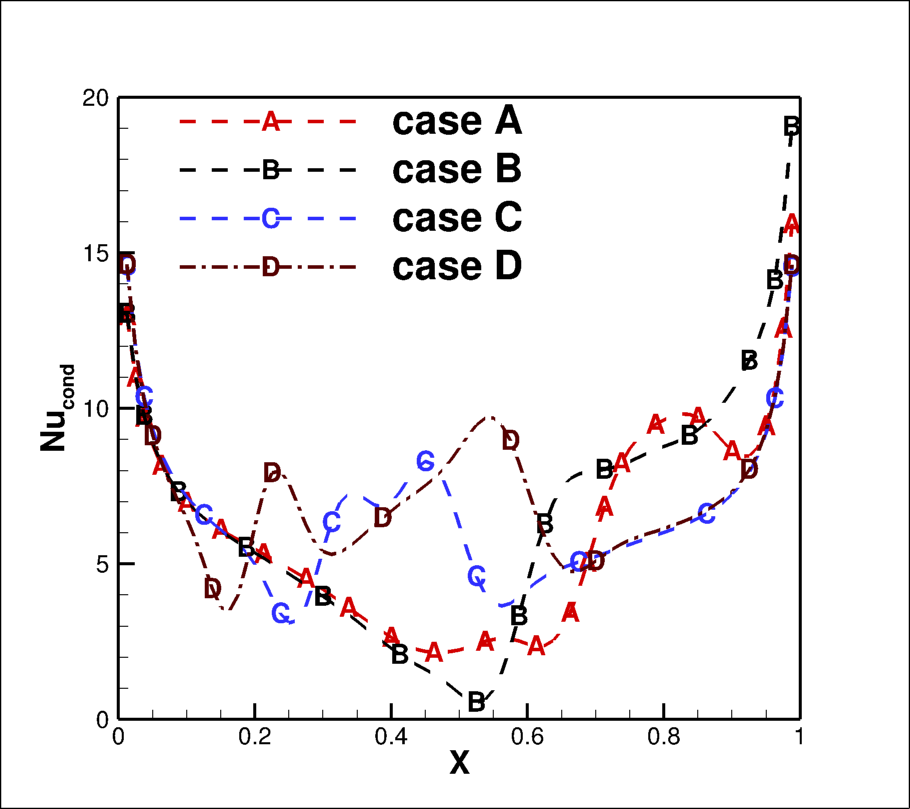}
    \caption{}
    \label{AsA_cond_NuN1}
  \end{subfigure}
  \begin{subfigure}{8cm}
    \centering\includegraphics[width=8cm]{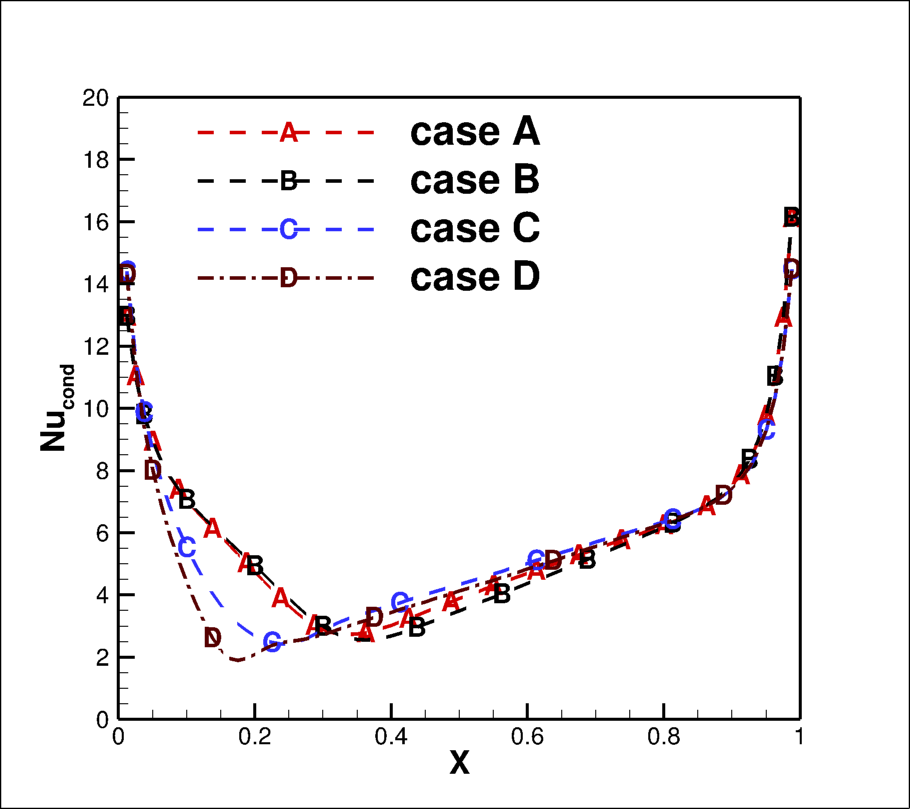}
    \caption{}
    \label{AsA_cond_NuN10}
  \end{subfigure}
  \hspace{1.1cm}
   \begin{subfigure}{8cm}
    \centering\includegraphics[width=8cm]{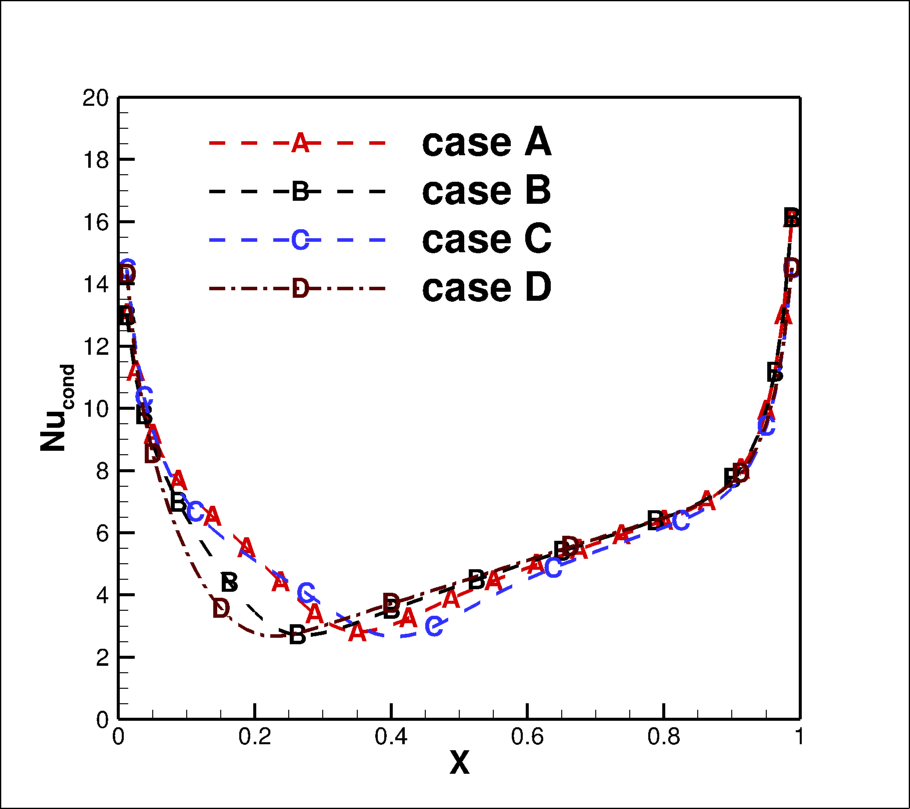}
    \caption{}
    \label{AsA_cond_NuN50}
  \end{subfigure}
  \caption{The variation of the conduction Nusselt number on the bottom wall for the different cases for (a) $Pl=0$  (b) $Pl=1$ (c) $Pl=10$ and (d) $Pl=50$}
\label{As_cond_bot_Nu}
\end{figure} 

Figure \ref{As_cond_bot_Nu} shows the variation of conduction Nusselt number on the bottom wall for different cases for a range of Planck numbers. The conduction Nusselt number goes to minimum (almost zero) before the strike length of collimated beam, rises to maximum ($Nu_{cond}$=16) in the strike length, decreases at the end of the strike length then goes up till end of the bottom wall for case A and for Planck number zero (see fig \ref{AsA_cond_NuN0}). Similarly, the minimum conduction Nusselt number ($Nu_{cond}$=1) is obtained just before strike point of the collimated beam for case B for $Pl=0$, afterwards, the sharp rise in the conduction Nusselt number curve appears at the start of the strike zone of collimated beam for case B, and afterward low rate of increase in conduction Nusselt number happens and finally the highest Nusselt number is achieved at the end. Moreover, the minimum conduction Nusselt number ($Nu_{cond}$=3) is obtained at the end of the strike point of the strike point for case C for $Pl=0$. The conduction Nusselt number curve has two peaks in the strike zone of the collimated beam for $Pl=0$ and for case D. The characteristics of conduction Nusselt number graph for cases and for $Pl=1$ are similar to non-participating medium ($Pl=0$) except low and high peaks of the curve in the strike zone of the collimated beam have reduced. These peaks have disappared for Planck number $Pl=10$. The minimum point in the conduction Nusselt number curve is same for case A and B, however, if shifts to the left for case C and further shifting happens for case D. Other than the inflection point, the conduction Nusselt number curve remain same for all cases. The similar behaviour is observed for Planck number 50. One interesting fact is to notice is that conduction Nusselt number behaviour for case A and C is similar and almost same for case B and D. The minimum Nusselt number obtained is 2 for $Pl=10$ and 50, and maximum is 16 which is obtained at the right end of the bottom wall.

\begin{figure}[!t]
\begin{subfigure}{8cm}
    \centering\includegraphics[width=8cm]{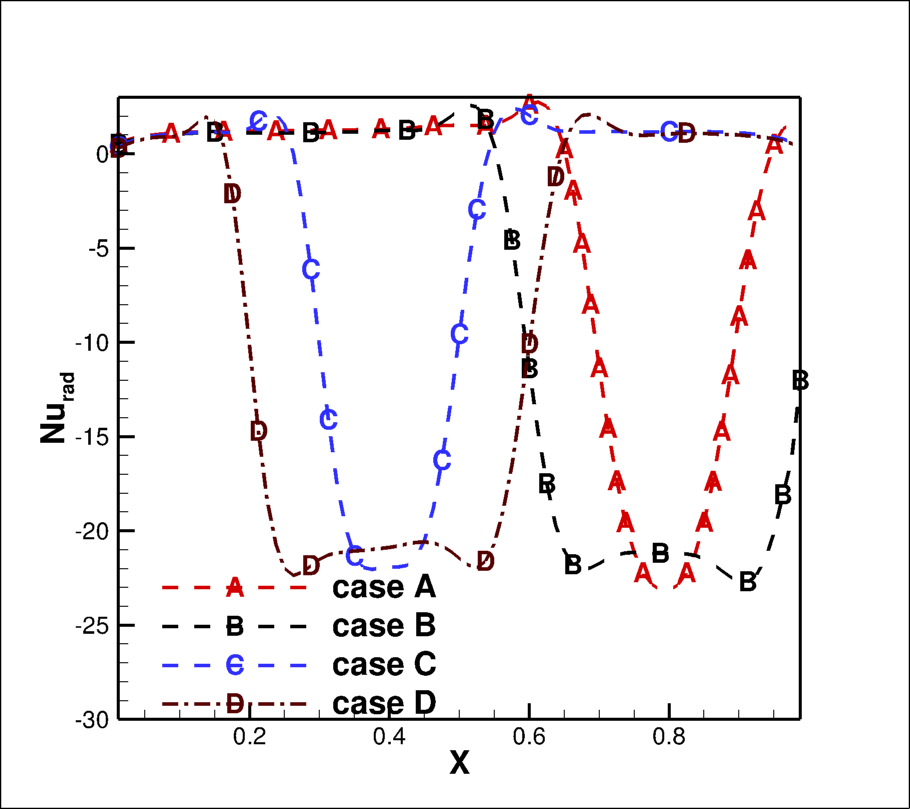}
    \caption{}
    \label{AsA_rad_NuN0}
  \end{subfigure}
   \begin{subfigure}{8cm}
    \centering\includegraphics[width=8cm]{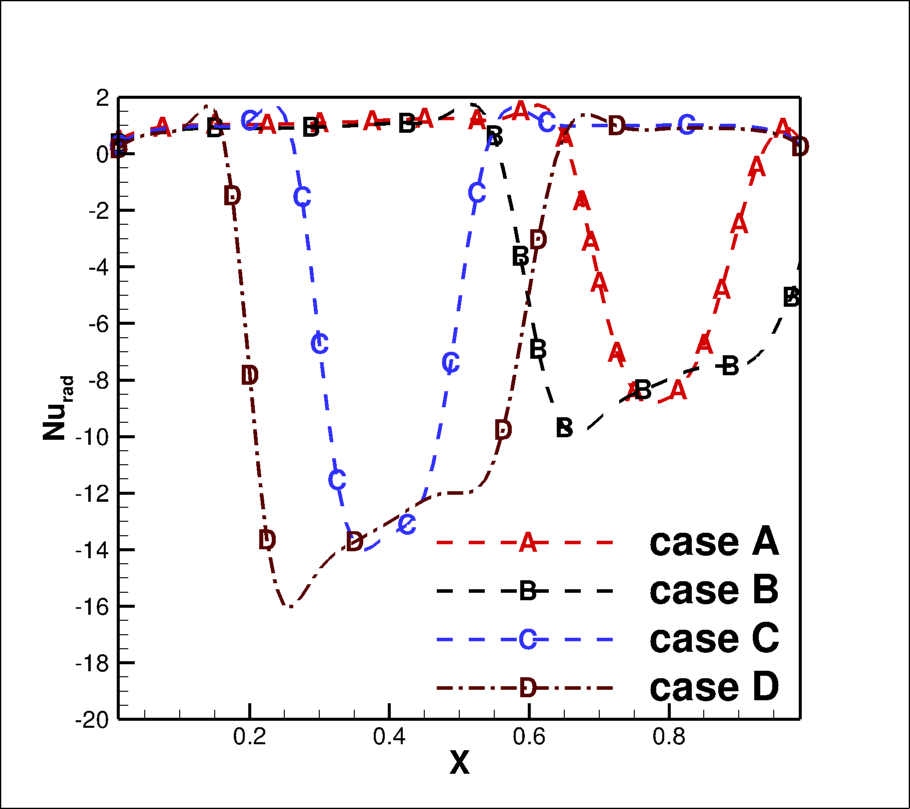}
    \caption{}
    \label{AsB_rad_NuN1}
  \end{subfigure}
  \begin{subfigure}{8cm}
    \centering\includegraphics[width=8cm]{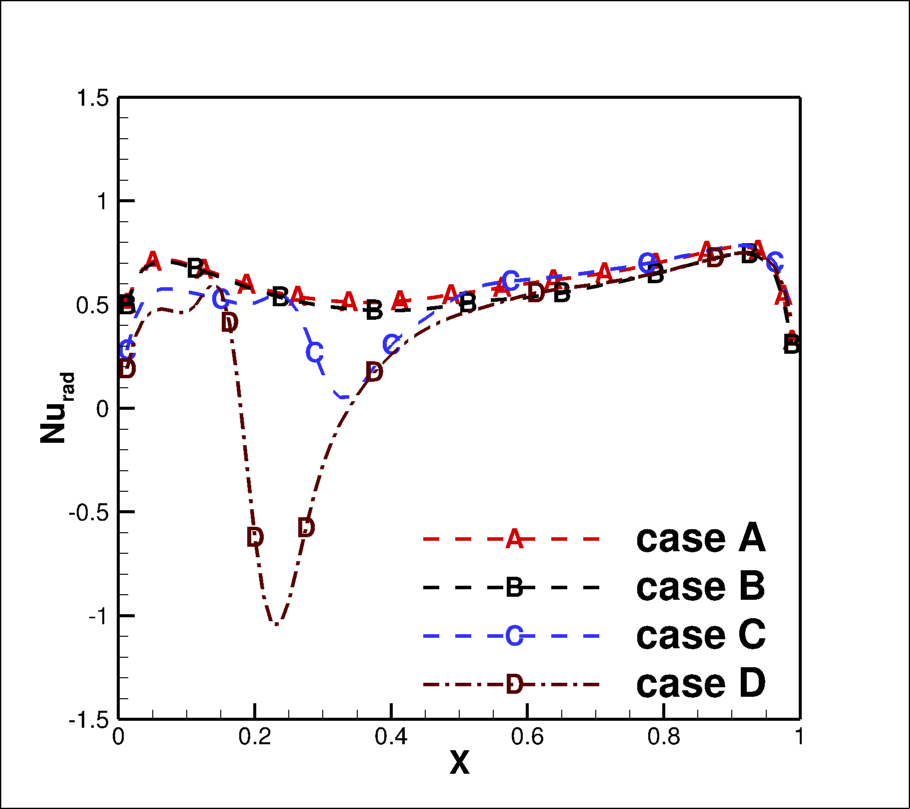}
    \caption{}
    \label{AsC_rad_NuN10}
  \end{subfigure}
  \hspace{1.1cm}
   \begin{subfigure}{8cm}
    \centering\includegraphics[width=8cm]{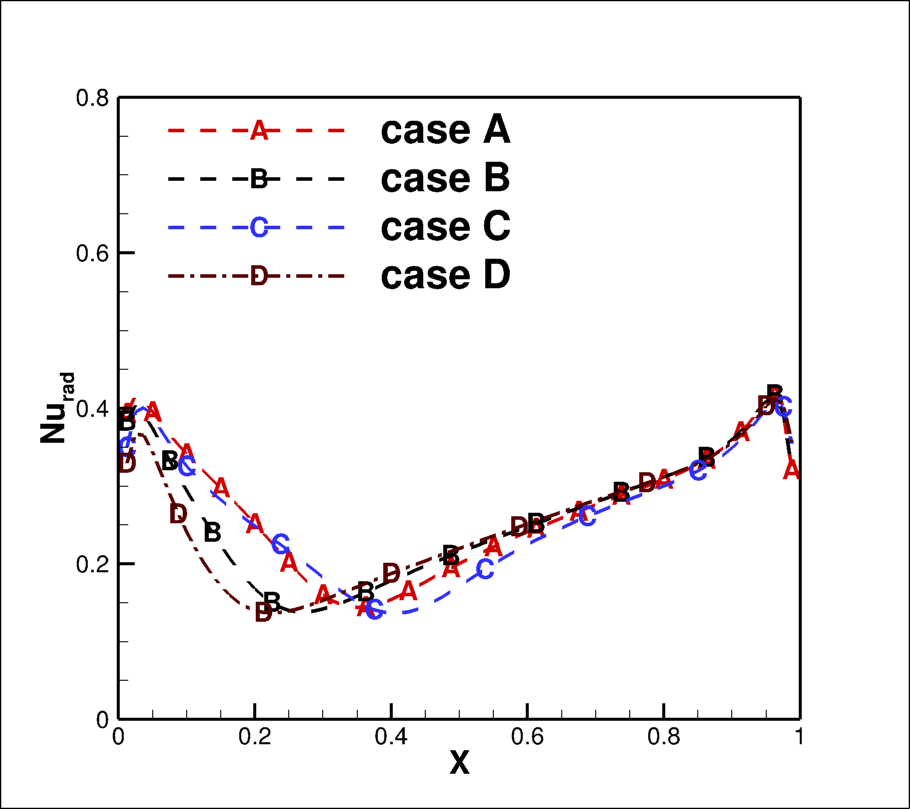}
    \caption{}
    \label{AsD_rad_NuN50}
  \end{subfigure}
  \caption{The variation of the radiation Nusselt number on the bottom wall for the different cases for (a) $Pl=0$  (b) $Pl=1$ (c) $Pl=10$ and (d) $Pl=50$}
\label{As_rad_bot_Nu}
\end{figure} 

The radiative Nusselt number distribution on the bottom wall for different cases and range of Planck numbers (Pl=0-50) is presented in fig \ref{As_rad_bot_Nu}. The radiative Nusselt number is almost zero over the length of the bottom wall except the length over which the collimated beam strikes. The shapes of radiative Nusselt number in the collimated beam strike zone is Gaussian for cases A and C, where it is almost square shape for cases B and D. The top of the square shape of radiative Nusselt number is not flat but wavy which has peak at the corner and trough at the middle. Moreover, the maximum value of radiative Nusselt number is almost same for all cases for Planck number zero (fig \ref{AsA_rad_NuN0}). The shapes of the radiative Nusselt number curve for different case are retained for Pl=1 (fig \ref{AsB_rad_NuN1}), however maximum value of radiative Nusselt number is monotonically increasing from case A to case D. One interesting fact to notice that head of these shapes are also tilted which has higher value at the start of the strike zone of the collimated beam and lower value at end of the beam. The peaks in the radiative Nusselt number curve disappear for cases A and B (fig \ref{AsC_rad_NuN10}) however, the shapes for case C and D becomes Gaussian where case D has higher value than the case C. Furthermore, these peaks disappear in cases C and D (see fig \ref{AsD_rad_NuN50}).

\begin{table}[!b]
\centering
\caption{The area average Nusselt number on the bottom wall for different cases for the range of Planck numbers}
\label{Avg_bot_Nu}
\begin{tabular}{|c|c|c|c|c|c|c|l|c|c|c|l|c|c|c|}
\hline
\multirow{3}{*}{Pl} 
& \multicolumn{3}{c|}{case A} & \multicolumn{4}{c|}{case B} & \multicolumn{4}{c|}{case C} & \multicolumn{3}{c|}{case D} \\ \cline{2-15} 
 & Cond & Rad & Tot & Cond & Rad & \multicolumn{2}{c|}{Tot} & Cond & Rad & \multicolumn{2}{c|}{Tot} & Cond & Rad & Tot \\ \hline
0 & 6.513 & -3.363 & 3.033 & 8.13 & -7.709 & \multicolumn{2}{c|}{0.421} & 7.126 & -3.296 & \multicolumn{2}{c|}{3.83} & 8.718 & -7.97 & 0.748 \\ \hline
1 & 5.712 & -0.712 & 5 & 6.332 & -2.835 & \multicolumn{2}{c|}{3.497} & 6.422 & -1.867 & \multicolumn{2}{c|}{4.555} & 7.021 & -4.882 & 2.139 \\ \hline
10 & 5.643 & 0.610 & 6.254 & 5.546 & 0.574 & \multicolumn{2}{c|}{6.12} & 5.632 & 0.118 & \multicolumn{2}{c|}{5.75} & 5.154 & 0.338 & 5.492 \\ \hline
50 & 5.845 & 0.264 & 6.109 & 5.653 & 0.254 & \multicolumn{2}{c|}{5.907} & 5.663 & 0.066 & \multicolumn{2}{c|}{5.699} & 5.513 & 0.248 & 5.761 \\ \hline
\end{tabular}
\end{table}

Table \ref{Avg_bot_Nu} shows the area average Nuselt number on the bottom wall of the cavity. The average conduction Nusselt number decreases on the bottom wall upto $Pl=10$ then there is slight increase in the average Nusselt number whereas area average radiative Nusselt number first being negative for ($Pl=0$ and 1) then becomes positive for ($Pl=10$ and 50) for all cases. Furthermore total average Nusselt number increases upto $Pl=10$ and then there is minimal decrease in the average Nusselt number for all cases. The minimum total average value found for $Pl=0$ and for case B and the maximum total Nusselt number is found for $Pl=10$ for case A on the bottom wall. 

\begin{table}[!t]
\centering
\caption{The area average Nusselt number for on the right wall for different cases for the range of Planck numbers}
\label{Avg_right_Nu}
\begin{tabular}{|c|c|c|c|c|c|c|l|c|c|c|l|c|c|c|}
\hline
\multirow{3}{*}{Pl} 
 & \multicolumn{3}{c|}{case A} & \multicolumn{4}{c|}{case B} & \multicolumn{4}{c|}{case C} & \multicolumn{3}{c|}{case D} \\ \cline{2-15} 
 & Cond & Rad & Tot & Cond & Rad & \multicolumn{2}{c|}{Tot} & Cond & Rad & \multicolumn{2}{c|}{Tot} & Cond & Rad & Tot \\ \hline
0 & -3.380 & -1.045 & -4.425 & -4.385 & -1.569 & \multicolumn{2}{c|}{-5.954} & -3.641 & -0.939 & \multicolumn{2}{c|}{-4.58} & -4.354 & -1.267 & -5.621 \\ \hline
1 & -3.831 & -0.940 & -4.771 & -5.265 & -1.287 & \multicolumn{2}{c|}{-6.552} & -3.899 & -0.892 & \multicolumn{2}{c|}{-4.791} & -4.863 & -1.097 & -5.96 \\ \hline
10 & -3.619 & -0.499 & -4.118 & -4.924 & -0.641 & \multicolumn{2}{c|}{-5.565} & -3.298 & -0.44 & \multicolumn{2}{c|}{-3.738} & -4.397 & -0.59 & -4.987 \\ \hline
50 & -3.046 & -0.142 & -3.188 & -3.033 & -0.141 & \multicolumn{2}{c|}{-3.174} & -3.18 & -0.144 & \multicolumn{2}{c|}{-3.324} & -2.956 & -0.135 & -3.091 \\ \hline
\end{tabular}
\end{table}

\begin{table}[!b]
\centering
\caption{The area average Nusselt number on the left wall for different Planck numbers}
\label{Avg_left_Nu}
\begin{tabular}{|c|c|c|c|c|c|c|l|c|c|c|l|c|c|c|}
\hline
\multirow{3}{*}{Pl} 
& \multicolumn{3}{c|}{case A} & \multicolumn{4}{c|}{case B} & \multicolumn{4}{c|}{case C} & \multicolumn{3}{c|}{case D} \\ \cline{2-15} 
 & Cond & Rad & Tot & Cond & Rad & \multicolumn{2}{c|}{Tot} & Cond & Rad & \multicolumn{2}{c|}{Tot} & Cond & Rad & Tot \\ \hline
0 & -2.258 & -0.688 & -2.946 & -1.875 & -0.669 & \multicolumn{2}{c|}{-2.544} & -3.046 & -0.764 & \multicolumn{2}{c|}{-3.81} & -3.136 & -0.694 & -3.83 \\ \hline
1 & -3.424 & -0.814 & -4.238 & -2.984 & -0.666 & \multicolumn{2}{c|}{-3.65} & -3.251 & -0.767 & \multicolumn{2}{c|}{-4.018} & -3.05 & -0.691 & -3.74 \\ \hline
10 & -3.92 & -0.574 & -4.494 & -3.313 & -0.407 & \multicolumn{2}{c|}{-3.72} & -3.621 & -0.443 & \multicolumn{2}{c|}{-4.055} & -3.433 & -0.429 & -3.862 \\ \hline
50 & -3.188 & -3.361 & -0.138 & -2.831 & -0.126 & \multicolumn{2}{c|}{-2.957} & -3.272 & -0.162 & \multicolumn{2}{c|}{-3.439} & -2.586 & -0.141 & -2.727 \\ \hline
\end{tabular}
\end{table}

The area average Nusselt number conduction, radiation and total on the right and left walls with exclusion of the semitransparent window are presented in Table \ref{Avg_right_Nu} and \ref{Avg_left_Nu}, respectively. Both the conduction and the raidiative Nusselt numbers are negative on both the wall for all cases and all Planck numbers. The average conduction Nusselt number increases upto Planck number $Pl=1$ on the right wall and $Pl=10$ on the left wall for all cases. However, average radiation Nusselt number monotonically decreases with Planck number for all the cases. The mximum total Nusselt number is found for Pl=1 and case B, whereas minimum total Nusselt number is found for $Pl=50$ and case D on the right wall, similar maximum and minimum total Nusselt numbers are found for $Pl=10$.

\begin{table}[!t]
\centering
\caption{The area average Nusselt number on the semitransparent window for different Planck numbers}
\label{Avg_semi_Nu}
\begin{tabular}{|c|c|c|c|c|c|c|l|c|c|c|l|c|c|c|}
\hline
\multirow{3}{*}{Pl} 
& \multicolumn{3}{c|}{case A} & \multicolumn{4}{c|}{case B} & \multicolumn{4}{c|}{case C} & \multicolumn{3}{c|}{case D} \\ \cline{2-15} 
 & Cond & Rad & Tot & Cond & Rad & \multicolumn{2}{c|}{Tot} & Cond & Rad & \multicolumn{2}{c|}{Tot} & Cond & Rad & Tot \\ \hline
0 & -0.816 & 5.152 & 4.336 & -1.848 & 9.94 & \multicolumn{2}{c|}{8.092} & -0.537 & 5.102 & \multicolumn{2}{c|}{4.565} & -1.351 & 10.055 & 8.704 \\ \hline
1 & -1.194 & 5.182 & 3.988 & -3.2. & 9.953 & \multicolumn{2}{c|}{6.723} & -0.862 & 5.12 & \multicolumn{2}{c|}{4.258} & -2.558 & 10.123 & 7.564 \\ \hline
10 & -2.738 & 5.195 & 2.457 & -6.757 & 9.943 & \multicolumn{2}{c|}{3.186} & -3.01 & 5.166 & \multicolumn{2}{c|}{2.036} & -6.87 & 10.23 & 3.36 \\ \hline
50 & -4.489 & 5.166 & 0.677 & -9.858 & 10.169 & \multicolumn{2}{c|}{0.311} & -4.197 & 5.265 & \multicolumn{2}{c|}{1.068} & -10.375 & 10.435 & 0.062 \\ \hline
\end{tabular}
\end{table}

The area average conduction, radiation and total Nusselt number on the semitransparent wall for all the cases and range of Planck numbers is depicted in Table \ref{Avg_semi_Nu}. The conduction Nusselt number is always negative and increases with increase of the Planck number of the medium for all cases. While the radiative Nusselt number is positive and almost constant for all Planck numbers in each case. One thing to notice that radiative Nusselt number is same for case A and C and similarly it is same for cases B and D. The maximum conduction Nusselt number is found for case D for Planck number $Pl=50$. The maximum and minimum total Nusselt number on the semitransparent wall are found for case D for $Pl=0$ and $Pl=50$, respectively.

\section{Conclusions}
The effects of the semitransparent window's aspect ratio on the interaction of collimated beam with natural convection have been studied numnerically in a square cavity which is heated from the bottom for Ra=$10^5$, Pr=0.7. The four combination of height ratio ($h_r$) and window width ratio ($w_r$) and range of Planck numbers have been considered. A collimated beam is irradiated on this semitransparent window at an azimuthal angle $135^0$ and the interaction of this collimated beam irradiation with natural convection is studies. The following conclusions are drawn:
\begin{enumerate}
     \item The left vortex is bigger than the right vortex for case A and B for transparent medium $Pl=0$, Furthermore, the fluid flow turns almost right angle turn happens for right vortex at the junction of two vortices for case C and D.
     \item The left vortex changes its dynamics with Planck number of the participating medium. It breaks into two parts for cases B, C and D for Planck number $Pl=10$. Moreover, the left vortex remains confined into lower left corner for case B and breaks into two part for case D for Planck number 50.
     \item The thermal plume flickers right to left for medium changes from non-participating to participating medium for all cases except Planck number $Pl=50$ for case C where thermal plume is bent to right.
     \item The local heating of the fluid occurs by collimated beam for Planck number $Pl=0$ and 10 for case B, and $Pl=10$ for case A.
     \item The temperature rise on the bottom wall at the location of collimated incidence happens in increasing order for case A to case D for transparent medium case and this rise diminishes also in the same order for Planck number $Pl=1$, and finally no major rise in the temperature at the bottom wall appear for $Pl=50$.
     \item The maximum non-dimensional temperature inside the cavity increase beyond, and maximum non-dimensional temperature is found for $Pl=0$ case D and minimum for $Pl=10$ case A.
     \item The conduction Nusselt number curve at the collimated incidence location on the bottom wall is Gaussian for case A and square for case D for transparent medium. The rise diminishes fast and no such rise in the conduction Nusselt number is seen for all cases for Pl=50.
     \item The radiative Nusselt number rise at the collimated beam incidence on the bottom wall is almost same for all cases for transparent medium case. However this curve is Gaussian for case A and C and square for B and D, nevertheless wavy with peak at the ends and trough at the middle for square distribution.
     \item The peak in radiation Nusselt number diminishes fast from case D to case A for $Pl=1$ and no peak appear for $Pl=50$.
     \item The maximum Nusselt number is 5.76 is found for $Pl=50$ for case D and minimum total Nusselt number i.e., 0.42 for $Pl=0$ for case B on the bottom wall.
\end{enumerate}

\bibliographystyle{asmems4}

\bibliography{asme2e}

\end{document}